\pdfoutput=1

\PassOptionsToPackage{final}{graphicx}
\documentclass[oribibl,envcountsame,final]{elsarticle}

\usepackage[utf8]{inputenc}
\usepackage[colorlinks,linkcolor={blue},citecolor={blue},urlcolor={red},breaklinks]{hyperref}

\newcommand{\diag}{\Delta}
\usepackage{etex}
\usepackage{amssymb,stmaryrd,amsthm} 
\usepackage[nosumlimits]{amsmath} 
\usepackage{xcolor} 
\usepackage{wasysym}
\usepackage{stackengine}
\usepackage{mathtools}

\usepackage{enumitem}

\setlist{itemsep=-.15ex}

\usepackage{lipsum}

\usepackage{makeidx} 

\usepackage{proof}
\usepackage{bussproofs}
\usepackage{fitch}

\nddim{4.5ex}{3.5ex}{1.5ex}{1em}{.5em}{.5em}{2.5em}{.2mm}

\usepackage{rotating}

\usepackage{caption}
\usepackage{subcaption}
\usepackage{wrapfig}

\usepackage{relsize}

\usepackage{tikz}

\usetikzlibrary{arrows,tikzmark}

\usetikzlibrary{decorations.pathreplacing}
\tikzset{%
  show curve controls/.style={
    postaction={
      decoration={
        show path construction,
        curveto code={
          \draw [blue]
            (\tikzinputsegmentfirst) -- (\tikzinputsegmentsupporta)
            (\tikzinputsegmentlast) -- (\tikzinputsegmentsupportb);
          \fill [red, opacity=0.5]
            (\tikzinputsegmentsupporta) circle [radius=.5ex]
            (\tikzinputsegmentsupportb) circle [radius=.5ex];
        }
      },
      decorate
}}}

\makeindex 

\let\oldemph\emph
\renewcommand{\emph}[1]{\oldemph{\index{#1}#1}}

\providecommand{\catname}{\mathbf} 
\providecommand{\clsname}{\mathcal}
\providecommand{\oname}[1]{\operatorname{\mathsf{#1}}}

\def\defcatname#1{\expandafter\def\csname B#1\endcsname{\catname{#1}}}
\def\defcatnames#1{\ifx#1\defcatnames\else\defcatname#1\expandafter\defcatnames\fi}
\defcatnames ABCDEFGHIJKLMNOPQRSTUVWXYZ\defcatnames

\def\defclsname#1{\expandafter\def\csname C#1\endcsname{\clsname{#1}}}
\def\defclsnames#1{\ifx#1\defclsnames\else\defclsname#1\expandafter\defclsnames\fi}
\defclsnames ABCDEFGHIJKLMNOPQRSTUVWXYZ\defclsnames

\def\defbbname#1{\expandafter\def\csname BB#1\endcsname{\mathbb{#1}}}
\def\defbbnames#1{\ifx#1\defbbnames\else\defbbname#1\expandafter\defbbnames\fi}
\defbbnames ABCDEFGHIJKLMNOPQRSTUVWXYZ\defbbnames

\def\Set{\catname{Set}}

\providecommand{\argument}{\operatorname{-\!-}}

\providecommand{\ul}{\underline}			                     %

\DeclareOldFontCommand{\bf}{\normalfont\bfseries}{\mathbf}
\providecommand{\mplus}{{\scriptscriptstyle\bf+}} 	       %

\providecommand{\PSet}{{\mathcal P}}			                 %
\providecommand{\FSet}{{\mathcal P}_{\omega}}		           %
\providecommand{\CSet}{{\mathcal P}_{\omega_1}}		         %
\providecommand{\NESet}{{\mathcal P}^{\mplus}}		         %
\providecommand{\Id}{\operatorname{Id}}

\providecommand{\Hom}{\mathsf{Hom}}
\providecommand{\id}{\mathsf{id}}
\providecommand{\op}{\mathsf{op}}
\providecommand{\comp}{\mathbin{\circ}}

\providecommand{\bang}{\operatorname!}				             %

\providecommand{\ito}{\hookrightarrow}				             %
\providecommand{\mto}{\mapsto}
\providecommand{\xto}[1]{\,\xrightarrow{#1}\,}

\providecommand{\dar}{\kern-1.2pt\operatorname{\downarrow}}	
\providecommand{\uar}{\kern-1.2pt\operatorname{\uparrow}}	

\providecommand{\fst}{\oname{fst}}
\providecommand{\snd}{\oname{snd}}
\providecommand{\pr}{\oname{pr}}

\providecommand{\brks}[1]{\langle #1\rangle}

\providecommand{\inl}{\oname{inl}}
\providecommand{\inr}{\oname{inr}}
\providecommand{\inj}{\oname{in}}

\DeclareSymbolFont{Symbols}{OMS}{cmsy}{m}{n}
\DeclareMathSymbol{\iobj}{\mathord}{Symbols}{"3B}
\providecommand{\curry}{\oname{curry}}
\providecommand{\uncurry}{\oname{uncurry}}

\usepackage{stmaryrd}

\providecommand{\lsem}{\llbracket}
\providecommand{\rsem}{\rrbracket}
\providecommand{\sem}[1]{\lsem #1 \rsem}

\providecommand{\comma}{,\operatorname{}\linebreak[1]}		 %

\providecommand{\by}[1]{\text{/\!\!/~#1}}			             %
\providecommand{\pacman}[1]{}					                     %

\newcommand{\undefine}[1]{\let #1\relax}					                       %

\providecommand{\noqed}{\def\qed{}}				                 %

\providecommand{\mone}{{\text{\kern.5pt\rmfamily-}\mathsf{\kern-.5pt1}}}

\makeatletter
\@ifpackageloaded{enumitem}{}{
  \usepackage[loadonly]{enumitem}                          %
}                                                          %
\makeatother

\newlist{citemize}{itemize}{1}
\setlist[citemize]{label=\labelitemi,wide} 

\newlist{cenumerate}{enumerate}{1}
\setlist[cenumerate,1]{label=\arabic*.~,ref={\arabic*},wide} 
\makeatletter
\def\mfix#1{\oname{#1}\@ifnextchar\bgroup\@mfix{}}	       %
\def\@mfix#1{#1\@ifnextchar\bgroup\mfix{}}			           %
\makeatother

\providecommand{\ift}[3]{\mfix{if}{\mathbin{}#1}{then}{\mathbin{}#2}{else}{\mathbin{}#3}}
\providecommand{\case}[3]{\mfix{case}{\mathbin{}#1}{of}{#2}{\kern-1pt;}{\mathbin{}#3}}

\def\defbbname#1{\expandafter\def\csname BB#1\endcsname{{\bm{\mathsf{#1}}}}}
\def\defbbnames#1{\ifx#1\defbbnames\else\defbbname#1\expandafter\defbbnames\fi}
\defbbnames ABCDEFGHIJKLMNOPQRSTUVWXYZ\defbbnames

\providecommand{\dist}{\oname{dist}}
\newcommand{\assoc}{\oname{assoc}}
\newcommand{\klstar}{\star}  				%
\newcommand{\istar}{\dagger}  				%
\newcommand{\iistar}{\ddagger}  			%
\newcommand{\out}{\operatorname{\mathsf{out}}}
\newcommand{\tuo}{\operatorname{\out^{\text{\kern.5pt\rmfamily-}\kern-.5pt1}\kern-1pt}}

\renewcommand{\comp}{\,}					%
\renewcommand{\c}{\colon}

\renewcommand{\fst}{\pr_1}	
\renewcommand{\snd}{\pr_2}

\renewcommand{\comma}{,\operatorname{}\linebreak[1]}		%

\newcommand{\VDash}{\vdash}
\newcommand{\ctx}[1]{\VDash_{#1}}

\newcommand{\cctx}{\ctx{\mathsf{c}}}
\newcommand{\vctx}{\ctx{\mathsf{v}}}

\DeclareRobustCommand{\pcase}[2]{\mfix{case\,}{#1}{\,of\,}{#2}}

\DeclareRobustCommand{\handle}[3]{\mfix{handle\,}{#1}{\,in\,}{#2}{\,with\,}{#3}}
\DeclareRobustCommand{\handleit}[3]{\mfix{handleit\,}{#2=#1}{\,in\,}{#3}}
\DeclareRobustCommand{\case}[3]{\mfix{case\,}{#1}{\,of\,}{#2}{\kern-1pt;\,}{#3}}

\usepackage{ifdraft} 

 \ifdraft
 {
 \usepackage[layout=footnote,draft]{fixme}
 \usepackage[notcite,notref]{showkeys}
 
 }{
\usepackage[layout=footnote,final]{fixme}
}

\FXRegisterAuthor{sg}{asg}{SG}	%
\FXRegisterAuthor{ls}{als}{LS}	%
\FXRegisterAuthor{cr}{acr}{CR}	%

\theoremstyle{plain}%

\newtheorem{theorem}{Theorem}
\newtheorem{lemma}[theorem]{Lemma}
\newtheorem{proposition}[theorem]{Proposition}

\theoremstyle{definition}
\newtheorem{definition}[theorem]{Definition}
\newtheorem{example}[theorem]{Example}

\newtheorem{notation}[theorem]{Notation}

\makeatletter
\newcommand{\superimpose}[2]{%
  {\ooalign{$#1\@firstoftwo#2$\cr\hfil$#1\@secondoftwo#2$\hfil\cr}}}
\makeatother

\newcommand{\grd}{\operatorname{\diamondsuit}}

\newcommand{\lrule}[3]{\textbf{#1}~~\frac{#2}{#3}}

\usepackage{bbold}

\newcommand{\Act}{\mathsf{Act}}
\newcommand{\nat}{\bm{\mathbb{N}}}

\newcommand{\Put}{\operatorname{\mathit{put}}}
\newcommand{\Succ}{\operatorname{\mathit{succ}}}
\newcommand{\Pred}{\operatorname{\mathit{pred}}}
\newcommand{\zero}{\operatorname{\mathit{zero}}}

\newcommand{\gcase}[6]{\oname{gcase} #1(#2)\,\oname{of \inl} {#3}\mto {#4};\,\inr {#5}\mto #6}

\newcommand\pp{
\mathchoice{\mathbin{\raisebox{1pt}{${\scriptstyle\bf+}\mkern-08mu{\scriptstyle\bf+}$}}}
           {\mathbin{\raisebox{1pt}{${\scriptstyle\bf+}\mkern-08mu{\scriptstyle\bf+}$}}}
           {\mathbin{\raisebox{1pt}{${\scriptscriptstyle\bf+}\mkern-07mu{\scriptscriptstyle\bf+}$}}}
           {\mathbin{\raisebox{1pt}{${\scriptscriptstyle\bf+}\mkern-07mu{\scriptscriptstyle\bf+}$}}}
}

\newcommand{\str}{\mathit{Str}}

\newcommand{\gtag}{{\oname{g}}}
\newcommand{\utag}{{\oname{u}}}

\newcommand{\infrule}[2]{\frac{#1}{#2}}
\newcommand{\anonrule}[3]{\infrule{#2}{#3}}

\usepackage{todos}

\usepackage{savesym}

\savesymbol{degree}
\savesymbol{leftmoon}
\savesymbol{rightmoon}
\savesymbol{fullmoon}
\savesymbol{newmoon}
\savesymbol{diameter}
\savesymbol{emptyset}
\savesymbol{bigtimes}
\savesymbol{triangleright}

\usepackage{bm}
\usepackage[matha]{mathabx}
\restoresymbol{other}{emptyset}
\restoresymbol{other}{triangleright}

\usepackage{tikz}
\usepackage{tkz-euclide}
\usepackage{tikz-cd}

\tikzset{
    commutative diagrams/.cd,
    arrow style=tikz,
    diagrams={>=stealth},
    row sep=large, 
    column sep = huge
}

\newcommand{\IHom}{\Hom^{\kern-.2pt\scalebox{.58}{$\grd$}}}
\newcommand{\GHom}{\Hom^{\kern-1pt\bullet}}

\newcommand{\mbind}[2]{\mfix{do}{#1;}{}{#2}}
\newcommand{\ret}{\oname{ret}}

\newcommand\rsmraise[1]{%
  \ifx#1\displaystyle .8\else
    \ifx#1\textstyle .8\else
      \ifx#1\scriptstyle .6\else
        .45%
      \fi
    \fi
  \fi}

\newcommand{\cpto}{
  \mathrel{\raisebox{0.5ex}{\kern3pt\ensuremath{\mathrel{\tikz{ \draw [-stealth,line width=0.4] (0.6ex,1ex) -- (0,1ex) -- (0,0.4ex) -- (2.2ex,0.4ex); }}}\kern3pt}}
}

\newcommand{\rept}[3]{\mfix{case\ inr}{\mathbin{}#1}{\,\gets\,}{#2}{\kern-1pt; repeat\kern1pt}{\mathbin{}#3}}

\usepackage[final]{listings}

\newcommand*{\Comment}[1]{\hfill\makebox[4.5cm][l]{\coco{/\!\!/ \textit{#1}}}}%

\usepackage{textcomp}

\definecolor{rules}{rgb}{0.8,0.4,0.23}
\definecolor{propositions}{rgb}{0.46,0.40,0.9}
\definecolor{assertions}{rgb}{0.10,0.5,0.36}

\newcommand{\ruco}{\textcolor{rules}}

\newcommand{\vaco}{\textcolor{magenta}}
\newcommand{\coco}{\textcolor{gray}}
\newcommand{\blco}{\textcolor{black}}
\newcommand{\asco}{\textcolor{assertions}}

\newcommand\csep{
\mathbin{
\mathchoice{\raisebox{.2ex}{\kern.5pt\scalebox{1}[.75]{$\mid$}}}
           {\raisebox{.2ex}{\kern.5pt\scalebox{1}[.75]{$\mid$}}}
           {\raisebox{.15ex}{\scalebox{1.5}[.5]{$\mid$}}}
           {\raisebox{.15ex}{\scalebox{1.5}[.5]{$\mid$}}}
}}

\renewcommand{\ul}[1]{\mkern2mu\underline{\mkern-2mu #1\mkern-2mu}\mkern2mu } %

\usepackage{scalerel}
\newcommand\scslash{\stretchrel*{$/$}{\textsc{H}}}

\newcommand{\ssl}[2]{#1 \mathbin{\scslash\hspace{-.3em}\scslash} #2}

\title{A Metalanguage for Guarded Iteration\tnoteref{dfg}\tnoteref{ext}}
\tnotetext[dfg]{Work forms part of the DFG-funded project \emph{A High Level Language for Programming and Specifying Multi-Effect Algorithms (HighMoon2, 
SCHR~1118/8-2, GO~2161/1-2)}}
\tnotetext[ext]{This article is a revised version of~\cite{GoncharovRauchEtAl18}.}

\author{Sergey Goncharov, Christoph Rauch and Lutz Schröder}
\address{%
  Friedrich-Alexander-Universität Erlangen-Nürnberg}

\begin{document}\allowdisplaybreaks{}
\begin{abstract}
Notions of guardedness serve to delineate admissible recursive
definitions in various settings in a compositional manner. In recent
work, we have introduced an axiomatic notion of guardedness in
symmetric monoidal categories, which serves as a unifying framework
for various examples from program semantics, process algebra, and
beyond. In the present paper, we propose a generic
\emph{metalanguage for guarded iteration} based on combining this
notion with the fine-grain call-by-value paradigm, which we intend
as a unifying programming language for guarded and unguarded
iteration in the presence of computational effects. We give a
generic (categorical) semantics of this language over a suitable
class of strong monads supporting guarded iteration, and show it to
be in touch with the standard operational behaviour of iteration by
giving a concrete big-step operational semantics for a certain
specific instance of the metalanguage and establishing soundness and 
(computational) adequacy for
this case.
\end{abstract}


\maketitle

\section{Introduction}\label{sec:intro}
\noindent Guardedness is a recurring theme in programming and
semantics, fundamentally distinguishing the view of computations as
processes unfolding in time from the view that identifies computations
with a final result they may eventually produce. Historically, the
first perspective is inherent to process algebra
(e.g.~\cite{Milner89}), where the main attribute of a process is its
\emph{behaviour}, while the second is inherent to classical
denotational semantics via domain theory~\cite{Winskel93}, where the
only information properly infinite computations may communicate to the
outer world is the mere fact of their divergence. This gives rise to a
distinction between \emph{intensional} and \emph{extensional}
paradigms in semantics~\cite{Abramsky14}.

For example, in CCS~\cite{Milner89} a process is guarded in a variable $x$ if
every occurrence of $x$ in this process is preceded by an action. One effect of
this constraint is that guarded recursive specifications can be solved uniquely,
e.g.\ the equation $x = \bar a.\, x$, whose right-hand side is guarded in~$x$,
has the infinite stream $\bar a.\bar a.\ldots$ as its unique solution. If we
view $\bar a$ as an action of producing an output, we can also view the process
specified by $x = \bar a.\, x$ as \emph{productive} and the respective solution
$\bar a.\bar a\ldots$ as a \emph{trace} obtained by collecting its outputs. The
view of guardedness as productivity is pervasive in programming and reasoning
with coinductive types~\cite{Coquand94,Gimenez95,Gimenez98,HancockSetzer05}
as implemented in dependent type environments such as Coq
and Agda. Semantic models accommodate this idea in various
ways, e.g.\ from a modal~\cite{Nakano00,AppelMelliesEtAl07,MiliusLitak17},
(ultra-)metric~\cite{Escardo99,KrishnaswamiBenton11}, and a unifying
topos-theoretic perspective~\cite{BirkedalMgelbergEtAl12,CloustonBizjakEtAl16}.

In recent work, we have proposed a new \emph{axiomatic} approach to unifying
notions of guardedness~\cite{GoncharovSchroderEtAl17,GoncharovSchroder18},
where the main idea is to provide an \emph{abstract} notion of guardedness
applicable to a wide range of (mutually irreducible) models, including, e.g.,
complete partial orders, complete metric spaces, and infinite-dimensional
Hilbert spaces, instead of designing a concrete model carrying a specific
notion of guardedness. A salient feature of axiomatic guardedness is that it
varies in a large spectrum starting from \emph{total guardedness} (everything
is guarded) and ending at \emph{vacuous guardedness} (very roughly, guardedness
in a variable means essentially non-occurrence of this variable in the defining
expression) with proper examples as discussed above lying between these two
extremes. The fact that axiomatic guardedness can be varied so broadly indicates
that it can be used for bridging the gap between the intensional and extensional
paradigms, which is indeed the perspective we are pursuing here by introducing a
\emph{metalanguage for guarded iteration}.

\begin{figure}[t!]
\begin{lstlisting}[columns=fullflexible, escapechar=\#] %

handle r in
  (handleit e = * in                                   #\Comment{start a loop}#
    print ("think of a number") &                      #\Comment{execute the loop guard}#
      (do y <- rand(); 
          z <- read();
          if (y = 42) then raise_r* else              #\Comment{42 is the ultimate answer}#
          if (z = y) then ret* else raise_e*))       #\Comment{continue, unless}#
    #\Comment{number guessed correctly}#
with print ("the answer!")  
\end{lstlisting}
\caption{Example of a guarded loop.}
\label{fig:intro}
\end{figure}

The developments in~\cite{GoncharovSchroder18} are couched in terms of
a special class of monoidal categories called \emph{guarded traced
  symmetric monoidal categories}, equipped with a monoidal notion of
guardedness and a monoidal notion of feedback allowing only such
cyclic computations that are guarded in the corresponding sense. In
the present work we explore a refinement of this notion by
instantiating guarded traces to \emph{Kleisli categories} of
computational monads in the sense of Moggi~\cite{Moggi91}, with coproduct
(inherited from the base category) as the monoidal structure. The
feedback operation is then equivalently given by \emph{guarded
  effectful iteration}, i.e.\ a (partial) operator
\begin{align}
  \label{eq:elgot-iter}
  \anonrule{}{f\c X\to T(Y+X)}{f^{\istar}\c X\to TY}    
\end{align}
to be thought of as iterating~$f$ over~$X$ until a result in~$Y$ is
reached~\cite{GoncharovSchroderEtAl17}.  As originally argued by
Moggi, strong monads can be regarded as representing computational
effects, such as nondeterminism, exceptions, or process algebra
actions, and thus the corresponding internal language of strong
monads, the \emph{computational metalanguage}~\cite{Moggi91}, can be
regarded as a generic programming language over these effects. We
extend this perspective by parametrizing such a language with a notion
of guardedness and equipping it with guarded iteration.  In doing so,
we follow the approach of Geron and Levy~\cite{GeronLevy16} who
already explored the case of unguarded iteration by suitably extending
a fine-grain call-by-value language~\cite{LevyPowerEtAl02}, a refined
variant of Moggi's original computational $\lambda$-calculus.

A key insight we borrow from~\cite{GeronLevy16} is that effectful iteration
can be efficiently organized via throwing and handling exceptions (also
called \emph{labels} in this context) in a loop, leading to a more convenient
programming style in comparison to the one directly inspired by the typing
of the iteration operator~\eqref{eq:elgot-iter}. We show that the exception
handling metaphor seamlessly extends to the guarded case and is compatible
with the axioms of guardedness. A quick illustration is presented in
Fig.~\ref{fig:intro} where the $\oname{handleit}$ command implements a loop
in which the $\oname{raise}$ command indexed with the corresponding exception $e$
identifies the tail call. The $\mathit{print}$ operation acts as a guard and
makes the resulting program well-typed. We also involve two operations
$\mathit{rand}$ and $\mathit{read}$ for random number generation and for
reading a user input from the console correspondingly. Apart from the non-standard 
use of exceptions via the $\oname{handleit}$ construct, they can be processed 
in a standard way with the $\oname{handle}$ command, and therefore in the example,
we can break from the loop by throwing exception $r$ when the random number 
appears to be $42$ (the answer to the ultimate question of life, the universe, 
and everything).

To interpret our metalanguage we derive and
explore a notion of \emph{strong guarded iteration} and give a generic
(categorical) denotational semantics, for which the main subtlety are
functional abstractions of guarded morphisms. We then define a
big-step operational semantics for a concrete (simplistic) instance of
our metalanguage and show an adequacy result w.r.t.\ a concrete choice
of the underlying category and the strong monad.
 
\paragraph{Related work}
We have already mentioned work by Geron and
Levy~\cite{GeronLevy16}. The instance of operational semantics we
explore here is chosen so as to give the simplest proper example of
guarded iteration, i.e.\ the one giving rise to infinite traces,
making the resulting semantics close to one explored in a line of work
by Nakata and Uustalu~\cite{NakataUustalu10,NakataUustalu10a,
  Nakata11,NakataUustalu15}. We regard our operational semantics as a
showcase for the denotational semantics, and do not mean to address
the notorious issue of undecidability of program termination, which is
the main theme of Nakata and Uustalu's work. We do however see our
work as a stepping stone both for deriving more sophisticated styles
of operational semantics and for developing concrete denotational
models for addressing the operational behaviour as discussed in
op.cit. The \emph{guarded
  $\lambda$-calculus}~\cite{CloustonBizjakEtAl16} is a recently
introduced language for guarded recursion (as apposed to guarded
iteration), on the one hand much more expressive than ours, but on the
other hand capturing a very concrete model, the \emph{topos of
  trees}~\cite{BirkedalMgelbergEtAl12}. 

This paper extends a previous conference
publication~\cite{GoncharovRauchEtAl18} by giving full proofs and additional
explanations and example material. Also, we consolidate the treatment of
iteration-in-context by showing the necessity of conditions relating
the strength to guardedness and iteration
(Theorem~\ref{thm:slice-converse}). The version of the metalanguage we present here
(Fig.~\ref{fig:lang.full}) improves slightly on the original conference
version by modifying the formation rules for $\oname{gcase}$ and $\oname{handle}$;
this, in particular, allows us to type more terms, and handle ``unguarded exceptions''.

\paragraph{Plan of the paper}
In Section~\ref{sec:monad} we give the necessary technical
preliminaries, and discuss and complement the semantic foundations for
guarded iteration~\cite{GoncharovSchroderEtAl17,GoncharovSchroder18}.
In Sections~\ref{sec:meta} and~\ref{sec:deno} we present our
metalanguage for guarded iteration (without functional types) and its
generic denotational semantics.  In Section~\ref{sec:expo} we identify
conditions for interpreting functional types and extend the
denotational semantics to this case. In Section~\ref{sec:adeq} we
consider an instance of our metalanguage (for a specific choice of
signature), give a big-step operational semantics and prove a
corresponding adequacy result. Conclusions are drawn in Section~\ref{sec:concl}.

\section{Monads for Effectful Guarded Iteration}\label{sec:monad}
\noindent We use the standard language of category theory~\cite{MacLane98}. Some conventions
regarding notation are in order. By $|\BC|$ we denote the class of objects 
of a category~$\BC$, and by $\Hom_{\BC}(A,B)$ (or~$\Hom(A,B)$, if no confusion arises)
the set of morphisms $f\c A\to B$ from $A\in |\BC|$ to $B\in|\BC|$. We tend to omit
object indices on natural transformations.

\paragraph{Coproduct summands and distributive categories} 
We call a pair
$\sigma=\brks{{\sigma_1\c Y_1\to X}\comma \sigma_2\c Y_2\to X}$ of
morphisms a \emph{summand} of $X$, denoted ${\sigma\c Y_1\cpto X}$, if
it forms a coproduct cospan, i.e.\ $X$ is a coproduct of $Y_1$
and~$Y_2$ with~$\sigma_1$ and~$\sigma_2$ as coproduct injections. Each
summand $\sigma=\brks{\sigma_1,\sigma_2}$ thus determines a
\emph{complement summand}
$\bar\sigma=\brks{\sigma_2,\sigma_1}\c Y_2\cpto X$.  We often identify a
summand $\brks{\sigma_1,\sigma_2}$ with its first component
when~$\sigma_2$ is predetermined canonically, clear from the context,
or irrelevant.  Summands of a given object~$X$ are naturally
preordered by taking $\brks{\sigma_1,\sigma_2}$ to be smaller than
$\brks{\theta_1,\theta_2}$ iff $\sigma_1$ factors
through~$\theta_1$. In the presence of an initial object $\iobj$, with
unique morphisms $\bang\c\iobj\to X$, this preorder has a greatest
element $\brks{\id_X,\bang}\c X \cpto X$ and a least element
$\brks{\bang,\id_X}\c \iobj \cpto X$. By writing $X_1+\ldots+X_n$ we designate the
latter as a coproduct of the $X_i$ and assign the canonical names
$\inj_i\c X_i\cpto X_1+\ldots+X_n$ to the corresponding
summands; if $\sigma\c Y_1 \cpto X_1, \vartheta\c Y_2 \cpto X_2$ are summands,
then so is $\sigma + \vartheta\c Y_1 + Y_2 \cpto X_1 + X_2$.
Dually to summands, we write $\pr_i\c X_1\times\ldots\times X_n\to X_i$
for canonical \emph{projections} (without introducing a special arrow
notation); by $\diag$ we abbreviate the diagonal natural
transformation $\brks{\id_A,\id_A}\c A\to A\times A$.
Note that in an \emph{extensive category}~\cite{CarboniLackEtAl93}, the second
component of any coproduct summand $\brks{\sigma_1,\sigma_2}$ is determined by
the first up to isomorphism. However, we do not generally assume extensiveness,
working instead with the weaker assumption of distributivity~\cite{Cockett93}: a
category with finite products and coproducts (including a final and an initial
object) is \emph{distributive} if the natural transformation
\begin{align*}
X\times Y + X\times Z \xrightarrow{~[\id\times\inj_1,\id\times\inj_2]~} X\times (Y+Z) 
\end{align*} 
is an isomorphism, whose inverse we denote by $\dist_{X,Y,Z}$, or
usually just~$\dist$. Then~$\dist$ is natural in $X,Y,Z$, and moreover
compatible with the coproduct structure in the expected sense; in
particular,
\begin{flalign*}
  \dist\comp(\id\times\inj_1)& =\inj_1\\
  \dist\comp(\id\times\inj_2)& =\inj_2\\
  [h\times f,h\times g]\comp\dist & = h\times[f,g]
\end{flalign*}
for $h\c X\to U$, $f\c Y\to W$, $g\c Z\to W$. In proofs, we summarily refer to such
properties by the keyword \emph{distributivity}.

\paragraph{Strong monads} Following Moggi~\cite{Moggi91}, we identify
a \emph{monad} $\BBT$ on a category~$\BC$ with the corresponding
\emph{Kleisli triple} $(T,\eta,(\argument)^\klstar)$ on $\BC$
consisting of an endomap~$T$ on $|\BC|$, a $|\BC|$-indexed class of
morphisms $\eta_X\c X\to TX$, called the \emph{unit} of $\BBT$, and the
\emph{Kleisli lifting} maps
$(\argument)^\klstar\c\Hom(X,TY)\to\Hom(TX,TY)$ such that
\begin{align*} 
\eta^{\klstar}=\id&& f^{\klstar}\eta=f&& (f^{\klstar} g)^{\klstar}=f^{\klstar}g^{\klstar} .
\end{align*}  
These definitions imply that $T$ is an endofunctor (with
$Tf=(\eta f)^\klstar$) and $\eta$ is a natural transformation.
Provided that $\BC$ has finite products, a monad~$\BBT$ on $\BC$ is
\emph{strong} if it is equipped with \emph{strength}, i.e.\ a
natural transformation
$
\tau_{X,Y}\c X\times TY\to T(X\times Y)
$
satisfying the following standard coherence conditions~(e.g.~\cite{Moggi91}):
\begin{equation*}
\begin{tikzcd}[column sep = 3ex,row sep = 4ex]
(X\times Y)\times TZ\dar["\assoc"']\ar[rr,"\tau"] &[3ex] &  T((X\times Y)\times Z)\dar["T\assoc"]\\
X\times (Y\times TY)\rar["\id\times\tau"] & X\times T(Y\times Z)\rar["\tau"] &  T(X\times (Y\times Z))
\end{tikzcd}\\[-.1ex]
\end{equation*}
\begin{equation*}
\begin{tikzcd}[column sep = 3ex,row sep = 4ex]
X\times TY\dar["\tau"']\rar["\snd"] &  TY\\
T(X\times Y)\urar["T\snd"'] &  
\end{tikzcd}
\hspace{1ex}
\begin{tikzcd}[column sep = 2ex,row sep = 4ex]
X\times Y\dar["\id\times\eta"']\rar["\eta"] &  T(X\times Y)\\
X\times TY\urar["\tau"'] &  
\end{tikzcd}
\hspace{1ex}
\begin{tikzcd}[column sep = 4ex,row sep = 4ex]
X\times TY\dar["\tau"']\ar[rr,"\id\times f^\klstar"] & &  X\times TZ\dar["\tau"]\\ %
T(X\times Y)\ar[rr,"(\tau(\id \times f))^\klstar"] & & T(X\times Z) %
\end{tikzcd}
\end{equation*}
where $f\c Y\to TZ$.

Morphisms of the form $f\c X\to TY$ constitute the \emph{Kleisli category} of
$\BBT$, which has the same objects as~$\BC$, units $\eta_X\c X\to TX$ as
identities, and composition $(f,g)\mto f^\klstar g$, also
called~\emph{Kleisli composition}.

In programming language semantics, both the strength~$\tau$ and the
distributivity transformation~$\dist$ essentially serve to propagate
context variables. We often need to combine them into
\begin{displaymath}
\delta=   
(T\dist)\comp\tau\c X\times T(Y+Z)\to T(X\times Y + X\times Z).
\end{displaymath}
In what follows we will make extensive use 
of the following simple property of~$\delta$:
\begin{align}\label{eq:d-delta}
  \delta\brks{\fst,\delta\brks{\fst,f}} = T(\brks{\fst,\id_{X\times Z}}+\brks{\fst,\id_{X\times W}})\comp \delta\comp\brks{\fst,f}.
\end{align}
for $f\c X\times Y\to T(Z+W)$ (where the morphisms in the equation have
type $X\times Y\to T(X\times X\times Z+X\times X\times W)$).

\begin{figure}[t!]
\begin{gather*}
\textbf{(trv)}~\frac{f\c X\to TY}{~(T\inj_1) f\c X\to_{\inj_2} T(Y+Z)~}\qquad
\textbf{(sum)}~\frac{~f\c X\to_\sigma TZ\quad~g\c Y\to_\sigma TZ}
{~[f,g]\c X+Y\to_\sigma TZ}\\[2.5ex]
\textbf{(cmp)}~\frac{~f\c X\to_{\inj_2} T(Y+Z)\qquad g\c Y\to_{\sigma} TV\qquad h\c Z\to TV~}{[g,h]^\klstar\comp f\c X\to_{\sigma} TV} \\[2ex]
\textbf{(str)}~\frac{~f\c X \to_{\sigma}TY %
~}{\tau \comp (\id_Z\times f)\c Z\times X \to_{\id \times \sigma} T(Z \times Y)} 
\end{gather*}
\caption{Axioms of abstract guardedness.}
\label{fig:guard}
\end{figure}

\paragraph{Guarded Iteration} Let us fix a distributive category $\BC$
and a strong monad $\BBT$ on~$\BC$. The monad $\BBT$ is
\emph{(abstractly) guarded} if it is equipped with a notion of
guardedness, i.e.\ with a relation between Kleisli morphisms
$f\c X\to TY$ and summands $\sigma\c Y'\cpto Y$ closed under the rules in
Fig.~\ref{fig:guard}, where $f\c X\to_\sigma TY$ denotes the fact that
$f$ and $\sigma$ are in the relation in question, in which case $f$ is also called
 \emph{$\sigma$-guarded}. We denote by
$\Hom_{\sigma}(X,TY)$ (or, more precisely, $\Hom_{\BC,\sigma}(X,TY)$)
the subset of $\Hom(X,TY)$ consisting of the morphisms
${X\to_\sigma TY}$. We also write $f\c X\to_i TY$ for
$f\c X\to_{\inj_i} TY$. More generally, we use the notation
$f\c X\to_{p,q,\ldots} TY$ to indicate guardedness in the union of
injections $\inj_p, \inj_q,\ldots$ where $p,q,\ldots$ are sequences
over $\{1,2\}$ identifying the corresponding coproduct summand
in~$Y$. For example, we write $f\c X\to_{12,2} T((Y+Z)+Z)$ to mean that
$f$ is $[\inj_1\inj_2,\inj_2]$-guarded.

The above formulation of the notion of guardedness is necessitated by the standard 
categorical view of (binary) coproducts as a \emph{property}: a binary coproduct
is any object that satisfies the corresponding universal property; therefore,
coproducts are defined up to isomorphism, and intrinsically refer to the specified coproduct 
injections. The alternative is to treat coproducts as a \emph{structure}, 
i.e.\ work with canonical coproducts. It is then also possible to adapt the formulation 
of guardedness and guarded iteration to comply with this view~\cite{LevyGoncharov19}.

The axioms~\textbf{(trv)}, \textbf{(sum)} and~\textbf{(cmp)} come
from~\cite{GoncharovSchroderEtAl17}. %
Intuitively, \textbf{(trv)} says that if a program does not output
anything via a summand of the output type then it is guarded in that
summand. Rule~\textbf{(cmp)} asserts that guardedness is preserved by
composition: if the unguarded part of the output of a program is
postcomposed with a $\sigma$-guarded program then the result is
$\sigma$-guarded, no matter how the guarded part is
transformed. Finally, rule~\textbf{(sum)} says that putting two
guarded equation systems side by side again produces a guarded system.
Here, we
also add the rule~\textbf{(str)} stating compatibility of guardedness
and strength.  Note that since~$\BC$ is distributive,
$\id_Z\times\sigma\c Z\times Y'\to Z\times Y$ is actually a summand whose 
canonical complement we take to be $\id_Z\times\bar\sigma$. 

Let us record some simple consequences of the axioms in Fig.~\ref{fig:guard}.
\begin{lemma}\label{lem:weak}
  The following rules are derivable:
\begin{equation*}
    \mathbf{(iso)}~\frac{f\c X \to_{\sigma} TY \qquad \vartheta\c Y \simeq
        Y'}{(T\vartheta) \comp f\c X \to_{\vartheta \comp \sigma} TY'}
\qquad 
    \mathbf{(wkn)}~\frac{~f\c X\to_{\sigma} TY~}{~f\c X\to_{\sigma\vartheta} TY~}\quad
\end{equation*}

\begin{equation*}
\mathbf{(cmp^\star)}~\frac{
f\c X\to_{\sigma+\id} T(Y+Z)\quad~ 
g\c Y\to TV\quad h\c Z\to TV\quad~ 
g\bar\sigma\c Y'\to_\vartheta TV}{[g,h]^\klstar f\c X\to_{\vartheta} TV}
\end{equation*}

\begin{equation*}
\mathbf{(cdm)}\quad\frac{~g\c X\to TY\qquad f\c Y\to_{\sigma} TZ~}{~f^\klstar g\c X\to_{\sigma} TZ~}
\end{equation*}
\end{lemma}
\begin{proof}
  The rule $\mathbf{(cdm)}$ is obtained from~$\mathbf{(cmp^\star)}$ by
  instantiating $Z$ with $\iobj$ and~$\sigma$ with~$\bang$.
  
  Let us show~\textbf{(iso)}. Let w.l.o.g.\ $Y_1 + Y_2 = Y$ and $f\c X \to_{\inj_2} T(Y_1 + Y_2)$,
  i.e.\ $\sigma = \inj_2$. Since $\vartheta$ is an isomorphism, we have
  $\vartheta = [\vartheta_1, \vartheta_2]\c Y_1 + Y_2 \to Y'$ and hence we
  derive
  \[
  \infer[\mathbf{(cmp)}]{[\eta \comp \vartheta_1, \eta \comp \vartheta_2]^{\klstar} \comp f :
    X \to_{\vartheta_2} TY'}{f\c  X \to_{\inj_2} T(Y_1 + Y_2)
    \quad\quad
    \eta \comp \vartheta_2\c Y_2 \to TY'
    \quad\quad
    \infer[\mathbf{(trv)}]{(T\vartheta_1) \comp \eta\c Y_1 \to_{\vartheta_2}
      TY'}{\eta\c Y_1 \to TY_1}
    \quad
  }
  \]
  Next, we check~$\mathbf{(cmp^\star)}$. Let w.l.o.g.\
  $\sigma = \inj_2$ and $\vartheta = \inj_2$.  Note that
  by~$\mathbf{(iso)}$,
  $(T\oname{assoc}^\mone)\comp f\c X\to_2 T(Y'+(Y''+Z))$ where
  $\oname{assoc}$ is the associativity isomorphism
  $Y'+(Y''+Z)\cong (Y'+Y'')+Z$.
  Then
  \[ [g,h]^\klstar f = [[g\comp\bar\sigma,g\comp\sigma],h]^\klstar f =
  [g\comp\bar\sigma,[g\comp\sigma,h]]^\klstar\comp
  (T\oname{assoc}^\mone)\comp f \]
  is $\vartheta$-guarded by $\mathbf{(cmp)}$.
  
  The rule~\textbf{(wkn)} is obtained from~$\mathbf{(cmp^\star)}$ by
  instantiating $Z$ with $\iobj$, $g$ with~$\eta$ and $\vartheta$ with
  $\sigma\vartheta$. The induced non-trivial premise becomes
  $\eta\bar\sigma\c Y'\to_{\sigma\vartheta} TV$, and it is verified as
  follows:
  $\eta\bar\sigma = (T\bar\sigma)\eta= (T\overline{\sigma\vartheta})
  (T\xi)\eta$
  which is $\sigma\vartheta$-guarded by~\textbf{(trv)}
  and~\textbf{(cdm)}. Here we used the fact that $\bar\sigma$ factors
  as $\overline{\sigma\vartheta}\comp\xi$ with some~$\xi$, for,
  dually, $\sigma\vartheta$ factors through $\sigma$.
\end{proof}
\begin{figure}[t!]
  \tikzset{
    font=\tiny,
    nonterminal/.style={
      rectangle,
      minimum size=6mm,
      very thick,
      draw=orange!50!black!50,         %
      fill=orange!50!white,
      font=\itshape
    },
    terminal/.style={
      scale=.5,
      circle,
      inner sep=0pt,
      thin,draw=black!50,
      top color=white,bottom color=black!20,
      font=\ttfamily
    },
    iterated/.style={
      fill=green!20,
      thick,
      draw=green!50
    },
    natural/.style={
      circle,
      minimum size=4mm,
      inner sep=2pt,
      thin,draw=black!50,
      top color=white,bottom color=black!20,
      font=\ttfamily},
    skip loop/.style={to path={-- ++(0,#1) -| (\tikztotarget)}},
    o/.style={
      shorten >=#1,
      decoration={
        markings,
        mark={
          at position 1
          with {
            \fill[black!55] circle [radius=#1];
          }
        }
      },
      postaction=decorate
    },
    o/.default=2.5pt,
    p/.style={
      shorten <=#1,
      decoration={
        markings,
        mark={
          at position 0
          with {
            \fill[black!55] circle [radius=#1];
          }
        }
      },
      postaction=decorate
    },
    p/.default=2.5pt
  }

  {
    \tikzset{nonterminal/.append style={text height=1.5ex,text depth=.25ex}}
    \tikzset{natural/.append style={text height=1.5ex,text depth=.25ex}}
  }
  \captionsetup[subfigure]{labelformat=empty,justification=justified,singlelinecheck=false}
  \pgfdeclarelayer{background}
  \pgfdeclarelayer{foreground}
  \pgfsetlayers{background,main,foreground}
    \begin{subfigure}{\textwidth}
    \centering
    \caption{Fixpoint:}
    \vspace{-2ex}
    \raisebox{-.5\height}{
    \begin{tikzpicture}[
      point/.style={coordinate},>=stealth',thick,draw=black!50,
      tip/.style={->,shorten >=0.007pt},every join/.style={rounded corners},
      hv path/.style={to path={-| (\tikztotarget)}},
      vh path/.style={to path={|- (\tikztotarget)}},
      text height=1.5ex,text depth=.25ex %
      ]
      \node [nonterminal] (f) {$f$};
      \draw [<-] (f.west) -- +(-1,0) node [midway,above] {$X$};
      \path [o,<-,draw] ($(f.west)+(-0.5,0)$) -- +(0,-0.8) -| ($(f.east)+(0.5,-0.15)$) node [pos=0.8,right] {$X$} -- +(-0.5,0);
      \draw [->] (f.east)++(0,0.15) -- +(1,0) node [midway,above] {$Y$};
      
      \begin{pgfonlayer}{background}
        \draw [iterated] ($(f.north west)+(-0.25,0.25)$) rectangle ($(f.south east)+(0.25,-0.25)$);
      \end{pgfonlayer}
    \end{tikzpicture}
    }
    ~~=~~
    \raisebox{-.5\height}{
    \begin{tikzpicture}[
      point/.style={coordinate},>=stealth',thick,draw=black!50,
      tip/.style={->,shorten >=0.007pt},every join/.style={rounded corners},
      hv path/.style={to path={-| (\tikztotarget)}},
      vh path/.style={to path={|- (\tikztotarget)}},
      text height=1.5ex,text depth=.25ex %
      ]
      \node [nonterminal] (f) {$f$};
      \node [nonterminal] (f2) at ($(f.east)+(1.5,-0.15)$) {$f$};
      \draw [<-] (f.west) -- +(-1,0) node [midway,above] {$X$};
      \draw [p,->] (f.east)++(0,-0.15) -- (f2.west) node [pos=0.25,below] {$X$};
      \path [o,<-,draw] (f2.west)++(-0.5,0) -- ++(0,-0.8) -| ($(f2.east)+(0.5,-0.15)$) node [near end,right] {$X$} -- +(-0.5,0);
      \draw [->] (f.east)++(0,0.15) -- node [midway,above] {$Y$} ++(0.5,0) -- ++(0,0.5) -- ++(3,0);
      \draw [->] (f2.east)++(0,0.15) -- ++(0.5,0) -- node [midway,right] {$Y$} ++(0,0.65);

      \begin{pgfonlayer}{background}
        \draw [iterated] ($(f2.north west)+(-0.25,0.25)$) rectangle ($(f2.south east)+(0.25,-0.25)$);
      \end{pgfonlayer}
    \end{tikzpicture}
    }
  \end{subfigure}
  \par%
  \begin{subfigure}{\textwidth}
    \centering
    \caption{Naturality:}
    \raisebox{-.5\height}{
    \begin{tikzpicture}[
      point/.style={coordinate},>=stealth',thick,draw=black!50,
      tip/.style={->,shorten >=0.007pt},every join/.style={rounded corners},
      hv path/.style={to path={-| (\tikztotarget)}},
      vh path/.style={to path={|- (\tikztotarget)}},
      text height=1.5ex,text depth=.25ex %
      ]
      \node [nonterminal] (f) {$f$};
      \node [nonterminal] (g) at ($(f.east)+(1.5,0.15)$) {$g$};
      \draw [<-] (f.west) -- +(-1,0) node [midway,above] {$X$};
      \path [o,<-,draw] (f.west)++(-0.5,0) -- ++(0,-0.8) -| ($(f.east)+(0.5,-0.15)$) node [near end,right] {$X$}  -- ++(-0.5,0);
      \draw [->] ($(f.east)+(0,0.15)$) -- (g) node [midway,above] {$Y$};
      \draw [->] (g.east) -- +(1,0) node [midway,above] {$Z$};
      \begin{pgfonlayer}{background}
        \draw [iterated] ($(f.north west)+(-0.25,0.25)$) rectangle ($(f.south east)+(0.25,-0.25)$);
      \end{pgfonlayer}
    \end{tikzpicture}
    }
    ~~=~~
    \raisebox{-.5\height}{
    \begin{tikzpicture}[
      point/.style={coordinate},>=stealth',thick,draw=black!50,
      tip/.style={->,shorten >=0.007pt},every join/.style={rounded corners},
      hv path/.style={to path={-| (\tikztotarget)}},
      vh path/.style={to path={|- (\tikztotarget)}},
      text height=1.5ex,text depth=.25ex %
      ]
      \node [nonterminal] (f) {$f$};
      \node [nonterminal] (g) at ($(f.east)+(1.5,0.15)$) {$g$};
      \draw [<-] (f.west) -- +(-1,0) node [midway,above] {$X$};
      \path [o,<-,draw] (f.west)++(-0.5,0) 
        -- ++(0,-0.75) 
        -| ($(g.east)+(0.5,-0.5)$) node [near end,right] {$X$} 
        -| ($(f.east)+(0.6,-0.15)$) 
        -- ++(-0.6,0);
      \draw [->] ($(f.east)+(0,0.15)$) -- (g) node [midway,above] {$Y$};
      \draw [->] (g.east) -- +(1,0) node [midway,above] {$Z$};
      \begin{pgfonlayer}{background}
        \draw [iterated] ($(f.north west)+(-0.25,0.25)$) rectangle ($(g.south east)+(0.25,-0.35)$);
      \end{pgfonlayer}
    \end{tikzpicture}
    }
  \end{subfigure}
  \par\medskip
  \begin{subfigure}{\textwidth}
    \centering
    \vspace{1ex}
    \caption{Codiagonal:}
    \vspace{-1ex}
    \raisebox{-.5\height}{
    \begin{tikzpicture}[
      point/.style={coordinate},>=stealth',thick,draw=black!50,
      tip/.style={->,shorten >=0.007pt},every join/.style={rounded corners},
      hv path/.style={to path={-| (\tikztotarget)}},
      vh path/.style={to path={|- (\tikztotarget)}},
      text height=1.5ex,text depth=.25ex %
      ]
      \node [nonterminal,minimum height=1.2cm] (f) {$g$};
      \draw [<-] (f.west) -- +(-1,0) node [midway,above] {$X$};
      \draw [->] (f.east)++(0,0.4) -- ++(1.5,0) node [pos=0.3,above] {$Y$};
      \draw [p,->] (f.east) -- ++(1,0) --node [midway,right] {$X$}  ++(0,-1.1) -| ($(f.west)+(-0.5,0)$);
      \draw [p,->] (f.east)++(0,-0.4) -- node [midway,below] {$X$} ++(0.5,0) -- ++(0,0.4);
      \begin{pgfonlayer}{background}
        \draw [iterated] ($(f.north west)+(-0.25,0.25)$) rectangle ($(f.south east)+(0.75,-0.25)$);
      \end{pgfonlayer}
    \end{tikzpicture}
    }
    ~~=~~
    \raisebox{-.5\height}{
    \begin{tikzpicture}[
      point/.style={coordinate},>=stealth',thick,draw=black!50,
      tip/.style={->,shorten >=0.007pt},every join/.style={rounded corners},
      hv path/.style={to path={-| (\tikztotarget)}},
      vh path/.style={to path={|- (\tikztotarget)}},
      text height=1.5ex,text depth=.25ex %
      ]
      \node [nonterminal,minimum height=1cm] (f) {$g$};
      \draw [<-] (f.west) -- +(-1.6,0) node [pos=0.7,above] {$X$};
      \draw [->] (f.east)++(0,0.3) -- ++(1.5,0) node [pos=0.3,above] {$Y$};
      \draw [p,->] (f.east) -- ++(1.1,0) -- node [midway,right] {$X$} ++(0,-1.35) -| ($(f.west)+(-0.95,0)$);
      \draw [p,->] (f.east)++(0,-0.3) -- ++(0.5,0) -- node [midway,right] {$X$} ++(0,-0.65) -| ($(f.west)+(-0.5,0)$);
      \begin{pgfonlayer}{background}
        \draw [iterated] ($(f.north west)+(-0.75,0.35)$) rectangle ($(f.south east)+(0.85,-0.6)$);
        \draw [iterated,fill=green!35] ($(f.north west)+(-0.25,0.2)$) rectangle ($(f.south east)+(0.25,-0.25)$);
      \end{pgfonlayer}
    \end{tikzpicture}
    }
  \end{subfigure}
  \par%
  \begin{subfigure}{\textwidth}
    \caption{Uniformity:}
    \centering
\begin{tabular}{rcl}
   \raisebox{-.5\height}{
    \begin{tikzpicture}[
      point/.style={coordinate},>=stealth',thick,draw=black!50,
      tip/.style={->,shorten >=0.007pt},every join/.style={rounded corners},
      hv path/.style={to path={-| (\tikztotarget)}},
      vh path/.style={to path={|- (\tikztotarget)}},
      text height=1.5ex,text depth=.25ex %
      ]
      \node [nonterminal,fill=blue!20,draw=blue!50] (h) {$h$};
      \node [nonterminal] (f) at ($(h.east)+(1.5,0)$) {$f$};
      \draw [<-] (h.west) -- +(-1,0) node [midway,above] {$Z$};
      \draw [->] (h.east) -- (f.west) node [midway,above] {$X$};
      \draw [->] (f.east)++(0,0.15) -- +(1,0) node [midway,above] {$Y$};
      \draw [p,->] (f.east)++(0,-0.15) -- +(1,0) node [midway,below] {$X$};
    \end{tikzpicture}
    }
    &$~~=~~$&
    \raisebox{-.5\height}{
    \begin{tikzpicture}[
      point/.style={coordinate},>=stealth',thick,draw=black!50,
      tip/.style={->,shorten >=0.007pt},every join/.style={rounded corners},
      hv path/.style={to path={-| (\tikztotarget)}},
      vh path/.style={to path={|- (\tikztotarget)}},
      text height=1.5ex,text depth=.25ex %
      ]
      \node [nonterminal] (f) {$g$};
      \node [nonterminal,fill=blue!20,draw=blue!50] (h) at ($(f.east)+(1.5,-0.15)$) {$h$};
      \draw [<-] (f.west) -- +(-1,0) node [midway,above] {$Z$};
      \draw [p,->] (f.east)++(0,-0.15) -- (h.west) node [midway,below] {$Z$};
      \draw [->] (f.east)++(0,0.15) -- ++(0.65,0) -- ++(0,0.4) node [midway,left] {$Y$} -- ++(2.35,0);
      \draw [->] (h.east) -- +(1,0) node [midway,above] {$X$};
    \end{tikzpicture}
    }\\
    &\Large{$\Downarrow$}&\\
    \raisebox{-.5\height}{
    \begin{tikzpicture}[
      point/.style={coordinate},>=stealth',thick,draw=black!50,
      tip/.style={->,shorten >=0.007pt},every join/.style={rounded corners},
      hv path/.style={to path={-| (\tikztotarget)}},
      vh path/.style={to path={|- (\tikztotarget)}},
      text height=1.5ex,text depth=.25ex %
      ]
      \node [nonterminal,fill=blue!20,draw=blue!50] (h) {$h$};
      \node [nonterminal] (f) at ($(h.east)+(1.5,0)$) {$f$};
      \draw [<-] (h.west) -- +(-1,0) node [midway,above] {$Z$};
      \draw [->] (h.east) -- (f.west) node [midway,above] {$X$};
      \draw [->] (f.east)++(0,0.15) -- +(1,0) node [midway,above] {$Y$};
      \path [o,<-,draw] (f.west)++(-0.5,0) -- ++(0,-0.8) -|
      ($(f.east)+(0.5,-0.15)$) node [pos=0.8,right] {$X$} -- ++(-0.5,0);
      \begin{pgfonlayer}{background}
        \draw [iterated] ($(f.north west)+(-0.25,0.25)$) rectangle ($(f.south east)+(0.25,-0.25)$);
      \end{pgfonlayer}
    \end{tikzpicture}
    }
    &$~~=~~$&
    \raisebox{-.5\height}{
    \begin{tikzpicture}[
      point/.style={coordinate},>=stealth',thick,draw=black!50,
      tip/.style={->,shorten >=0.007pt},every join/.style={rounded corners},
      hv path/.style={to path={-| (\tikztotarget)}},
      vh path/.style={to path={|- (\tikztotarget)}},
      text height=1.5ex,text depth=.25ex %
      ]
      \node [nonterminal] (f) {$g$};
      \draw [<-] (f.west) -- +(-1,0) node [midway,above] {$Z$};
      \path [o,<-,draw] ($(f.west)+(-0.5,0)$) -- +(0,-0.8) -| ($(f.east)+(0.5,-0.15)$) node [pos=0.8,right] {$Z$} -- +(-0.5,0);
      \draw [->] (f.east)++(0,0.15) -- +(1,0) node [midway,above] {$Y$};
      
      \begin{pgfonlayer}{background}
        \draw [iterated] ($(f.north west)+(-0.25,0.25)$) rectangle ($(f.south east)+(0.25,-0.25)$);
      \end{pgfonlayer}
    \end{tikzpicture}
    }
  \end{tabular}
  \end{subfigure}
\vspace{2ex}
\caption{Axioms of guarded Elgot iteration.}
\label{fig:ax}
\end{figure}

\begin{definition}[Guarded (pre-)iterative/Elgot monads]\label{def:strong_guard_monad}
A strong mo\-nad~$\BBT$ on a distributive category is \emph{guarded pre-iterative}
if it is equipped with a guarded iteration operator
\begin{align}\label{eq:g-iter}
  \anonrule{}{f\c X\to_2 T(Y+X)}{f^\istar\c X\to TY}
\end{align}
satisfying the  
\begin{itemize}
  \item\emph{fixpoint law}: $f^\istar = [\eta,f^\istar]^\klstar  f$.
\end{itemize}
We call a pre-iterative monad $\BBT$ \emph{guarded Elgot}~\cite{LevyGoncharov19} if it satisfies
\begin{itemize}
  \item\emph{naturality:} $g^{\klstar} f^{\istar} = ([(T\inj_1) \comp g, \eta\inj_2]^{\klstar}  f)^{\istar}$ for $f\c X\to_2 T(Y+X)$, ${g\c Y \to TZ}$;
  \item\emph{codiagonal:} $(T[\id,\inj_2] \comp f)^{\istar} = f^{\istar\istar}$ for  $f\c X \to_{12,2} T((Y + X) + X)$;
  \item\emph{uniformity:} $f \comp h = T(\id+ h) \comp g$ implies
	$f^{\istar} \comp h = g^{\istar}$ for $f\c  X \to_2 T(Y + X)$, $g\c Z \to_2 T(Y + Z)$ and
	$h\c Z \to X$;
  \item\emph{strength:} $\tau\comp(\id_W\times f^\istar) =(\delta\comp(\id_W\times f))^\istar$ for  $f\c X\to_2 T(Y+X)$.
\end{itemize}
and \emph{guarded iterative} if $f^\istar$ is a unique solution of the fixpoint law 
(the remaining axioms then are granted~\cite{GoncharovSchroderEtAl17}).
\end{definition}
\noindent The above axioms of iteration are standard
(cf.~\cite{BloomEsik93}), except \emph{strength}, which we need here
for the semantics of computations in multivariable contexts. To
understand the axiom, observe that the right-hand side iterates over
$W\times X$ leaving the $W$-component unchanged, and eventually
returns the $W$-component as part of the result, while the left-hand
side iterates over~$X$ and subsequently pairs the result with the
originally given element of~$W$. These axioms, again except strength,
can be presented in an intuitive graphical form as equations of
flowchart diagrams -- see Fig.~\ref{fig:ax}. Here, the orange boxes
identify Kleisli morphisms and blue boxes identify morphisms of the
underlying category~$\BC$. We indicate the scopes of feedback loops,
representing applications of the iteration operator, by shaded green
frames.  Finally, we indicate by black bullets those outputs in which
a corresponding Kleisli morphism is guarded.

The notion of (abstract) guardedness is a common generalization of
various special cases occurring in practice. Every monad can be
equipped with a least notion of guardedness, called~\emph{vacuous
  guardedness} and defined as follows: $f\c X\to_2 T(Y+Z)$ iff $f$
factors through $T\inj_1\c Y\to T(Y+Z)$; that is, intuitively speaking,
the definitions of elements of~$X$ given by~$f$ do not mention
variables in~$Z$, or more precisely speaking can be rewritten to
ensure this.
On the other hand, the greatest notion of guardedness is \emph{total guardedness},
defined by taking $f\c X\to_2 T(Y+Z)$ for every $f\c X\to T(Y+Z)$.
This addresses \emph{total iteration} operators on $\BBT$, whose existence depends
on special properties of~$\BBT$, such as being enriched over complete partial orders. 
Our motivating examples
are mainly those that lie properly between these two extreme situations, e.g.\
\emph{completely iterative monads} for which guardedness is defined 
via monad modules and the iteration operator is partial, but uniquely satisfies the 
fixpoint law~\cite{Milius05}.
For illustration, we consider several instances of guarded iteration.
\begin{example}\label{expl:monad} We fix the category of sets and functions $\Set$ as an ambient 
  distributive category in the following examples.
\begin{cenumerate} 
\item\emph{(Finitely branching processes)} Let
  $TX = \nu\gamma.\,\PSet_{\omega}(X+\Act\times\gamma)$, the final
  $\PSet_{\omega}(X+\Act\times\argument)$-coalgebra with $\FSet$ being
  the finite powerset functor. Thus,~$TX$ is equivalently described as
  the set of finitely branching nondeterministic trees with edges
  labelled by elements of $\Act$ and with terminal nodes possibly
  labelled by elements of $X$ (otherwise regarded as nullary
  nondeterminism, i.e.\ \emph{deadlock}), taken modulo
  bisimilarity. Every $f\c X\to {T(Y+X)}$ can be viewed as a family
  $(f(x)\in T(Y+X))_{x\in X}$ of trees whose terminal nodes are
  labelled in the disjoint union of~$X$ and~$Y$. Each tree $f(x)$ thus
  can be seen as a recursive process definition for the process
  name~$x$ relative to the names in $X+Y$.  The notion of guardedness
  borrowed from process algebra requires that every $x'\in X$
  occurring in $f(x)$ must be preceded by a transition, and if this
  condition is satisfied, we can calculate a unique \emph{solution}
  $f^\istar\c X\to TY$ of the system of definitions
  $(f(x)\c T(Y+X))_{x\in X}$. In other words, $\BBT$ is guarded
  iterative with $f\c X\to_2 T(Y+Z)$~iff
\begin{displaymath}
  \out f\c X\to \PSet_{\omega}((Y+Z)+\Act\times T(Y+Z))
\end{displaymath}
factors through $\FSet(\inj_1+\id)$ where $\out\c TX\cong\PSet_{\omega}(X+\Act\times TX)$
is the canonical final coalgebra isomorphism. As a result, $\BBT$
is a guarded iterative monad (more specifically~\emph{completely iterative}~\cite{Milius05}). 
\item\emph{(Countably branching processes)} A variation of the
  previous example is obtained by replacing finite nondeterminism with countable
  nondeterminism, i.e.\ by replacing~$\PSet_{\omega}$ with the
  countable powerset functor $\CSet$. Note that in the previous
  example we could not extend the iteration operator to a total one,
  because unguarded systems of recursive process equations may define
  infinitely branching processes~\cite{BergstraPonseEtAl01}. The monad
  $TX = \nu\gamma.\,\CSet(X+\Act\times\gamma)$ does however support
  both partial guarded iteration in the sense of the previous example,
  and total iteration extending the former.
This monad is therefore both
  guarded iterative in the former sense, but only guarded Elgot in
  the latter sense, for 
  under total iteration, the fixpoints~$f^\istar$ are no longer
  unique. This setup is analysed more generally in detail
  in previous work~\cite{GoncharovSchroderEtAl18,GoncharovSchroderEtAl17}.
\item A very simple example of total guarded iteration is obtained
  from the (full) powerset monad $T=\PSet$. The corresponding Kleisli
  category is enriched over complete partial orders and continuous
  functions and therefore admits total iteration calculated via least
  fixpoints. This yields an example of a guarded Elgot monad which is
  not guarded iterative.
\item\emph{(Complete finite traces)} Let
  $TX = \PSet(\Act^\star\times X)$ be the monad obtained from~$\PSet$
  by an obvious modification ensuring that the first elements of the
  pairs from $\Act^\star\times X$, i.e.\ \emph{finite traces}, are
  concatenated along Kleisli
  composition~\cite{CenciarelliMoggi93}. Like~$\PSet$, this monad is
  \emph{order-enriched} and thus supports a total iteration operator
  via least fixpoints (see e.g.~\cite{GoncharovMiliusEtAl16}). From
  this, a guarded iteration operator is obtained by restricting to the
  guarded category with $f\c X\to_2\PSet(\Act^\star\times( Y + Z))$ iff
  $f$ factors through the map
  \begin{displaymath}
    \PSet(\Act^\star\times Y + \Act^\mplus\times Z)\xto{\PSet(\id+\iota\times\id)}
    \PSet(\Act^\star\times Y + \Act^\star\times Z)\cong
    \PSet(\Act^\star\times( Y + Z))
  \end{displaymath}
  induced by the inclusion $\iota\c\Act^\mplus\ito\Act^\star$. Like in Clause~3,
  we obtain a guarded Elgot monad with a total iteration operator.
\item Finally, an example of partial guarded iteration can be obtained
  from Clause~3 above by replacing $\PSet$ with the \emph{non-empty
    powerset monad}~$\PSet^{\mplus}$. Total iteration as defined in
  Clause~3 does not restrict to total iteration on $\NESet$, because
  empty sets can arise from solving systems not involving empty sets,
  e.g.\ $\eta\inj_2\c 1\to\PSet^{\mplus}(1+1)$ would not have a solution
  in this sense. However, it is easy to see that total iteration does
  restrict to guarded iteration for $\PSet^{\mplus}$ with the notion of
  guardedness defined as follows: $f\c X\to_2\PSet^{\mplus}(Y+Z)$ iff
  for every~$x$, $f(x)$ contains at least one element from $Y$. Therefore,
  $\NESet$ is a guarded Elgot monad, which is not guarded iterative and 
  with properly partial iteration.
\end{cenumerate}
\end{example}
\noindent For a pre-iterative monad $\BBT$, we derive a \emph{strong
  iteration operator}:
\begin{equation}\label{eq:iter_par}%
\anonrule{}{f\c W\times X\to_2 T(Y+X)}{f^\iistar= \bigl(T(\snd+\id)\comp\delta\comp\brks{\fst,f}\bigr)^\istar\c W\times X\to TY}
\end{equation}
which essentially generalizes the original operator $(\argument)^\istar$ to 
morphisms extended with a context via $W\times (\argument)$. This will become 
essential in Section~\ref{sec:meta} for the semantics of our metalanguage. 
\begin{lemma}\label{lem:iistar}
For every strong guarded Elgot monad $\BBT$, strong iteration~\eqref{eq:iter_par} 
satisfies $\tau\comp\brks{\fst, f^\iistar} = (\delta\brks{\fst,f})^\istar$ for every
$f\c W\times X\to_2 T(Y+X)$.
\end{lemma}
\begin{proof}
Let us rewrite the left hand side as follows:
\begin{flalign*}
&&\tau\comp\brks{\fst, f^\iistar} 
=&\; T(\fst\times\id)\comp\tau\comp (\id\times f^\iistar)\comp\diag\\*
&&=&\; T(\fst\times\id)\comp\tau\comp (\id\times (T(\snd+\id)\comp\delta\brks{\fst,f})^\istar)\comp\diag&\!\!\by{defn.~of~$(\argument)^\iistar$}\\ 
&&=&\; T(\fst\times\id)\comp\tau\comp (\id\times (T\snd)  (\delta\brks{\fst, f})^\istar)\comp\diag&\by{naturality}\\ 
&&=&\; T(\fst\times\snd)\comp\tau\comp (\id\times (\delta\brks{\fst, f})^\istar)\comp\diag\\ 
&&=&\; T(\fst\times\snd)\comp (\delta\comp (\id\times \delta\brks{\fst, f}))^\istar\comp\diag&\by{strength}\\ 
&&=&\; (T(\fst\times\snd+\id)\comp\delta\comp (\id\times \delta\brks{\fst, f}))^\istar\comp\diag.&\by{naturality}
\intertext{
Note that 
}
&&T(\id\times&\snd+\id)\comp\delta\comp (\id\times \delta\brks{\fst, f})\comp (\fst\times\id)\\
&&=&\;T(\id\times\snd+\id)\comp\delta\comp (\fst\times \delta\brks{\fst, f})\\
&&=&\;T(\fst\times\snd+\fst\times\id)\comp\delta\comp (\id\times \delta\brks{\fst, f})\\
&&=&\;T(\id+\fst\times\id)\comp T(\fst\times\snd+\id)\comp\delta\comp (\id\times \delta\brks{\fst, f}),
\intertext{
and therefore, by \emph{uniformity} (instantiating the equation from
Definition~\ref{def:strong_guard_monad} with $f = T(\id \times \snd + \id)
\comp \delta \comp (\id \times \delta \comp \brks{\fst, f})$, $g = T(\fst
\times \snd + \id) \comp \delta \comp (\id \times \delta\comp\brks{\fst,f})$,
and $h = \fst \times \id$), }
&&\tau\comp\brks{\fst, f^\iistar} =&\; (T(\id\times\snd+\id)\comp\delta\comp (\id\times \delta\brks{\fst, f}))^\istar
(\fst\times\id)\comp\diag
\\
&&=&\; (T(\id\times\snd+\id)\comp\delta\comp (\id\times \delta\brks{\fst, f}))^\istar
 \brks{\fst,\id}.
\intertext{
Finally, observe that
}
  &&T(\id\times&\snd+\id)\comp\delta\comp (\id\times \delta\brks{\fst, f})\comp
 \brks{\fst,\id}\\ 
  &&=&\; T(\id\times\snd+\id)\comp\delta\comp\brks{\fst, \delta\brks{\fst, f}}\\
  &&=&\; T(\id\times\snd+\id)\comp T(\brks{\fst,\id}+\brks{\fst,\id})\comp \delta\brks{\fst,f}&\by{\eqref{eq:d-delta}}\\
  &&=&\; T(\brks{\fst,\snd}+\brks{\fst,\id})\comp \delta\brks{\fst,f}\\
  &&=&\; T(\id+\brks{\fst,\id})\comp \delta\brks{\fst,f}
\intertext{and therefore, by~\emph{uniformity},} 
&&\tau\comp\brks{\fst, f^\iistar} 
=&\; (T(\id\times\snd+\id)\comp\delta\comp (\id\times \delta\brks{\fst, f}))^\istar\brks{\fst,\id}\\
&&=&\; (\delta\brks{\fst,f})^\istar
\end{flalign*}
as desired.
\end{proof}

\paragraph{Strength and simple slices} To clarify the role of strong
iteration~\eqref{eq:iter_par}, we characterize it as iteration in a
\emph{simple slice category}~\cite{Jacobs99} $\ssl{\BC}{W}$ arising for
every fixed $W\in |\BC|$ as the co-Kleisli category of the
\emph{product comonad}~\cite{BrookesVanStone93} $W\times\argument$;
that is, $|\ssl{\BC}{W}| = |\BC|$,
$\Hom_{\ssl{\BC}{W}}(X,Y)=\Hom_{\BC}(W\times X,Y)$, identities in $\ssl{\BC}{W}$
are projections $\snd\c W\times X\to X$, and the composite of
$g\c W\times X\to Y$ and $f\c W\times Y\to Z$ is
$f\comp\brks{\fst,g}\c W\times X\to Z$. We often indicate composition in~$\ssl{\BC}{W}$ 
by $\circ^W$ for clarity. We note that $\ssl{\BC}{1}$ is isomorphic
to $\BC$, and $\ssl{(\ssl{\BC}{W})}{V}$ is isomorphic to $\ssl{\BC}{W\times V}$, where the
isomorphism just rebrackets products. The assignment $W\mapsto\ssl{\BC}{W}$
in fact extends to a strict indexed category: A morphism $k\c W\to V$
induces a functor $\ssl{\BC}{k}\c\ssl{\BC}{V}\to\ssl{\BC}{W}$ which acts as identity on
objects and maps $f\in\Hom_{\ssl{\BC}{V}}(X,Y)=\Hom_\BC(X\times V, Y)$ to
$f(k\times \id)\in\Hom_{\ssl{\BC}{W}}(X,Y)=\Hom_\BC(W\times X, Y)$.
Moreover, we have embeddings $J^W\c\BC\to\ssl{\BC}{W}$, given by~$J^WX=X$ and
$J^Wf=f\snd$, which commute with the functors $\ssl{\BC}{k}$, i.e.\
$(\ssl{\BC}{k})J^V=J^W$. In particular, up to the isomorphism
$J^1\c\BC\cong\ssl{\BC}{1}$ the functor $J^W$ coincides with $\ssl{\BC}{\bang}$ where
$\bang$ is the unique $\BC$-morphism $W\to
1$. %
Of course,~$J^W$ is the right adjoint to the forgetful functor
$U^W\c\ssl{\BC}{W}\to\BC$, which acts on objects as $U^WX=W\times X$ and on
morphisms $f\c X\to Y$ as $U^Wf=\brks{\fst,f}$. To avoid confusion with
the unit of~$\BBT$, we write $j_X\in\Hom_{\ssl{\BC}{W}}(X, W\times X)$ for
the unit of this adjunction, which is the $\BC$-morphism
$\id\c W\times X\to W\times X$. Like in all co-Kleisli categories, the
adjoint transpose map
$\Hom_{\ssl{\BC}{W}}(X,Y)=\Hom_{\ssl{\BC}{W}}(X,J^WY)\cong\Hom_\BC(U^WX,Y)=\Hom_\BC(W\times
X,Y)$ is just identity; we thus have
\begin{equation}
  \label{eq:slice-universal}
  f=(J^Wf)\circ^W j_X
\end{equation}
(as also easily verified directly) for each $f\c X\to Y$ in~$\ssl{\BC}{W}$,
i.e.\ $f\c W\times X\to Y$ in~$\BC$.

The monad~$\BBT$ being strong means in particular that for every
$W\in |\BC|$,~$\tau$ yields a distributive law of the monad $\BBT$
over the comonad $W\times\argument$, which extends~$\BBT$ from~$\BC$
to $\ssl{\BC}{W}$~\cite{BrookesVanStone93}. We state this more
precisely, and complement it with similar statements on propagation of
guardedness and iteration:%

\begin{theorem}\label{thm:context-comonad}
  Let $\BBT$ be a strong monad on a distributive category~$\BC$. Then
  the following hold.
\begin{enumerate}
\item\label{item:monad-extension} For every $W\in|\BC|$, $\ssl{\BC}{W}$ is
  distributive, and $\BBT$ \emph{coherently} extends to a strong monad
  over~$\ssl{\BC}{W}$: For every $k\c W\to V$, the functor $\ssl{\BC}{k}$ strictly
  preserves the monad structure, i.e.\ if $\BBT^W$ and $\BBT^V$ denote
  the extensions of~$\BBT$ to $\ssl{\BC}{W}$ and $\ssl{\BC}{V}$ respectively, then
  $(\ssl{\BC}{k})\comp T^V=T^W\ssl{\BC}{k}$, and the pair consisting of $\ssl{\BC}{k}$ and the
  identity natural transformation on $(\ssl{\BC}{k})\comp T^V$ is a monad
  morphism. The same holds for the functors~$J^W$.
\item\label{item:context-guarded} If $\BBT$ is guarded, then so is its
  extension~$\BBT^W$ to~$\ssl{\BC}{W}$, with the same notion of guardedness
  (i.e.\
  $\Hom_{\ssl{\BC}{W},\sigma}(X,T^WY)=\Hom_{\BC,\sigma}(W\times X,TY)$), and
  all functors $\ssl{\BC}{k}$, as well as the functors~$J^W$, preserve
  guardedness.
\item\label{item:pre-iterative} If\/ $\BBT$ is guarded pre-iterative
  on $\BC$ then so is the extension of~$\BBT$ to~$\ssl{\BC}{W}$, under the
  same definition of guardedness and with iteration defined as strong
  iteration~\eqref{eq:iter_par}. If moreover~$\BBT$ satisfies
  \emph{uniformity}, then all functors~$\ssl{\BC}{k}$, as well as the
  functors~$J^W$, preserve iteration.
\item\label{item:elgot} If\/ $\BBT$ is guarded Elgot on $\BC$ then so
  is the extension of $\BBT$ to $\ssl{\BC}{W}$.
  \item If\/ $\BBT$ is guarded iterative then so is the extension of $\BBT$ to $\ssl{\BC}{W}$. 
\end{enumerate}
\end{theorem}
\noindent Moreover, we have partial converses to the above claims,
which further justify the axioms and definitions regarding strength,
specifically the \textbf{(str)} rule for guardedness, the definition
of strong iteration, and the strength law for iteration.  Only for
purposes of the statement and proof of the following theorem, we
introduce notions of guardedness, guarded iterativity etc.\ for monads
that are \emph{not} assumed to be strong; these are axiomatized in the
expected way, i.e.\ by just removing the axioms and rules referring to
strength. We designate these notions as \emph{weak}, and the standard
versions as \emph{strong} for clarity. E.g.\ a \emph{weakly guarded
  monad} is a monad~$\BBT$ equipped with distinguished subsets
$\Hom_\sigma(X,TY)$, indexed over summands~$\sigma\c Y'\cpto Y$, that
satisfy axioms~\textbf{(trv)}, \textbf{(sum)} and~\textbf{(cmp)}, and a
\emph{strongly guarded monad} is a weakly guarded strong monad
satisfying axiom~\textbf{(str)}.
\begin{theorem}\label{thm:slice-converse}
  Let~$\BBT$ be a monad on~$\BC$, and assume that $\BBT$ extends
  coherently to monads~$\BBT^W$ on all~$\ssl{\BC}{W}$, in the sense that
  $(\ssl{\BC}{k})\comp T^V=T^W\comp(\ssl{\BC}{k})$ for every $k\c W\to V$. Then the following hold
  \begin{enumerate}
  \item\label{item:monad-converse} The monad $\BBT$ is strong. In
    fact, the construction of the strength and the opposite
    construction from
    Theorem~\ref{thm:context-comonad}.\ref{item:monad-extension}
    (which induces the~$\BBT^W$ from a given strength) are mutually
    inverse.
  \item If $\BBT$ is weakly guarded, $\BBT^W$ is weakly guarded, and
    $J^W$ preserves guardedness, then
    $\Hom_{\ssl{\BC}{W},\sigma}(X,T^WY)\supseteq\Hom_{\BC,\sigma}(W\times
    X,TY)$.
  \item\label{item:strength-guarded} If $\BBT$ is weakly guarded, and
    putting
    \begin{math}
      \Hom_{\ssl{\BC}{W},\sigma}(X,T^WY)=\Hom_{\BC,\sigma}(W\times X\comma TY)
    \end{math}
    makes each~$\BBT^W$ into a weakly guarded monad, then~$\BBT$ is a
    strongly guarded monad, i.e.\ satisfies~\textbf{(str)}.
  \item\label{item:converse-strong-it} If $\BBT$ is strongly guarded
    and pre-iterative, each $\BBT^W$ is pre-iterative and satisfies
    uniformity, and $J^W\c\BC\to\ssl{\BC}{W}$ preserves iteration, then
    iteration on~$\BBT^W$ is strong iteration~\eqref{eq:iter_par}.
  \item\label{item:converse-strength-law} If $\BBT$ is strongly
    guarded and pre-iterative, each $\BBT^W$, made into a
    pre-iterative monad by equipping it with iteration defined as
    strong iteration on~$\BBT$, satisfies \emph{naturality}, and $J^W$
    preserves iteration, then $\BBT$ satisfies the \emph{strength} law
    (Definition~\ref{def:strong_guard_monad}).
  \end{enumerate}
\end{theorem}
\noindent We prove Theorem~\ref{thm:context-comonad} first but in fact
occasionally make use of the converse statements recorded in
Theorem~\ref{thm:slice-converse} (whose proof will not depend on
Theorem~\ref{thm:context-comonad}). Specifically, to establish a
property regarding strength, we apply the current implication to
conclude a weak (i.e.\ strength-free) property of
$\ssl{\BC}{W\times V}\cong\ssl{(\ssl{\BC}{W})}{V}$, and then apply
Theorem~\ref{thm:slice-converse} to obtain a property of $\ssl{\BC}{W}$
referring to strength.

\begin{proof}[Proof (Theorem~\ref{thm:context-comonad})]
\begin{cenumerate}
\item Being a co-Kleisli category, $\ssl{\BC}{W}$ inherits finite products
  from~$\BC$. Finite coproducts are inherited thanks to~$\BC$ being
  distributive;~e.g.\
  \begin{align*}
    \Hom_{\ssl{\BC}{W}}(X+Y,Z)
    =&\;\Hom_\BC(W\times (X+Y),Z)\\
    \cong&\;\Hom_\BC(W\times X+W\times Y,Z)\\
    \cong&\;\Hom_\BC(W\times X,Z)\times\Hom_\BC(W\times Y,Z)\\
    =&\; \Hom_{\ssl{\BC}{W}}(X,Z)\times\Hom_{\ssl{\BC}{W}}(Y,Z).
  \end{align*}
  Since both products and coproducts in $\ssl{\BC}{W}$ are inherited
  from~$\BC$, so is distributivity. We have already noted that~$\BBT$
  lifts to $\ssl{\BC}{W}$ because the strength yields a distributive law
  of~$\BBT$ over the product comonad~\cite{BrookesVanStone93}. The
  lifted monad is explicitly described as follows. The unit is just
  $\eta\snd\c W\times X\to TX$ where~$\eta$ is the unit of~$\BBT$
  in~$\BC$, and the Kleisli lifting of $f\in\Hom_{\ssl{\BC}{W}}(X,TY)$ is
  $f^\klstar\tau$ where $f^\klstar\c T(W\times X)\to TY$ is the Kleisli
  lifting of~$f\c W\times X\to TY$ in~$\BC$ and~$\tau$ is the strength
  of~$\BBT$ in~$\BC$. We note in particular that this implies
  $T^Wf=Tf \tau$ for $f\c X\to Y$ in~$\ssl{\BC}{W}$ (hence $f\c W\times X\to Y$
  in~$\BC$). We defer consideration of the strength, and tackle
  coherence first.

  We need to show that $(\ssl{\BC}{k})\comp T^V=T^W\comp(\ssl{\BC}{k})$. So let $f\c X\to Y$
  in~$\ssl{\BC}{W}$, i.e.\ $f\c W\times X\to Y$ in~$\BC$. Then
  \begin{flalign*}
    && (\ssl{\BC}{k})\comp (T^Vf) 
    &\;= (Tf)\comp \tau_{V,X}\comp (k\times\id_{TX})    &\by{definitions}\\
    &&&\; = (Tf)\comp T(k\times\id_X)\comp\tau_{W,X}  &\by{naturality of~$\tau$}\\
    &&&\; = T^W(\ssl{\BC}{k})f.                   &\by{definitions} 
  \end{flalign*}
  Preservation of the monad structure is then clear by the above
  description of this structure. The claim for~$J^W$ follows as a
  special case, since the isomorphism of $\BC$ and $\ssl{\BC}{1}$ clearly
  extends to the corresponding monads.
  
  We conclude by the initially mentioned strategy that~$\BBT^W$ is
  strong: By the above,~$\BBT$ extends to a monad~$\BBT^{W\times V}$
  on $\ssl{\BC}{W\times V}$, which transfers along the isomorphism
  $\ssl{\BC}{W\times V}\cong\ssl{(\ssl{\BC}{W})}{V}$ to a monad $(\BBT^W)^V$ on
  $\ssl{(\ssl{\BC}{W})}{V}$ acting on morphisms
  $f\in\Hom_{\ssl{(\ssl{\BC}{W})}{V}}(X,Y)=\Hom_\BC(W\times V\times X,Y)$ by
  $(T^W)^Vf=(Tf)  \tau_{W\times V,X}$. Since under the isomorphism
  $\ssl{\BC}{W\times V}\cong\ssl{(\ssl{\BC}{W})}{V}$, the embedding of $\ssl{\BC}{W}$ into
  $\ssl{(\ssl{\BC}{W})}{V}$ corresponds to $\ssl{\BC}{\fst}\c\ssl{\BC}{W}\to\ssl{\BC}{W\times V}$, the
  above preservation property for functors~$\ssl{\BC}{k}$ implies that
  $(\BBT^W)^V$ extends $\BBT^W$; again by the preservation properties
  already established, these extensions are coherent. By
  Theorem~\ref{thm:slice-converse}.\ref{item:monad-converse}, it
  follows that $\BBT^W$ is strong. The strength
  $V\times T^WX\to T^W(V\times X)$ constructed in the proof of
  Theorem~\ref{thm:slice-converse}.\ref{item:monad-converse} is
  $(T^W)^Vk_X$ (understood as a $\ssl{\BC}{W}$-morphism), where $k_X$ is the
  $\ssl{\BC}{W}$-identity on $V\times X$ taken as a $\ssl{(\ssl{\BC}{W})}{V}$-morphism
  $X\to V\times X$, which as a $\BC$-morphism
  $W\times V\times X\to V\times X$ projects to the second and third
  component. By the above description of $(T^W)^V$, we have, eliding
  associativity isomorphisms,
  $(T^W)^Vk_X=(Tk_X)\comp\tau_{W\times V,X}=\tau\snd$ where
  $\snd\c W\times (V\times TX)\to V\times TX$, using standard coherence
  properties of~$\tau$. It follows that the $\ssl{\BC}{k}$ preserve also the
  strength.

\item We need to verify that the extension of $\BBT$ to $\ssl{\BC}{W}$
  satisfies the axioms of guardedness from Fig.~\ref{fig:guard}.
\begin{citemize}
  \item\textbf{(trv)} Given $f\c W\times X\to TY$, we need to check that 
$T(\inj_1\snd)\comp\tau\brks{\fst,f}\c W\times X\to_2 T(Y+Z)$. Indeed, 
$T(\inj_1\snd)\comp\tau\brks{\fst,f}$ reduces to $(T\inj_1)\comp f$ and we are done by 
the original~\textbf{(trv)} for $\BC$.
  \item\textbf{(sum)} Given $f\c W\times X\to_\sigma TZ$, $g\c W\times Y\to_\sigma TZ$,
by~\textbf{(sum)} for~$\BC$, $[f,g]\c W\times X+W\times Y\to_\sigma TZ$. After precomposing 
the result with the isomorphism $\dist$, we are done by Proposition~\ref{lem:weak}.  
  \item\textbf{(cmp)} Let $f\c W\times X\to_{\inj_2} T(Y+Z)$, $g\c W\times Y\to_{\sigma} TV$,
  $h\c W\times Z\to TV$ and we need to show that $[g,h]^\klstar\comp\delta\brks{\fst,f}\c W\times X\to_{\sigma} TV$.
The latter morphism equals the composite
\begin{align*}
  W\times X&\,\xto{\tau\brks{\id,f}} T((W\times X)\times (Y+Z))\\
           &\,\xto{(\eta\dist(\fst\times\id))^\klstar} T(W\times Y+ W\times Z) 
           \xto{[g,h]^\klstar} TV. 
\end{align*}
By~\textbf{(cmp)}, we reduce to the problem of showing 
\begin{displaymath}
  (\eta\dist\comp(\fst\times\id))^\klstar\tau\brks{\id,f}\c W\times X\to_{2} T(W\times Y+W\times Z).
\end{displaymath}
Note that by~\textbf{(str)}, $\tau\brks{\id,f}\c W\times
X\to_{\id\times\inj_2}T((W\times X)\times (Y+Z))$. Now $(W\times
X)\times (Y+Z)$ is a coproduct of $(W\times X)\times Y$ and
$(W\times X)\times Z$, and $\eta\dist(\fst\times\id)$, regarded as
a universal morphism induced by this coproduct structure, yields
$\eta\inj_1(\fst\times\id)=(T\inj_1)\comp\eta(\fst\times\id)$ by composition with
the corresponding left coproduct injection; the latter morphism is 
$\inj_2$-guarded by~\textbf{(trv)}. We are therefore done by~\textbf{(cmp)}.
\item\textbf{(str)} As indicated above, we go via
  Theorem~\ref{thm:slice-converse}: The guardedness structure of the
  monad $(\BBT^W)^V$ on $\ssl{(\ssl{\BC}{W})}{V}$ is clearly the same as the one of
  the monad $\BBT^{W\times V}$ on $\ssl{\BC}{W\times V}$, hence satisfies
  (\textbf{trv}), (\textbf{sum}), and (\textbf{cmp}) by the above. By
  Theorem~\ref{thm:slice-converse}.\ref{item:strength-guarded}, it
  follows that $\BBT^W$ satisfies (\textbf{str}).
\end{citemize}
It remains to show that given $k\c W\to V$, $\ssl{\BC}{k}\c\ssl{\BC}{V}\to\ssl{\BC}{W}$
preserves guardedness: If $f\c X\to_\sigma T^VY$ in~$\ssl{\BC}{V}$, then by
definition $f\c V\times X\to TY$ in~$\BC$. By \textbf{(cdm)}, it follows
that $f(k\times id)\c W\times X\to_\sigma TY$, so by definition
$(\ssl{\BC}{k})\comp f\c X\to_\sigma T^WY$ in~$\ssl{\BC}{W}$. The claim for~$J^W$ follows as
a special case, since~$\BC$ and~$\ssl{\BC}{1}$ clearly remain isomorphic as
guarded monads.

\item We have to verify the fixpoint law. Suppose that
  $f\c W\times X\to_2 T(Y+X)$ and check that
  $f^\iistar=[\eta\snd, f^\iistar]^\klstar\delta
  \brks{\fst,f}$. Indeed,
\begin{flalign*}
&& f^\iistar
&  \;= (T(\snd+\id)\comp \delta\brks{\fst,f})^\istar &\by{definition}\\
&&&\;= [\eta,(T(\snd+\id)\comp \delta\brks{\fst,f})^\istar]^\klstar\comp T(\snd+\id)\comp \delta\brks{\fst,f} &\by{fixpoint}\\
&&&\;= [\eta\snd,f^\iistar]^\klstar \delta\brks{\fst,f}. &
\end{flalign*}
It remains to show that for $k\c W\to V$, $\ssl{\BC}{k}\c\ssl{\BC}{V}\to\ssl{\BC}{W}$
preserves iteration, so let $f\c V\times X\to T(Y+X)$ in~$\BC$;
expanding the definition of $\ssl{\BC}{k}$ and strong iteration, we have to
show that
\begin{equation*}
  (T(\snd+\id_{V\times X})\delta\brks{\fst,f})^\dagger(k\times\id_X)=
  (T(\snd+\id_{W\times X})\delta\brks{\fst,f(k\times \id)})^\dagger.
\end{equation*}
By uniformity, this equation follows from commutativity of (the outer
frame in) the following diagram
\begin{equation*}
  \begin{tikzcd}[column sep = 3cm]
    W\times X \ar{d}[left]{\brks{\fst,f(k\times\id)}}
    \ar{r}{k\times\id} & V\times X\ar{d}[right]{\brks{\fst,f}}\\
    W\times T(Y+X)\ar{r}{k\times\id}\ar{d}[left]{\delta}
    & V\times T(Y+X)\ar{d}[right]{\delta}\\
    T(W\times Y+W\times X)\ar{r}{T(k\times\id+k\times\id)}
    \ar{d}[left]{T(\snd+\id)} 
    & T(V\times Y+V\times X)\ar{d}[right]{T(\snd+\id)}\\
    T(Y+W\times X)\ar{r}[above]{T(\id+k\times\id)}
    & T(Y+V\times X)
  \end{tikzcd}
\end{equation*}
in which the middle square commutes by naturality of~$\delta$ and
commutativity of the other two squares is obvious.

The claim for~$J^W$ follows as a special case as soon as we show
that~$\BBT$ and~$\BBT^1$ are isomorphic as guarded pre-iterative monads.
This is by uniformity w.r.t.\ the isomorphisms
$\brks{\bang,\id_X}\c X\to 1\times X$ (composition with which defines the
isomorphism $\ssl{\BC}{1}\to\BC$), with the application condition checked
in a very similar calculation as above.

  \item %
We check the laws one by one. 
\begin{citemize}
  \item\emph{(naturality)} We have to show that 
\begin{displaymath}
  g^\klstar\tau \brks{\fst,f^{\iistar}} = ([(T\inj_1) \comp g, \eta\inj_2\snd]^{\klstar} \delta\brks{\fst, f})^{\iistar}
\end{displaymath} 
with $f\c W\times X\to_2 T(Y+X)$, $g\c W\times Y \to TZ$. Let us rewrite
the left-hand side as follows, using the definition of
$(\argument)^\iistar$ and \emph{naturality} of $(\argument)^\istar$,
with steps marked by capital letters explained in detail afterwards:
\begin{flalign*}
&&g^\klstar\tau \brks{&\!\fst,f^{\iistar}} \\
&&=&\; g^\klstar(\delta\brks{\fst,f})^\istar &\by{Lemma~\ref{lem:iistar}}\\
&&=&\; ([(T\inj_1)g,\eta\inj_2 ]^\klstar\delta\brks{\fst,f})^\istar &\by{naturality}\\
&&=&\; ([(T\inj_1)\comp g \snd , \eta\inj_2(\id\times\snd)]^{\klstar}\comp\\
&&  &\qquad\qquad T(\brks{\fst,\id}+\brks{\fst,\id})\comp \delta\brks{\fst,f})^\istar&\by{(A)}\\
&&=&\; ([(T\inj_1)\comp g \snd , \eta\inj_2(\id\times\snd)]^{\klstar}\comp \delta\brks{\fst,\delta\brks{\fst, f}})^\istar & \by{\eqref{eq:d-delta}}
  \\
&&=&\; \bigl(T(\snd+\id)\comp [(T\inj_1) \tau(\id\times g), (T\inj_2)\tau(\id\times\eta\snd)]^{\klstar}\\
&&                    &\qquad\qquad\comp  \delta\brks{\fst,\delta\brks{\fst, f}}\bigr)^\istar&\by{(B)}\\
&&=&\; \bigl(T(\snd+\id)\comp [(T\inj_1) \tau, (T\inj_2)\tau]^{\klstar}\comp \\
&&                    &\qquad\qquad  T(\id\times g+\id\times\eta\snd)\comp\delta\brks{\fst,\delta\brks{\fst, f}}\bigr)^\istar&\by{coproducts}\\
&&=&\; \bigl(T(\snd+\id)\comp [(T\inj_1) \tau, (T\inj_2)\tau]^{\klstar}\comp\\
&&  &\qquad\qquad\delta\brks{\fst,T(g+\eta\snd)\delta\brks{\fst, f}}\bigr)^\istar&\by{naturality of~$\delta$}\\
&&=&\; \bigl(T(\snd+\id)\comp\delta\comp(\id\times[T\inj_1,T\inj_2]^\klstar) \\
&&  &\qquad\qquad\brks{\fst,T(g+\eta\snd)\delta\brks{\fst, f}}\bigr)^\istar&\by{(C)}\\
&&=&\; \bigl(T(\snd+\id)\comp\delta\comp\brks{\fst,[(T\inj_1) \comp g, (T\inj_2)\eta\snd]^{\klstar}\comp \delta\brks{\fst, f}}\bigr)^\istar&\by{(co-)products}\\ 
&&=&\;\bigl([(T\inj_1) \comp g, \eta\inj_2\snd]^{\klstar}\comp \delta\brks{\fst, f}\bigr)^{\iistar}
     &\by{definition}
\end{flalign*}
In step~(A), we use that generally, $k^\klstar\comp (Th)=(k\comp h)^\klstar$
and that
\begin{equation*}
  (\snd+\id\times\snd)\comp(\brks{\fst,\id}+\brks{\fst,\id})=\id+\id=\id.
\end{equation*}
In step~(B), we use that generally,
$(Tk)\comp [h,u]^\klstar=(Tk\comp[h,u])^\klstar=[(Tk)\comp h,(Tk)\comp u]^\klstar$ and that
\begin{flalign*}
&&  T(\snd+\id)&\comp(T\inj_1)\comp\tau\comp(\id\times g) \\*
&&  & = (T\inj_1)\comp (T\snd)\comp\tau\comp(\id\times g)&\by{coproducts}\\
&&  & = (T\inj_1) \comp\snd\comp(\id\times g) & \by{coherence of~$\tau$}\\
&&  & = (T\inj_1) \comp g\comp\snd & \by{products}
\intertext{
as well as
}
&&  T(\snd+\id)&\comp(T\inj_2)\comp\tau\comp(\id\times\eta\snd) \\
&&  & =(T\inj_2)\comp\tau\comp(\id\times\eta\snd) &\by{coproducts}\\
&&  & =(T\inj_2)\comp\eta\comp(\id\times\snd)& \by{coherence of~$\tau$}\\
&&  & = \eta\comp\inj_2\comp(\id\times\snd)&\by{naturality of~$\eta$}
\end{flalign*}
Finally, we justify step~(C) as follows. First, we note that
\begin{equation}
  \label{eq:tau-delta}
  [T\inj_1,T\inj_2]^\klstar\comp T(\tau + \tau) = \delta^\klstar\comp T[\id \times T\inj_1, \id \times T\inj_2],
\end{equation}
as seen by the following calculation:
\begin{flalign*}
&&\delta^\klstar \comp& T[\id \times T \inj_1, \id \times T \inj_2]\\
&&&= ((T \dist) \comp \tau \comp [\id \times T \inj_1, \id \times T \inj_2])^\klstar & \by{definition}\\
&&&= ((T \dist) \comp [\tau \comp (\id \times T \inj_1), \tau \comp (\id \times T \inj_2)])^\klstar&\by{coproducts}\\
&&&= ((T \dist) \comp [T(\id \times \inj_1),T(\id \times \inj_2)] \comp (\tau + \tau))^\klstar&\by{naturality of~$\tau$}\\
&&&= ([T \inj_1, T \inj_2] \comp (\tau + \tau))^\klstar&\by{distributivity}\\
&&&= [T\inj_1, T \inj_2]^\klstar \comp T(\tau + \tau),
\intertext{
Using~\eqref{eq:tau-delta}, we now calculate
}
&& [(T&\inj_1)\comp\tau, (T\inj_2)\comp\tau]^{\klstar}\comp\delta \\
&&& = [T\inj_1 , T\inj_2]^{\klstar}\comp T(\tau+\tau)\comp T\dist\comp\tau &\by{definition} \\
&&& =  \delta^\klstar\comp T[\id \times T \inj_1, \id \times T \inj_2] \comp T\dist\comp\tau
  &\by{\eqref{eq:tau-delta}}\\
&&& =   \delta^\klstar\comp T(\id\times[T\inj_1,T\inj_2]) \comp\tau &\by{distributivity}\\
&&& = (T\dist)\comp\tau^\klstar\comp T(\id\times[T\inj_1,T\inj_2]) \comp\tau &\by{definition}\\
&&& = (T\dist)\comp\tau\comp(\id\times[T\inj_1,T\inj_2]^\klstar) &\by{coherence of~$\tau$}\\
&&& = \delta\comp(\id\times[T\inj_1,T\inj_2]^\klstar) &\by{definition}
\end{flalign*}
as used in~(C).

  \item\emph{(codiagonal)} We have to show that
\begin{displaymath}
  (T[\id,\inj_2] \comp f)^{\iistar} = f^{\iistar\iistar}
\end{displaymath} 
for $f\c W\times X\to_{12,2} T((Y+X)+X)$. We have the following
straightforward identity (which we prove after the main argument)
between two morphisms from $W\times T((Y+X)+X)$ to
$T(W\times Y+W\times X)$:
\begin{equation}\label{eq:delta-codiag}
  \delta (\id\times T[\id,\inj_2]) = T[\dist,\inj_2]\comp \delta .
\end{equation}
Using this equation, we obtain on the one hand, using
\emph{codiagonal} for~$(\argument)^\istar$:
\begin{flalign*}
&&(T[\id&,\inj_2]\comp f)^{\iistar}\\*
&&&= (T(\snd+\id)\comp\delta\comp\brks{\fst,T[\id,\inj_2]\comp f})^\istar&\by{definition}\\
&&&= (T(\snd+\id)\comp\delta\comp(\id\times T[\id,\inj_2])\comp \brks{\fst,f})^\istar&\by{products}\\
&&&= (T(\snd+\id)\comp T[\dist,\inj_2]\comp \delta \comp \brks{\fst,f})^{\istar}&\by{\eqref{eq:delta-codiag}}\\
&&&= (T[(\snd+\id) \dist,\inj_2]\comp \delta\comp \brks{\fst,f})^{\istar}& \by{coproducts}\\
&&&= (T[\id,\inj_2]\comp T((\snd+\id) \dist+\id)\comp\delta\comp \brks{\fst,f})^{\istar}& \by{coproducts}\\
&&&= (T((\snd+\id)\comp \dist+\id)\comp\delta \comp\brks{\fst,f})^{\istar\istar}&\by{codiagonal}
\intertext{and on the other hand:}%
&&f^{\iistar\iistar}
&= (T(\snd+\id)\comp\delta\comp\brks{\fst,f^\iistar})^\istar &\by{definition}\\
&&&= (T(\snd+\id)\comp(T\dist) \comp \tau \comp\brks{\fst,f^\iistar})^\istar&\by{defn.\ of $\delta$}\\
&&&= (T(\snd+\id)\comp (T\dist) \comp(\delta\comp \brks{\fst,f})^{\istar})^{\istar} &\by{Lemma~\ref{lem:iistar}} \\
&&&= (T((\snd+\id)\comp \dist+\id)\comp\delta \comp\brks{\fst,f})^{\istar\istar}. &\by{naturality}
\intertext{\noindent
It remains to prove~\eqref{eq:delta-codiag}: We have}
&&\delta \comp(\id &\times T[\id,\inj_2]) \\
&&&= (T\dist) \comp\tau\comp (\id \times T[\id,\inj_2])  &\by{definition}\\
&&&=(T\dist) \comp T(\id \times [\id,\inj_2])\comp \tau  &\by{naturality of~$\tau$}\\
&&&= (T\dist) \comp T([\id \times \id, \id \times \inj_2])\comp (T\dist) \comp\tau  &\by{distributivity}\\
&&&= T[\dist,\dist \comp(\id \times \inj_2)] \comp\delta &\by{coproducts, definition}\\
&&&= T[\dist,\inj_2] \comp\delta  &\by{distributivity}
\end{flalign*}

\item\emph{(uniformity)} For $f\c  W\times X \to_2 {T(Y + X)}$,
  $g\c W\times Z \to_2 T(Y + Z)$, and $h\c  W\times Z \to X$, the premise
  of the uniformity law expands by the definition of the structure
  of~$\ssl{\BC}{W}$ to the equation
  \begin{equation}\label{eq:unif-prem}
    f \comp \brks{\fst, h} = T(\snd+ h)\comp
    \delta \brks{\fst, g}.
  \end{equation}
  Then we derive the conclusion of the uniformity law,
  \begin{align*}
    f^{\iistar} \comp \brks{\fst, h} = (T(\snd+\id)\comp\delta\comp\brks{\fst,f})^\istar \comp \brks{\fst, h}= (T(\snd+\id)\comp\delta\comp\brks{\fst,g})^\istar = g^\iistar, 
  \end{align*} 
  using the definition of~$(\argument)^\iistar$ and uniformity of
  $(\argument)^\istar$, whose premise is verified as follows:
\begin{flalign*}
&&(T(\snd\,&+\id)\comp\delta\comp\brks{\fst,f}) \comp \brks{\fst, h}\\
&&=&\; T(\snd+\id)\comp\delta\comp\brks{\fst,f\comp \brks{\fst, h}}&\by{products}\\
&&=&\; T(\snd+\id)\comp\delta\comp\brks{\fst,T(\snd+ h)\comp \delta\comp\brks{\fst, g}}&\by{\eqref{eq:unif-prem}}\\
&&=&\; T(\snd+\id)\comp \delta\comp (\id\times T(\snd+ h))\comp\brks{\fst,\comp \delta\brks{\fst, g}}&\by{products}\\
&&=&\; T(\snd+\id)\comp T(\id\times\snd+\id\times h) \comp\delta\comp\brks{\fst,\comp \delta\brks{\fst, g}} & \by{naturality of~$\delta$}\\
&&=&\; T(\snd\snd+\id\times h) \delta\brks{\fst,\comp \delta\comp\brks{\fst, g}}\\
&&=&\; T(\snd\snd+\id\times h)\comp T(\brks{\fst,\id}+\brks{\fst,\id}) \comp \delta\comp\brks{\fst, g}&\by{\eqref{eq:d-delta}}\\
&&=&\; T(\snd+\brks{\fst, h})\comp\delta\comp\brks{\fst,g}. 
\end{flalign*} 
\item\emph{(strength)} Again, we go via
  Theorem~\ref{thm:slice-converse}. By the above, the guarded monad
  $\BBT^{W\times V}$ on $\ssl{\BC}{W\times V}$ is pre-iterative and satisfies
  \emph{uniformity}. The same thus transfers to the isomorphic guarded
  monad $(\BBT^W)^V$ on $\ssl{(\ssl{\BC}{W})}{V}$. Moreover, the embedding
  $\ssl{\BC}{W}\to\ssl{(\ssl{\BC}{W})}{V}$ corresponds to $\ssl{\BC}{\fst}$ under the
  isomorphism $\ssl{(\ssl{\BC}{W})}{V}\cong\ssl{\BC}{W\times V}$, and thus preserves
  iteration by item~\ref{item:pre-iterative}. By
  Theorem~\ref{thm:slice-converse}.\ref{item:converse-strong-it}, it
  follows that iteration on $(\BBT^W)^V$ is strong iteration
  on~$\BBT^W$. Moreover, again by the above, $(\BBT^W)^V$ satisfies
  \emph{naturality}. By
  Theorem~\ref{thm:slice-converse}.\ref{item:converse-strength-law},
  it follows that $\BBT^W$ satisfies \emph{strength}.

\end{citemize}
\item Suppose that $\BBT$ is guarded iterative, hence guarded
  Elgot. By the previous clause we know that given
  $f\c W\times X\to_2 T(Y+X)$, $f^\iistar$ satisfies the fixpoint law;
  by unfolding the definitions of the coproduct and monad structures on
  $\ssl{\BC}{W}$, we obtain
  $f^\iistar=[\eta\snd, f^\iistar]^\klstar\delta \brks{\fst,f}$. We
  are left to show that this equation is satisfied by $f^\iistar$
  uniquely. Indeed, suppose that for some $g\c W\times X\to T(Y+X)$,
  $g=[\eta\snd, g]^\klstar\delta \brks{\fst,f}$. Hence
  $g = [\eta, g]^\klstar\comp T(\snd+\id)\comp \delta \brks{\fst,f}$, and
  therefore $g = (T(\snd+\id)\comp \delta\comp \brks{\fst,f})^\istar$, using the
  fact that $\BBT$ is guarded iterative; but the right hand side is
  just the definition~\eqref{eq:iter_par} of $f^\iistar$.
  \qed
\end{cenumerate}
\noqed\end{proof}
\noindent The proof of the converse statements then runs as follows: 
\begin{proof}[Proof (Theorem~\ref{thm:slice-converse})]
  \begin{enumerate}[wide]
  \item Most of the claim is immediate from the known fact that
    giving a lifting of a monad to the Kleisli category of a comonad
    is equivalent to giving a comonad-over-monad distributive
    law~\cite{PowerWatanabe02}; that is, for each $W$ we have a
    distributive law $\tau_{W,-}$ of $W\times(\argument)$ over~$\BBT$, defined
    as
    \begin{equation}\label{eq:strength-from-lifting}
      (\tau_{W,X}\c W\times TX\to T(W\times X))=T^Wj_X,
    \end{equation}
    where the $\ssl{\BC}{W}$-morphism $T^Wj_X\c TX\to T(W\times X)$ is
    converted into a $\BC$-morphism $W\times TX\to T(W\times X)$, and
    this construction is inverse to the construction of a lifting
    of~$\BBT$ from a strength of~$\BBT$ given in the proof of
    Theorem~\ref{thm:context-comonad}.\ref{item:monad-extension}. Explicitly,
    this means that throughout the remainder of the proof, we can
    assume that strength and lifting relate to each other via
    Equation~\eqref{eq:strength-from-lifting} above and the
    description of~$\BBT^W$ in the proof of
    Theorem~\ref{thm:context-comonad}.\ref{item:monad-extension}. In
    particular,
    \begin{equation}\label{eq:lifting-from-strength}
      T^Wf = (Tf)\comp \tau
    \end{equation}
    for $f\c X\to Y$ in~$\ssl{\BC}{W}$ (i.e.\ $f\c W\times X\to Y$ in~$\BC$).
    Of course, $\tau_{W,X}$ will serve as the strength; it remains
    only to verify those axioms that do not already feature among the
    properties of $\tau_{W,-}$ as a distributive law
    (cf.~\cite{BrookesVanStone93}) -- that is, we need to verify
    naturality of~$\tau_{W,X}$ in~$W$ and compatibility with the
    associator, which both involve two different instances
    $W\times(\argument)$, $V\times(\argument)$ of the product comonad.
   
    \emph{Naturality in~$W$:} Let $k\c V\to W$; we have to show that
    \begin{equation*}
      (T^Wj^W_X)\comp (k\times T\id_X)=T(k\times\id_X)\comp (T^Vj^V_X) 
    \end{equation*}
    in~$\BC$, where we have decorated the unit~$j$ of the co-Kleisli
    adjunction with additional superscripts to indicate the relevant
    simple slice. We calculate as follows:
    \begin{flalign*}
      &&(T^Wj^W_X)\comp (k\times T\id_X)
      & =\; (\ssl{\BC}{k})\comp(T^Wj^W_X)                          &\by{definition}\\
      &&& =\; T^V((\ssl{\BC}{k})\comp j^W_X)                        & \by{coherence}\\
      &&& =\; T^V(j^W_X(k\times\id_X))                & \by{definition of~$\ssl{\BC}{k}$}\\
      &&& =\; T^V(k\times\id_X)                       &\by{$j_X^W$ is $\id$ in~$\BC$}\\
      &&& =\; T^V(J^V(k\times\id_X)\circ^Vj^V_X)      & \by{\eqref{eq:slice-universal}}\\
      &&& =\; T^V(J^V(k\times\id_X))\circ^VT^Vj^V_X   &\by{functoriality}\\
      &&& =\; J^V(T(k\times\id_X))\circ^VT^Vj^V_X     &\by{extension}\\
      &&& =\; T(k\times\id_X)\comp(T^Vj^V_X).               & \by{definitions of~$J^V$, $\circ^V$}
    \end{flalign*}
    
    \emph{Compatibility with the associator:} Eliding the actual
    associator $V\times (W\times X)\cong (V\times W)\times X$, we have
    to show that the diagram
    \begin{equation}\label{diag:assoc-strength}
      \begin{tikzcd}[column sep = huge]
        V\times W\times TX \ar{r}{\id_V\times\tau_{W,X}}\ar{dr}[below
        left]{\tau_{V\times W,X}}
        & V\times T(W\times X)\ar{d}{\tau_{V,W\times X}}\\
        & T(V\times W\times X)
      \end{tikzcd}
    \end{equation}
    commutes. Since each $\tau_{U,-}$ is a distributive law, it is
    compatible with the comultiplication of $U\times(\argument)$, which is
    $\Delta_U\times\id_X:U\times X\to U\times U\times X$; explicitly,
    all diagrams
    \begin{equation}\label{diag:diag-strength}
      \begin{tikzcd}[column sep = 3cm]
        U\times TX\ar{r}{\diag_U\times\id_{TX}}\ar{dd}[left]{\tau_{U,X}}
        & U\times U\times TX\ar{d}[right]{\id_U\times\tau_{U,X}}\\
        & U\times T(U\times X) \ar{d}[right]{\tau_{U,U\times X}}\\
        T(U\times X)\ar{r}[above]{T(\diag_U\times\id_X)}
          & T(U\times U\times X)
      \end{tikzcd}
    \end{equation}
    commute. We apply this to $U=V\times W$ in the following
    calculation proving commutation of~\eqref{diag:assoc-strength},
    using moreover naturality of~$\tau$ in both variables:
    \begin{flalign*}
      &&&\kern-1em\tau_{V\times W,X}\\ 
      &&& = T(\fst\times\snd\times\id_X)T(\diag_{V\times W}\times\id_X)\comp\tau_{V\times W,X}\\
      &&& = T(\fst\times\snd\times\id_X)\tau_{U,U\times X}
        (\id_{U}\times\tau_{U,X})(\diag_{U}\times\id_{TX})
      &\by{\eqref{diag:diag-strength}}\\
      &&& = \tau_{V,W\times X}(\fst\times T(\snd\times\id_X))
        (\id_U\times\tau_{U,X})(\diag_U\times\id_{TX}) &\by{naturality}\\
      &&& = \tau_{V,W\times X}(\fst\times T(\snd\times\id_X)\tau_{U,X})\brks{\fst,\id_{U\times TX}}\\
      &&& = \tau_{V,W\times X}(\fst\times \tau_{W,X}(\snd\times\id_{TX}))\comp\brks{\fst,\id_{U\times TX}} &\by{naturality}\\
      &&& = \tau_{V,W\times X}(\id_V\times  \tau_{W,X}).
    \end{flalign*}
  
  \item Let $f\c W\times X\to_\sigma TY$ in~$\BC$. Then
    $J^Wf\c W\times X\to T^WY$ is $\sigma$-guarded in~$\ssl{\BC}{W}$ since $J^W$ preserves
    guardedness. By~\eqref{eq:slice-universal},
    $f=(J^Wf)\circ^W j_X\c X\to T^WY$ in~$\ssl{\BC}{W}$, so
    $f\c X\to_\sigma T^WY$ in~$\ssl{\BC}{W}$ by \textbf{(cdm)}.
    
  \item We have to show that the guardedness structure on $\BBT$
    satisfies the axiom \textbf{(str)}. So let $\sigma\c Y'\cpto Y$,
    with complement $\sigma'\c Y''\cpto Y$, and let $f\c X\to_\sigma TY$
    in $\BC$.  We have to show that
    $\tau(\id_W\times f)\c W\times X\to_{\id_W\times \sigma} T(W\times
    X)$. By~\eqref{eq:strength-from-lifting} and the definition of
    composition in~$\ssl{\BC}{W}$, $\tau(\id_W\times f)$ is the morphism
    \begin{equation*}
      (T^Wj)\circ^W (J^Wf)\c X\to T^W(W\times Y)
    \end{equation*}
    in~$\ssl{\BC}{W}$. By
    Theorem~\ref{thm:context-comonad}.\ref{item:context-guarded},~$J^W$
    preserves guardedness, so we have
    $J^Wf\c X\to_\sigma T^W(W\times Y)$.  Since $W\times Y$ is a
    coproduct of $W\times Y'$ and $W\times Y''$, and~$j$ then has the
    form $j'+j''$ with $j'\c Y'\to W\times Y'$,
    $j''\c Y''\to W\times Y''$, it follows by \textbf{(cmp)} that
    $(T^Wj)\circ^W (J^Wf)$ is $\id_W\times\sigma$-guarded. By the
    definition of guardedness in~$\ssl{\BC}{W}$, the required guardedness
    of~$\tau(\id_W\times f)$ follows.
  \item We denote iteration in~$\ssl{\BC}{W}$ by~$(\argument)^\iistar$, and show that
    the equality~\eqref{eq:iter_par} holds. Let
    $f\c W\times X\to T(Y+X)$ in~$\BC$, i.e.\ $f\c X\to T^W(Y+X)$
    in~$\ssl{\BC}{W}$. The square
    \begin{equation*}
      \begin{tikzcd}
        X \ar{r}{j_X} \ar{d}[left]{f} &[3em] 
        W\times X\ar{d}[right]{J^W(T(\snd+\id)\delta\brks{\fst,f})}\\
        T^W(Y+X) \ar{r}[below]{T^W(\id+j_X)} &T^W(Y+W\times X)
      \end{tikzcd}
    \end{equation*}
    commutes in~$\ssl{\BC}{W}$: By \eqref{eq:slice-universal}, the upper
    right composite equals
    $T(\snd+\id)\delta\brks{\fst,f}\c W\times X\to T(Y + W\times X)$
    in~$\BC$, which is precisely the term obtained by unfolding the
    definition of the structure of~$\ssl{\BC}{W}$ in terms of that of~$\BC$
    in the lower left composite, in particular
    using~\eqref{eq:lifting-from-strength}. By uniformity in~$\ssl{\BC}{W}$,
    it follows that
    \begin{flalign*}
      &&f^\iistar & = (J^W(T(\snd+\id)\delta\brks{\fst,f}))^\iistar\circ^W j \\
      &&& = J^W((T(\snd+\id)\delta\brks{\fst,f})^\dagger)\circ^W j & \by{$J^W$ preserves iteration}\\
      &&& = (T(\snd+\id)\delta\brks{\fst,f})^\dagger. &\by{\eqref{eq:slice-universal}}
    \end{flalign*}
  \item Let $f\c X\to_2 T(Y+X)$. Like in the proof of
    Claim~\ref{item:strength-guarded}, we have that the left-hand side
    of the \emph{strength} law for~$f$ is written within $\ssl{\BC}{W}$ as
    $(T^Wj)\circ^W (J^Wf^\dagger)\c X\to T^W(W\times Y)$ with~$j$ as
    above, which we rewrite using preservation of iteration by~$J^W$
    and naturality in~$\ssl{\BC}{W}$ as
    \begin{align*}
      (T^Wj)\circ^W (J^Wf^\dagger) & = (T^Wj)\circ^W (J^Wf)^\iistar\\
      & = (T^W(j+\id_X)\circ^W (J^W f))^\iistar
    \end{align*}
    with~$(\argument)^\iistar$ denoting iteration in~$\ssl{\BC}{W}$ and all further
    data, including~$+$ and identities, read in~$\ssl{\BC}{W}$ as
    well. Expanding definitions, we have
    \begin{align*}
      T^W(j+\id_X)&\circ^W (J^W f) \\
      & = T(\id_{W\times Y} + \snd)\delta\brks{\fst,f\snd}\\
      & = T(\id_{W\times Y} + \snd)\delta(\id_W\times f)
    \intertext{in~$\BC$. Since~$(\argument)^\iistar$ is assumed to be strong iteration, we
    further have}
        (T^W(j+\id_X)&\circ^W (J^W f))^\iistar \\
      & = (T(\id_{W\times Y} + \snd)\delta(\id_W\times f))^\iistar\\
       &=  (T(\snd+\id_{W\times X})\delta\brks{\fst,T(\id_{W\times Y}+\snd)\delta(\id_W\times f)})^\dagger\\
      & = (\delta(\id_W\times f))^\dagger,
    \end{align*}
    which is the right-hand side of the \emph{strength} law.\qedhere
  \end{enumerate}
\end{proof}

\section{A Metalanguage for Guarded Iteration}\label{sec:meta}

\noindent We proceed to define a variant of fine-grain
call-by-value~\cite{LevyPowerEtAl02} following the ideas
from~\cite{GeronLevy16} on \emph{labelled iteration}. For our purposes
we extend the standard setup by allowing a custom signature of
operations $\Sigma$, but restrict the expressiveness of the language
being defined slightly, mainly by excluding function spaces for the
moment. The latter require some additional treatment, and we return to
this point in Section~\ref{sec:expo}.
We fix a supply $\oname{Base}$ of \emph{base types} and define
(composite) \emph{types} $A$, $B$ by the grammar
\begin{align}\label{eq:sum-types}
A,B,\ldots          &\Coloneqq C\mid 0\mid 1\mid A + B \mid A\times B && (C\in\oname{Base})
\end{align}
The signature $\Sigma$ consists of two disjoint parts: a \emph{value
  signature} $\Sigma_v$ containing signature symbols of the form
$f\c A\to B$, and an \emph{effect signature} $\Sigma_c$ containing
signature symbols of the form $f\c A\to B [C]$. While the former symbols
represent pure functions, the latter capture morphisms of type
$A\to_{2} T(B+C)$; in particular they carry side-effects from~$T$.
The term language over these data is given in
Fig.~\ref{fig:lang.full}.  We use a syntax inspired by Haskell's
$\oname{do}$-notation~\cite{JonesHughesEtAl99}.  The metalanguage
features two kinds of judgements:
\begin{align}\label{eq:judg}
\Gamma\vctx v\c A  &&\text{and}&& \Delta\csep\Gamma\cctx p\c A  
\end{align}
for \emph{values} and \emph{computations}, respectively. These involve
two kinds of contexts:~$\Gamma$ denotes the usual context of typed
\emph{variables} $x\c A$, and $\Delta$ denotes the context of typed
\emph{exceptions} $e\c E^{\alpha}$ with $E$ being a type
from~\eqref{eq:sum-types} and $\alpha$ being a tag from the
two-element set $\{\gtag,\utag\}$ to distinguish the exceptions raised
in a guarded context~($\gtag$) from those raised in an unguarded
context ($\utag$) of the program code. Let us denote by $|\Delta|$ the
list of pairs $e\c E$ obtained from an exception context~$\Delta$ by
removing the $\gtag$ and $\utag$ tags.  Variable and exception names
are drawn from the same infinite stock of symbols; they are required
to occur non-repetitively in~$\Gamma$ and in $\Delta$ separately, but
the same symbol may occur in $\Gamma$ and in $\Delta$ at the
same~time.

\begin{notation}\label{notation}
As usual, we use the dash $(\argument)$ to denote a fresh variable in binding 
expressions, e.g.\ $\mbind{\argument\gets p}{q}$, and use the standard conventions
of shortening $\mbind{\argument\gets p}{q}$ to $\mbind{p}{q}$ and $\mbind{x\gets p}{(\mbind{y\gets q}{r})}$
to $\mbind{x\gets p;y\gets q}{r}$. Moreover, we encode the if-then-else construct 
$\ift{b}{p}{q}$ as $\case{b}{{\inl\argument\mto p}}{}$ ${\inr\argument\mto q}$,
and also use the notation 
\begin{displaymath}
  f(v)\, \&\, p\qquad\text{for}\qquad \gcase{f}{v}{x}{\oname{init} x}{\argument}{p}
\end{displaymath}
whenever $f\c X\to 0[1]\in\Sigma_c$.
\end{notation}

\noindent The language constructs relating to products, coproducts,
and the monad structure are standard (except maybe $\oname{init}$,
which forms unique morphisms from the null type~$0$ into any type~$A$)
and should be largely self-explanatory. The key features of our
metalanguage, discussed next, concern algebraic operations on the one
hand, and exception-based iteration on the other hand.
\begin{figure}[t!]
\begin{center}%
{\parbox{\textwidth}{%
\small%
  \begin{subfigure}{\textwidth}
\begin{flalign*}
\quad\anonrule{var}{%
		x\c A \text{~~in~~} \Gamma
	}{%
		\Gamma \vctx x\c A
	}
&&\anonrule{u-sig}{%
		f\c A\to B\in\Sigma_v\quad \Gamma\vctx v\c A
	}{%
		\Gamma \vctx f(v)\c B
	}
&&
\anonrule{unit}{}{\Gamma\vctx\star\c 1}
\quad
\end{flalign*}
\\[-4ex]
\begin{flalign*}
\quad
\anonrule{prod}{%
		\Gamma\vctx v\c A\qquad \Gamma\vctx w\c B 
	}{%
		\Gamma \vctx \brks{v,w}\c A\times B
	}
&&
\anonrule{inl}{%
		\Gamma\vctx v\c A 
	}{%
		\Gamma \vctx \inl v\c A+B
	}
&&
\anonrule{inr}{%
		\Gamma\vctx w\c B 
	}{%
		\Gamma \vctx\inr w\c A+B
	}
\quad
\end{flalign*}
\medskip
\end{subfigure}

\dotfill

\medskip
\begin{subfigure}{\textwidth}
\begin{flalign*}
  \anonrule{prod}{%
    \Gamma \vctx p\c A\times B \qquad
    \Delta\csep \Gamma,x\c A,y\c B \cctx q\c C
  }{%
    \Delta\csep \Gamma \cctx \pcase{p}{\brks{x,y} \mto q}\c C
  }
\end{flalign*}\\[-4ex]
\begin{flalign*}
&&
  \anonrule{do}{%
    \Delta\csep \Gamma \cctx p\c A \qquad
    \Delta\csep \Gamma, x\c A \cctx q\c B
  }{%
    \Delta\csep \Gamma \cctx \mbind{x \gets p}{q}\c B
  }
&&
\anonrule{ret}{%
		\Gamma\vctx v\c A 
	}{%
		\Delta\csep \Gamma \cctx \ret v\c A
	}
&&
\end{flalign*}\\[-4ex]
\begin{flalign*}
\anonrule{g-sig}{%
    f\c A\to B [C]\in\Sigma_c \qquad 
    \Gamma \vctx v\c A\qquad 
    \begin{array}[b]{r@{\ }l}\Delta\csep\Gamma,x\c B &\cctx p\c D\\[.5ex]
    \Delta'\csep\Gamma,y\c C&\cctx q\c D
    \end{array}\qquad
    |\Delta| = |\Delta'|
	}{%
		\Delta\csep \Gamma \cctx \gcase{f}{v}{x}{p}{y}{q}\c D
	}
\end{flalign*}\\[-4ex]
\begin{flalign*}
\anonrule{void}{\Gamma\vctx t\c 0}{\Delta\csep\Gamma\cctx\oname{init} t\c A}
&&
  \anonrule{case}{%
    \Gamma \vctx v\c A+B \qquad
    \Delta\csep \Gamma,x\c A \cctx p\c C \qquad \Delta\csep \Gamma,y\c B \cctx q\c C
  }{%
    \Delta\csep \Gamma \cctx \case{v}{\inl x \mto p}{
      \inr y \mto q}\c C
  }
\end{flalign*}\\[-4ex]
\begin{flalign*}
  \anonrule{handle}{%
    \Delta, e\c E^\utag \csep\Gamma \cctx p\c A \qquad \Delta\csep \Gamma,e\c E \cctx q\c A%
  }{%
    \Delta\csep \Gamma \cctx \handle{e}{p}{q}\c A
  }
\end{flalign*}\\[-4ex]
\begin{flalign*}
&&
\anonrule{handle}{%
     e\c E^\utag ~\text{in}~\Delta\qquad~ \Gamma \vctx q\c E
  }{%
    \Delta\csep \Gamma \cctx\oname{raise}_e q\c D
  }
&&
  \anonrule{itcase}{%
    \Gamma \vctx v\c E \qquad \Delta, e\c E^\gtag \csep\Gamma,e\c E \cctx q\c A 
  }{%
    \Delta\csep \Gamma \cctx \handleit{v}{e}{q}\c A
  }
&&
\end{flalign*}
\end{subfigure}
}}
\end{center}
\caption{Term formation rules for values (top) and computations (bottom).}
\label{fig:lang.full}
\end{figure}

\paragraph{Algebraic operations via Generic effects} The signature
symbols $f\c A\to B[0]$ from $\Sigma_c$ have Kleisli morphisms $A\to TB$
as their intended semantics, specifically, if $A = n$ and $B=m$, with
$n$ and $m$ being identified with the corresponding $n$-fold and
$m$-fold coproducts of $1$, the respective morphisms $n\to Tm$ dually
correspond to \emph{algebraic operations}, i.e.\ certain natural
transformations ${T^m\to T^n}$, as elaborated by Plotkin and
Power~\cite{PlotkinPower01}. In context of this duality the Kleisli
morphisms of type $n\to Tm$ are also called \emph{generic
  effects}. Hence we regard $\Sigma_c$ as a stock of generic effects
declared to be available to the language. The respective algebraic
operations thus become automatically available -- for a brief example
consider the binary algebraic operation of nondeterministic choice
$\oplus\c T^2\to T^1$, which is modelled by a generic effect
$\oname{toss}\c 1\to T2$ as follows:
\begin{displaymath}
  p\oplus q = \mbind{c\gets\oname{toss}}{\case{c}{\inl\argument\mto p}{\inr\argument\mto q}}.
\end{displaymath}  
\paragraph{Exception raising} Following~\cite{GeronLevy16}, we involve
an exception raising/handling mechanism for organizing loops (we make
the connection to exceptions more explicit, in particular, we use the
term `\emph{exceptions}' and not `\emph{labels}', as the underlying
semantics does indeed accurately match the standard exception
semantics). Note that the design of the syntax presented here deviates
slightly from the conference version~\cite{GoncharovRauchEtAl18}. We allow
raising of a standard \emph{unguarded exception} $e\c E^\utag$ with
$\oname{raise}_e q$. More importantly, \emph{guarded exceptions}
$e\c E^\gtag$ can be arbitrarily introduced into the context by the typing
rule for $\ret$, in which~$\Delta$ is completely unspecified. The
\emph{guarded case} command
\begin{displaymath}
  \gcase{f}{v}{x}{p}{y}{q}.
\end{displaymath}
then works as follows: The $f(v)$ part acts as a \emph{guard}
partitioning the control flow into the left (unguarded) part in which
a computation $p$ is executed, and the right (guarded) part, in which
execution continues with $q$. Since~$\Delta'$ need not agree
with~$\Delta$ on guardedness tags, the exceptions occurring in $q$ and
therefore recorded in $\Delta'$ may be promoted from unguarded to
guarded.

The guarded case operator $\oname{gcase}$ also allows us to expose the
guarded part of an operation to the result; i.e.\ for
$f\c A\to B [C]\in\Sigma_c$ we can take $f(v)$ to be an abbreviation for
\begin{equation*}
  \gcase{f}{v}{x}{\ret \inl x}{y}{\ret \inr y}
\end{equation*}
and then derive a typing rule
\[
\anonrule{u-sig}{%
		f\c A\to B [C]\in\Sigma_c\qquad \Gamma\vctx v\c A
	}{%
		\Delta\csep \Gamma \cctx f(v)\c B+C
	}
\]
This is particularly useful for performing operations without
considering their guardedness properties, e.g.\ the final $print$ in
Fig.~\ref{fig:intro}.

\paragraph{(Iterated) exception handling} The syntax for exception handing via 
$\handle{e}{p}{q}$ is meant to be understood as follows: $p$ is a program possibly raising the exception 
$e$ and $q$ is a \emph{handling term} for it. This can be compared to the richer 
exception handling syntax of Benton and Kennedy~\cite{BentonKennedy01} whose 
construct $\mfix{try\,}{x}{\,\Leftarrow\,}{p}{\,in\,}{q}{\,unless\,}{\{e\mto r\}_{e\in E}}$
we can encode as:
\begin{align*}
  \mbind{z\gets\, &\handle{e}{(\mbind{x\gets p}{\ret\inl x})}{(\mbind{y\gets r}{\ret\inr y})}}{\\&
  \case{z}{\inl x\mto q}{\inr y\mto\ret y}}
\end{align*}
where $p$, $q$ and $r$ come from the judgements 
\begin{displaymath}
  \Delta,e\c E^\utag\csep\Gamma\cctx p\c A,\qquad  
  \Delta\csep\Gamma,x\c A\cctx q\c B,\qquad
  \Delta\csep\Gamma,e\c E\cctx r\c B,    
\end{displaymath}
and the idea is to capture the following behaviour: unless $p$ raises
exception $e\c E^\utag$, the result is bound to $x$ and passed to $q$
(which may itself raise $e$), and otherwise the exception is handled
by $r$. An analogous encoding is already discussed
in~\cite{BentonKennedy01} where the richer syntax is advocated and
motivated by tasks in compiler optimization, but since these
considerations are not relevant for our present developments, we
stick to the minimalist syntax as above.

Note that we restrict to handling unguarded exceptions only; since all
exceptions are introduced as unguarded ones, and promoted to guarded
exceptions only for the purpose of iteration, this clearly
suffices. %

The idea of the new construct $\handleit{p}{e}{q}$ is to handle the exception
in~$q$ recursively using $q$ itself as the handling term, so that if $q$
raises~$e$, handling continues repetitively. The value $p$ is substituted into
$q$ to initialize the iteration. For $\mfix{handleit}$, it is crucial that the
exception comes from a guarded context, as required by the relevant typing rule.

\begin{example}
We illustrate a type derivation process in Fig.~\ref{fig:dev-tree}, using the example 
in Fig.~\ref{fig:intro} from the introduction. Due to the page width limitations
the complete derivation tree is cut into five pieces with the curved 
arrows indicating how conclusions are further used as premises of subsequent derivations; 
additionally, we indicate by dots `$\ldots$' the repeated program fragments
taken from the premises.
\end{example}

\begin{figure}
    \small
  \begin{prooftree}
    \AxiomC{}
    \UnaryInfC{$\vaco{r}\c\coco{1^\utag},\vaco{e}\c\coco{1^\utag}\mid \vaco{y,z}\c\coco{\nat} \cctx \ruco{\oname{raise}}_{\vaco{r}}\,\star\c\coco{1}$}
    \AxiomC{}
    \UnaryInfC{$\vaco{r}\c\coco{1^\utag},\vaco{e}\c\coco{1^\utag}\mid \vaco{y,z}\c\coco{\nat} \cctx \ruco{\ret}\,\blco{\star}\c\coco{1}$}
    \BinaryInfC{$\vaco{r}\c\coco{1^\utag},\vaco{e}\c\coco{1^\utag}\mid \vaco{y,z}\c\coco{\nat} \cctx
      \ruco{\ift{\vaco{z} \asco{=} \blco{42}}{\oname{raise}_r 
          \blco{\star}}{\ret \blco{\star}}}\c\coco{1}  %
          $\tikzmark{b1}}
  \end{prooftree}

  \begin{prooftree}
    \AxiomC{}
    \UnaryInfC{$\vaco{r}\c\coco{1^\utag},\vaco{e}\c\coco{1^\utag}\mid \vaco{y,z}\c\coco{\nat} \cctx \ruco{\ret}\,\blco{\star}\c\coco{1}$}
    \AxiomC{}
    \UnaryInfC{$\vaco{r}\c\coco{1^\utag},\vaco{e}\c\coco{1^\utag}\mid \vaco{y,z}\c\coco{\nat} \cctx \ruco{\oname{raise}}_{\vaco{e}}\,\star\c\coco{1}$}
    \BinaryInfC{$\vaco{r}\c\coco{1^\utag},\vaco{e}\c\coco{1^\utag}\mid \vaco{y,z}\c\coco{\nat} \cctx
      \ruco{\ift{\vaco{y} \asco{=} \vaco{z}}{\ret
          \blco{\star}}{\oname{raise}_{\vaco{e}}\,\blco{\star}}}\c\coco{1}  %
$ \tikzmark{b2}}
  \end{prooftree}

  \begin{prooftree}
    \AxiomC{$read\c \coco{1 \to \nat[0]}$}
    \UnaryInfC{$\vaco{r}\c\coco{1^\utag},\vaco{e}\c\coco{1^\utag}\mid
      \vaco{y,z}\c\coco{\nat}\cctx read ()\c\coco{\nat}$}
    \AxiomC{$rand: \coco{1 \to\nat[0]}$}
    \UnaryInfC{$\vaco{r}\c\coco{1^\utag},\vaco{e}\c\coco{1^\utag}\mid
      \vaco{y,z}\c\coco{\nat}\cctx rand ()\c\coco{\nat}$}
    \AxiomC{\tikzmark{e2}\,\,\tikzmark{e1}}
    \TrinaryInfC{$\vaco{r}\c\coco{1^\utag},\vaco{e}\c\coco{1^\utag}\mid\coco{\emptyset}
      \cctx \ruco{\mbind{\vaco{y} \gets \blco{rand ()}}{\vaco{z} \gets
          \blco{read ()}; \blco{\ldots}\;;
          \blco{\ldots}\;}} %
  \c\coco{1} 
$ \tikzmark{b3}}
  \end{prooftree}

  \begin{prooftree}
    \AxiomC{$print\c\coco{\str \to 0[1]}$}
    \AxiomC{\qquad\tikzmark{e3}}
    \BinaryInfC{$\vaco{r}\c\coco{1^\utag},\vaco{e}\c\coco{1^\gtag}\mid\coco{\emptyset} \cctx \blco{print\, (
        \text{\texttt{"think of a number"}})}\,\&\,\ldots\; %
$}
    \UnaryInfC{$\vaco{r}\c\coco{1^\utag}\mid\coco{\emptyset}\cctx\ruco{\oname{handleit}\,\vaco{e} = \blco{\star}\,\oname{in}}\,\blco{\ldots}
      \c\coco{1} %
$ \tikzmark{b4}}
  \end{prooftree}

  \begin{prooftree}
    \AxiomC{\tikzmark{e4}}
    \AxiomC{}
    \UnaryInfC{$\coco{\coco{\emptyset}}\mid\vaco{r}\c\coco{1} \cctx \ruco{\ret}\ \star\c
      \coco{1}$}
    \BinaryInfC{$\vaco{r}\c\coco{1^\utag}\mid\coco{\emptyset}\cctx\ruco{\oname{handle}}\,\vaco{r}
        \,\oname{in}\, \ldots\;
\ruco{\oname{with}\, \ret\ \blco{\star}}\c\coco{1}$}
  \end{prooftree}
  \caption{Typing derivation for the example in Fig.~\ref{fig:intro}.}\label{fig:dev-tree}

\begin{tikzpicture}[overlay,>=latex,shorten >=1pt,->,remember picture]
   \draw[purple,>->] ([xshift=1ex,yshift=.5ex]pic cs:b1) .. controls +(right:3cm) and +(up:1cm) .. ([xshift=0ex,yshift=.5ex]pic cs:e1);
   \draw[purple,>->] ([xshift=1ex,yshift=.5ex]pic cs:b2) .. controls +(right:1.5cm) and +(up:.5cm) .. ([xshift=-1em,yshift=.5ex]pic cs:e2);
   \draw[purple,>->] ([xshift=1ex,yshift=.5ex]pic cs:b3) 
        .. controls +(right:.5cm) and +(right:.8cm) .. +(-.3cm,-.5cm) 
        .. controls +(left:1cm) and +(up:.6cm)   .. ([xshift=1ex,yshift=.5ex]pic cs:e3);
   \draw[purple,>->] ([xshift=1ex,yshift=.5ex]pic cs:b4) 
        .. controls +(right:.6cm) and +(right:1cm) 
        .. +(-.3cm,-.5cm) 
        .. controls +(left:3cm) and +(up:.8cm) 
        .. ([xshift=1ex,yshift=.5ex]pic cs:e4);
\end{tikzpicture}

\end{figure}

\section{Generic Denotational Semantics}\label{sec:deno}

\noindent We proceed to give a denotational semantics of the guarded
metalanguage assuming the following:
\
\begin{itemize}
 \item a distributive category $\BC$ (with initial objects);
 \item a strong guarded pre-iterative monad $\BBT$ on $\BC$.
\end{itemize} 
Supposing that every base type $A\in\oname{Base}$ is interpreted as an
object $\ul A$ in $|\BC|$, we define $\ul{A}$ for types $A$
(see~\eqref{eq:sum-types}) inductively by
\begin{gather*}
\ul{0} = \iobj,\qquad \ul{1} = 1,\qquad \ul{A+B} = \ul{A}+\ul{B},\qquad
\ul{A\times B} = \ul{A}\times\ul{B}.
\end{gather*}
To every $f\c A\to B\in\Sigma_v$ we associate an interpretation
$\sem{f}\in\Hom(\ul{A},\ul{B})$ in~$\BC$ and to every
$f\c A\to B [C]\in\Sigma_c$ an interpretation
$\sem{f}\in\Hom_{\inj_2}(\ul{A}\comma T(\ul{B}+\ul{C}))$.
Based on these we define the semantics of the term language from
Fig.~\ref{fig:lang.full}. The semantics of a value judgment
$\Gamma\vctx p\c A$ is a morphism
${\sem{\Gamma\vctx p\c A}}\in\Hom(\ul{\Gamma},\ul{A})$, and the semantics
of a computation judgment $\Delta\csep\Gamma\cctx p\c A$ is a morphism
$\sem{\Delta\csep\Gamma\cctx
  p\c A}\in\Hom_{\bang+\sigma_{\Delta}}(\ul{\Gamma},T(\ul{A}+\ul{\Delta}))$ where
\begin{flalign*}
&&\ul{\Gamma} =\;& \ul{A_1}\times \ldots\times \ul{A_n}&& \text{for~~} \Gamma=(x_1\c A_1\comma\ldots\comma x_n\c A_n)\\
&&\ul{\Delta} =\;& \ul{E_1}+ \ldots+ \ul{E_m}&& \text{for~~} \Delta=(e_1\c E_1^{\alpha_1}\comma\ldots\comma e_m\c E_m^{\alpha_m})
\end{flalign*}
and $\sigma_\Delta\c\ul\Delta'\cpto\ul\Delta$ is the summand induced by removal
of unguarded exceptions $e\c E^\utag$ from $\Delta$ with $\Delta'$ denoting the result. 

\begin{figure}\small
\vspace{-3ex}
\begin{flushleft}
\begin{align*}
&\lrule{(gcase)}{
  \begin{array}[b]{r@{~}l}
  \sem{\Gamma\vctx v\c A} =& h\c\ul{\Gamma}\to \ul{A}\\[1ex]
  \sem{f\c A\to B[C]} =& g\c\ul A\to_2 T(\ul B+\ul C)\\[1ex]
  \sem{\Delta\csep\Gamma,x\c B\cctx p:D} =& u\c\ul{\Gamma}\times\ul{B}\to_{\bang+\sigma_{\Delta}} T(\ul{D}+\ul\Delta)\\[1ex]
  \sem{\Delta'\csep\Gamma,y:C\cctx q:D} =& w\c\ul{\Gamma}\times\ul{C}\to T(\ul{D}+\ul\Delta)
  \end{array}
}
{
  \begin{aligned}
    &\sem{\Delta\csep \Gamma \cctx \gcase{f}{v}{x}{p}{y}{q}\c D} =\\ 
    &\qquad\ul\Gamma\xto{{\delta}\!\brks{\id,\,g h}}  T(\ul\Gamma\times\ul B+\ul\Gamma\times\ul C)\xto{[u, w]^\klstar} T(\ul D+\ul\Delta)
  \end{aligned}
}\\[2.4ex]
&\lrule{(prod)}{
   \begin{array}{r@{~}l}
      \sem{\Gamma \vctx p\c A\times B}=&g\c\ul{\Gamma}\to\ul{A}\times\ul{B}\\[1ex]
      \sem{\Delta\csep \Gamma,x\c A,y\c B \cctx q\c C}=&h\c\ul{\Gamma}\times\ul{A}\times\ul{B}\to_{\bang+\sigma_{\Delta}} T(\ul{C}+\ul{\Delta})
   \end{array}
}{
   \sem{\Delta\csep \Gamma \cctx \pcase{p}{\brks{x,y} \mto q}\c C}=h\brks{\id_{\ul{\Gamma}},g}\c\ul{\Gamma}\to T(\ul{C}+\ul{\Delta})
}
\\[2.4ex]
&\lrule{(ret)}{
  \sem{\Gamma\vctx t\c A}=g\c\ul{\Gamma}\to\ul{A}
}{
  \sem{\Delta\csep \Gamma \cctx \ret t\c A}=\eta \inj_1 g\c\ul{\Gamma}\to T(\ul{A}+\ul{\Delta})
}
\\[2.4ex]
&\lrule{(do)}{
   \begin{array}{r@{~}l}
      \sem{\Delta\csep \Gamma \cctx p\c A}=&g\c\ul{\Gamma}\to_{\bang+\sigma_{\Delta}} T(\ul{A}+\ul{\Delta})\\[1ex]
      \sem{\Delta\csep \Gamma, x\c A \cctx q\c B}=&h\c\ul{\Gamma}\times\ul{A}\to_{\bang+\sigma_{\Delta}} T(\ul{B}+\ul{\Delta})
   \end{array}
}{
  \sem{\Delta\csep \Gamma \cctx \mbind{x \gets p}{q}}=[h,\eta\inj_2\snd]^\klstar\comp\delta\!\comp
  \brks{\id_{\ul{\Gamma}},g}\c\ul{\Gamma}\to T(\ul{B}+\ul{\Delta})
}
\\[2.4ex]
&\lrule{(init)}{
  \sem{\Gamma\vctx t:0}=g\c\ul{\Gamma}\to \iobj
}{
  \sem{\Delta\csep \Gamma \cctx \oname{init} t\c A}= \bang\comp g\c\ul\Gamma\to T(\ul A+\ul\Delta)
}\\[2.4ex]
&\lrule{(case)}{
   \begin{array}{r@{~}l}
     \sem{\Gamma \vctx p\c A+B}=&g\c\ul{\Gamma}\to\ul{A}+\ul{B}\\[1ex]
     \sem{\Delta\csep \Gamma,x\c A \cctx q\c C}=&h\c\ul{\Gamma}\times\ul{A}
   \to_{\bang+\sigma_{\Delta}} T(\ul{C}+\ul{\Delta})\\[1ex]
   \sem{ \Delta\csep \Gamma,y\c B \cctx r\c C}=&u\c\ul{\Gamma}\times\ul{B}
   \to_{\bang+\sigma_{\Delta}} T(\ul{C}+\ul{\Delta})
   \end{array}
 }
 {
  \begin{aligned}
  &\sem{ \Delta\csep \Gamma \cctx \case{p}{\inl x \mto q}{ \inr y \mto r}\c C }=\\ 
    &\qquad[h,u]\dist\brks{\id_{\ul{\Gamma}},g}\c\ul{\Gamma}\to T(\ul{C}+\ul{\Delta})
  \end{aligned}
 }\\[2.4ex]
 &\lrule{(raise)}{
  \sem{\Gamma\vctx q\c E}=g\c\ul{\Gamma}\to\ul{E}
}{
  \sem{\Delta\csep \Gamma \cctx \oname{raise}_e q\c D}= \eta\inj_2\inj_e g\c\ul{\Gamma}\to T(\ul{D}+\ul{\Delta})
} 
\\[2.4ex]
&\lrule{(handle)}{
  \begin{array}{r@{~}l}
    \sem{\Delta, e\c E^\utag \csep\Gamma \cctx p\c A}=&g\c\ul{\Gamma}\to_{\bang+(\sigma_{\Delta}+\bang)} T(\ul{A}+(\ul{\Delta}+\ul{E}))\\[1ex]
    \sem{\Delta\csep \Gamma,e\c E \cctx q\c A} =&h\c\ul{\Gamma}\times\ul{E}\to_{\bang+\sigma_{\Delta}} T(\ul{A}+\ul{\Delta})
  \end{array}
}{
  \begin{aligned}
    \sem{\Delta\csep&\, \Gamma \cctx \handle{e}{p}{q}\c A}=\\
    \ul\Gamma&\xto{T(\id_{\ul{\Gamma}\times\ul{A}}+\dist)\delta\!\brks{\id_{\ul{\Gamma}},g}} T(\ul{\Gamma}\times\ul{A}+(\ul{\Gamma}\times\ul{\Delta}+\ul{\Gamma}\times\ul{E}))\\
    &\xto{[\eta\inj_1\pr_2, [\eta\inj_2\pr_2,h]]^\klstar}  T(\ul{A}+\ul{\Delta})
  \end{aligned}
}\\[2.4ex]
&\lrule{(iter)}
{
  \begin{array}{r@{~}l}
    \sem{\Gamma \vctx v\c E}=&g\c\ul{\Gamma}\to \ul{E}\\[1ex]
    \sem{\Delta,e\c E^\gtag \csep\Gamma,e\c E \cctx q\c A}=&h\c\ul{\Gamma}\times\ul{E}\to_{\bang+(\sigma_{\Delta}+\id)}
    T(\ul{A}+(\ul{\Delta}+\ul{E}))
   \end{array}
}{ 
  \sem{\Delta\csep \Gamma \cctx \handleit{v}{e}{q}\c A}=
   ((T\oname{assoc})\comp h)^\iistar\comp \brks{\id_{\ul{\Gamma}},g}\c\ul{\Gamma}\to T(\ul{A}+\ul\Delta)
}
\end{align*}
\caption{Denotational semantics.}
\label{fig:sem}
\end{flushleft}
\end{figure}

The semantic assignments for computation judgments are given in
Fig.~\ref{fig:sem} (we skip the obvious standard rules for values) where
$\inj_e\c\ul E\to\ul\Delta$ is the obvious coproduct injection of $\ul E$ to
$\ul\Delta$ identified by $e$, $\oname{assoc}$ is the associativity isomorphism
$X+(Y+Z)\cong (X+Y)+Z$, and $(\argument)^\iistar$ is the strong iteration operator
from~\eqref{eq:iter_par}.
The correctness of our semantic assignments is established
by the following claim:
\begin{proposition}
For every rule in Fig.~\ref{fig:lang.full}, assuming the premises, the morphism
in the conclusion is $(\bang+\sigma_{\Delta})$-guarded.
\end{proposition} 
\begin{proof}
  First note that each  $f\c X \to_{\bang +\sigma_{\Delta}}
  T(\ul A+ \ul\Delta)$ is isomorphic to some $\hat{f}\c X \to_{\bang +\id} T((\ul A+
  \ul\Delta_{\utag}) + \ul \Delta_{\gtag})$, where $\Delta \cong \Delta_{\utag}
  + \Delta_{\gtag}$ separates unguarded from guarded exceptions.
  Consider the rule~\textbf{(handle)} in detail. By regarding~$g$ and~$h$ as
  morphisms in $\ssl{\BC}{\ul\Gamma}$, we reformulate the goal as follows: assuming
  $g:1\to_{\bang+(\sigma_\Delta+\bang)} T(\ul A+((\ul\Delta_{\utag} + \ul \Delta_{\gtag}) + \ul E))$ and 
  $h\c\ul E\to_{\bang+\sigma_\Delta} T(\ul A+(\ul\Delta_{\utag} + \ul \Delta_{\gtag}))$, show that
  $[\eta\comp\inj_1,[\eta\comp\inj_2,h]]^\klstar\comp g:1\to_{\bang+\sigma_\Delta}
  T(\ul A+(\ul\Delta_{\utag} + \ul \Delta_{\gtag}))$. Let w.l.o.g.\ $\sigma_{\Delta} = \inj_2\c \ul\Delta_{\gtag} \to
  \ul\Delta_{\utag} + \ul\Delta_{\gtag}$. We obtain
\begin{align*}
    T\bigl[\inj_1\inj_1\inj_1,
    [\inj_1\inj_2+\id,\inj_1\inj_2]\bigr]\comp g\c 1
    \to_{\inj_2} T(((\ul A + \ul\Delta_{\utag}) + E) +\ul\Delta_{\gtag})
\end{align*}
by~\textbf{(iso)}, and 
\begin{align*}
    [\eta (\id+\inj_1), h]\c (\ul A + \ul\Delta_{\utag}) + E\to_{\bang+\sigma_\Delta}
  T(\ul A+(\ul\Delta_{\utag} + \ul \Delta_{\gtag}))
\end{align*}
by~\textbf{(trv)} and~\textbf{(sum)}. Then by~\textbf{(cmp)},
\begin{align*}
  \bigl[[\eta (&\id+\inj_1), h],\eta\comp\inj_2\comp\inj_2 \bigr]^\klstar\\
     &T\bigl[\inj_1\inj_1\inj_1,
    [\inj_1\inj_2+\id,\inj_1\inj_2]\bigr]\comp g :1\to_{\bang+\sigma_\Delta}
  T(\ul A+(\ul\Delta_{\utag} + \ul \Delta_{\gtag})),
\end{align*}
which further reduces down to the goal.

For%
~\textbf{(prod)},~\textbf{(ret)},~\textbf{(case)} and~\textbf{(init)},
the verification is straightforward by the axioms of guardedness
in~$\BC$.  For~\textbf{(gcase)} and~\textbf{(do)}, we proceed
analogously to~\textbf{(handle)} using the axioms of guardedness
in~$\ssl{\BC}{\ul\Gamma}$ and Theorem~\ref{thm:context-comonad}. Strong
iteration as figuring in~\textbf{(iter)} satisfies the fixpoint law by
Theorem~\ref{thm:context-comonad}, and the problem in question amounts
to verifying that $f^\istar\c X\to_{\sigma} TY$ whenever
$f\c X\to_{\sigma+\id} T(Y+X)$. This is already shown
in~\cite{GoncharovSchroderEtAl17}, using only the fixpoint~law.
\end{proof}

\section{Functional Types}\label{sec:expo}
\noindent In order to interpret functional types in fine-grain
call-by-value, it normally suffices to assume existence of
\emph{Kleisli exponentials}, i.e.\ objects $TB^A$ such that
$\Hom(C\comma TB^A)$ and $\Hom({C\times A}\comma TB)$ are naturally
isomorphic, or equivalently that all presheaves
$\Hom(\argument\times A, TB)\c\BC^{\op}\to\Set$ are representable. In
order to add functional types to our metalanguage we additionally need
to assume that all presheaves
$\Hom_{\sigma}(\argument,TA)\c\BC^{\op}\to\Set$ are representable, i.e.\
for every $A$ and $\sigma\c A'\cpto A$ there is $A_\sigma\in |\BC|$ such
that
\begin{align}\label{eq:s-hom}
\xi: \Hom(X,A_{\sigma}) \cong \Hom_{\sigma}(X,TA) 
\end{align}    
naturally in $X$.
By the Yoneda lemma, this requirement is equivalent to the following.  

\begin{definition}[Greatest $\sigma$-algebra]
Given $\sigma\c A'\cpto A$, a pair $(A_{\sigma}, \iota_{\sigma})$ consisting  
of an object $A_{\sigma}\in |\BC|$ and a morphism $\iota_{\sigma}\c A_{\sigma}\to_{\sigma} TA$
is called a \emph{greatest $\sigma$-algebra} if for every $f\c X\to_{\sigma} TA$ 
there is a unique $\hat f\c X\to A_{\sigma}$ with the property that $f=\iota_{\sigma}\comp \hat f$.
\end{definition}
\begin{wrapfigure}[4]{r}{3cm}
\vspace{-8ex} 
 \centering
\begin{minipage}{3cm}
\begin{equation*}
\begin{tikzcd}
X
	\ar[r, "f"] 	\ar[d, "\hat f"', dotted]
& TA\\
A_\sigma\ar[ur,"\iota_{\sigma}"'] &
\end{tikzcd}
\end{equation*}
\end{minipage}
\end{wrapfigure}

\noindent By the usual arguments, $(A_\sigma,\iota_\sigma)$ is defined
uniquely up to isomorphism. The connection between $\iota_{\sigma}$
and $\xi$ in~\eqref{eq:s-hom} is as follows:
$\iota_{\sigma} = \xi(\id\c A_{\sigma}\to A_{\sigma})$ and
$\xi(f\c X\to A_\sigma) = \iota_{\sigma}\comp f$.

It immediately follows by definition that $\iota_\sigma$ is a monomorphism. 
The name `$\sigma$-algebra' for $(A,\iota_\sigma)$ is justified as follows. 
\begin{proposition}\label{prop:s-alg}
  Suppose that $(A,\iota_\sigma)$ is a greatest $\sigma$-algebra.
  Then there is a unique $\alpha_\sigma\c TA_{\sigma}\to A_{\sigma}$ such
  that $\iota_\sigma\comp\alpha_\sigma = \iota_\sigma^\klstar$.  The
  pair $(A_\sigma,\alpha_\sigma)$ is a\/ $\BBT$-subalgebra of\/
  $(TA,\mu)$.
\end{proposition}
\begin{proof}
  Since $\iota_\sigma^{\klstar}\c TA_\sigma\to TA$ is the Kleisli
  composite of $\iota_\sigma\c A_\sigma\to_\sigma TA$ and
  $\id\c TA_\sigma\to TA_\sigma$, $\iota_\sigma^{\klstar}$ is
  $\sigma$-guarded by~\textbf{(cmp)}, so we obtain $\alpha_{\sigma}$
  such that $\iota_\sigma\alpha_\sigma=\iota_\sigma^\klstar$ by the
  universal property of $(A_\sigma,\iota_{\sigma})$. Since
  $\iota_\sigma^\klstar=\mu_A(T\iota_\sigma)$, it follows that
  $\iota_\sigma\c (A_\sigma,\alpha_\sigma)\to(A,\mu_A)$ is a morphism of
  functor algebras. Since monad algebras are closed under taking
  functor subalgebras and $\iota_\sigma$ is monic as observed above,
  it follows that $(A_\sigma,\alpha_\sigma)$ is a $\BBT$-subalgebra of
  $(A,\mu_A)$.
\end{proof}

\begin{proposition}\label{prop:s-great}
\begin{enumerate}
\item\label{item:greatest} Suppose that a greatest $\sigma$-algebra $(A_{\sigma},\iota_{\sigma})$ exists. Then 
\begin{enumerate}
 \item\label{item:ordering} $\iota_{\sigma}$ is the greatest element in the class of all $\sigma$-guarded 
subobjects of~$TA$; 
 \item\label{item:epi-cancellation} for every regular epic $e\c X\to Y$ and every morphism $f\c Y\to TA$, 
$f\comp e\c X\to_{\sigma} TA$ implies that $f\c Y\to_{\sigma} TA$.
\end{enumerate}
\item\label{item:converse} Assuming that every morphism in $\BC$ admits a factorization into a regular epic and
a monic, the converse of~(1) is true: If~(a) and~(b)
hold for $(A_{\sigma},\iota_{\sigma})$, then $(A_{\sigma},\iota_{\sigma})$ is a greatest $\sigma$-algebra. 
\end{enumerate}
\end{proposition}  
\begin{proof}
  \emph{\ref{item:greatest}.:} Part~\ref{item:ordering} is immediate;
  we show~\ref{item:epi-cancellation}.
  Given a regular epic $e\c X\to Y$ and a morphism $f\c Y\to TA$ such
  that $f\comp e\c X\to_{\sigma} TA$, consider the diagram
\begin{equation*}
\begin{tikzcd}[column sep = 6ex]
Z 
	\ar[r,shift right=.5ex,"g"']
    \ar[r,shift left=.5ex,"h"]
&
X
	\ar[r, "e"]\ar[d, "w"', dotted]
&
Y
	\ar[r, "f"]
&
TA\\
& A_\sigma\ar[rru, "\iota_{\sigma}"'] & &
\end{tikzcd}
\end{equation*}
where $e$ is the coequalizer of$~h$ and~$g$, and~$w$ exists uniquely
by the universal property of $\iota_{\sigma}$. Since
$\iota_{\sigma}\comp w\comp h = f\comp e\comp h = f\comp e\comp
g=\iota_{\sigma}\comp w\comp g$
and $\iota_{\sigma}$ is monic, $w\comp h = w\comp g$. Hence, there is
$u\c Y\to A_{\sigma}$ such that $w=u\comp e$.  Therefore we have
$\iota_{\sigma}\comp u\comp e= \iota_{\sigma}\comp w = f\comp e$.
Since $e$ is epi, this implies $f= \iota_{\sigma}\comp u$. Since
$\iota_{\sigma}$ is $\sigma$-guarded, so is $f$ by~\textbf{(cmp)}.

\emph{\ref{item:converse}.:} Let $f\c X\to_{\sigma} TA$, with
factorization $f = m\comp e$ into a mono~$m$ and a regular
epi~$e$. By~\ref{item:epi-cancellation},~$m$ is $\sigma$-guarded;
by~\ref{item:ordering}, it follows that~$m$, and hence~$f$, factor
through $\iota_{\sigma}$, necessarily uniquely since~$\iota_\sigma$ is
monic.  
\end{proof}

\begin{example}\label{expl:resump}
  Let $\BBT$ be a strong monad on a distributive category $\BC$ and let
  $\Sigma\c\BC\to\BC$ be an endofunctor such that all the fixpoints $T_\Sigma
  X = \nu \gamma.\,T(X + \Sigma \gamma)$ exist. These extend to a strong monad
  $\BBT_{\Sigma}$, called the \emph{generalized coalgebraic resumption monad
  transform} of $\BBT$~\cite{GoncharovSchroderEtAl17}. Moreover, $\BBT_{\Sigma}$ is guarded iterative 
  with $f\c X\to_\sigma T_{\Sigma}A$ iff $\out f\c X\to T(A+\Sigma T_{\Sigma}A)$
  factors as $T(\bar\sigma+\id)\comp g$ for some $g\c X\to T(A'+\Sigma T_{\Sigma}A)$. 
  Suppose that coproduct injections in $\BC$ are monic and~$T$ preserves monics. 
  Then for every $A\in |\BC|$ and $\sigma$ there 
  is at most one~$g$ such that $\out f = T(\bar\sigma+\id)\comp g$. This entails 
  an isomorphism
\begin{displaymath}
  \Hom(X,T(A'+\Sigma T_{\Sigma}A)) \cong \Hom_{\sigma}(X,T_\Sigma A) 
\end{displaymath}
obviously natural in $X$, from which we obtain by comparison with~\eqref{eq:s-hom}
that $A_\sigma = T(A' + \Sigma T_\Sigma A)$.
\end{example}
\begin{example}
Let $\sigma\c A''\cpto A$, whose complement is $\bar\sigma\c A'\cpto A$ and let us revisit Example~\ref{expl:monad}.
\begin{enumerate} 
  \item $T=\nu\gamma.\,\FSet(\argument+\Act\times\gamma)$ is an instance of Example~\ref{expl:resump},
and thus $A_\sigma = \FSet(A' + \Act\times T A)$.
  \item For $T=\nu\gamma.\,\CSet(\argument+\Act\times\gamma)$ under total guardedness,
 $A_\sigma = TA$ independently of $\sigma$. For the other notion of guardedness on 
  $\BBT$,  $A_\sigma$ is constructed in analogy to Clause~1.
  \item For $T=\PSet$ being totally guarded, again $A_\sigma = \PSet A$.
  \item For $T=\PSet(\Act^\star\times\argument)$,
  it follows that $A_{\sigma} = \PSet(\Act^\star\times A'+\Act^\mplus\times A'')$.  
  \item Finally, for $T=\NESet$,
  it follows by definition that $A_\sigma = \PSet A'\times\NESet A''$.
\end{enumerate} 
\end{example}

\begin{figure}[t!]
\begin{center}%
{\parbox{\textwidth}{%
\small%
  \begin{subfigure}{\textwidth}
\begin{flalign*}
&&
\anonrule{lambda}{%
		\Delta\csep \Gamma, x\c A\cctx p\c B
	}{%
		\Gamma \vctx \lambda x.\, p\c A\to_{\Delta} B
	}
&&
  \anonrule{app}{%
    \Gamma \vctx w\c A\qquad \Gamma \vctx v\c A\to_{\Delta} B
  }{%
    \Delta\csep \Gamma \cctx v w\c B
  }
&&
\end{flalign*}
\medskip\dotfill
\end{subfigure}
\begin{subfigure}{\textwidth}
\begin{gather*}
\anonrule{lambda}{%
		\sem{\Delta\csep \Gamma, x\c A\cctx p\c B}  = g\c \ul\Gamma\times\ul A\to_{\bang+\sigma_{\Delta}} T(\ul B+\ul\Delta) 
	}{%
		\sem{\Gamma \vctx \lambda x.\, p\c A\to_{\Delta} B} = \curry(\xi^\mone(g)): \ul{\Gamma}\to \ul{A}\to (\ul B+\ul\Delta)_{\bang+\sigma_{\Delta}}
	}
\\[2ex]
  \anonrule{app}{%
    \sem{\Gamma \vctx w\c A} = g\c\ul\Gamma\to\ul A \qquad 
    \sem{\Gamma \vctx v\c A\to_{\Delta} B} = h\c\ul\Gamma\to\ul A\to (\ul B+\ul\Delta)_{\bang+\sigma_{\Delta}}
  }{%
    \sem{\Delta\csep \Gamma \cctx v w\c B} = \xi(\uncurry h)\comp\brks{\id, g}\c\Gamma\to_{\bang+\sigma_{\Delta}} T(\ul B+\ul\Delta)
  }
\end{gather*}
\end{subfigure}
}}
\end{center}
\caption{Syntax (top) and semantics (bottom) of functional types.}
\label{fig:expo}
\end{figure}
\noindent Assuming that greatest $\sigma$-algebras exist, we
complement our metalanguage with functional types $A\to_\Delta B$
where the index~$\Delta$ serves to store information about (guarded)
exceptions of the curried function. Formally, these types are interpreted as
$\ul{A\to_\Delta B}=\ul{A}\to (\ul
B+\ul\Delta)_{\bang+\sigma_{\Delta}}$.
In the term language, this is reflected by the introduction of
$\lambda$-abstraction and application, with syntax and semantics as
shown in Fig.~\ref{fig:expo}, where $\xi$ is the isomorphism
from~\eqref{eq:s-hom}.

\section{Operational Semantics and Adequacy}\label{sec:adeq}
\noindent We proceed to complement our denotational semantics from
Sections~\ref{sec:deno} and~\ref{sec:expo} with a big-step operational
semantics. Following Geron and Levy~\cite{GeronLevy16}, we choose the
simplest concrete monad $\BBT$ sensibly illustrating all the main
features and model it operationally.  In~\cite{GeronLevy16} this is
the \emph{maybe monad} $TX=X+1$ on $\Set$, which suffices to give a
sensible account of total iteration. The $+1$ part is necessary for
modelling divergence.  Since total iteration is subsumed by guarded
iteration, we could formulate an adequate operational semantics over
this monad too. To that end we would need to assume that the only
operation $f\c A\to B[C]$ in $\Sigma_c$ with $C\neq 0$ is a
distinguished element $\mathit{tick}\c 1\to 0[1]$ whose denotation is the unit of
the monad (regarded as totally guarded). However, total iteration is
only a degenerate instance of guarded iteration, and here we therefore
replace $X+1$ with the guarded iterative monad freely generated by
an operation $put\c\nat\to 0[1]$ of outputting a natural number (say,
to console), explicitly (on~$\Set$):
$TX=(X\times\nat^{\star}) \cup \nat^{\omega}$.  More abstractly, $TX$
is the final $(X+\nat\times\argument)$-coalgebra. The denotations in
$TX$ are of two types: pairs $(x,\tau)\in X\times\nat^{\star}$ of a
value $x$ and a \emph{finite trace} $\tau$ of outputs (for terminating
iteration) and \emph{infinite traces} $\pi\in\nat^{\omega}$ of outputs
(for non-terminating iteration).

\begin{figure}[t!]
    {\parbox{\textwidth}{%
        \small%
\textit{Values, Computations, Terminals:}
        \begin{align*}
           v,w &\Coloneqq x\mid \star\mid \zero\mid \Succ v\mid\inl v\mid\inr v\mid\brks{v,w}\mid\lambda x.\,p  \\
           p,q &\Coloneqq \ret v\mid\Pred(v)\mid\Put(v)\mid\oname{raise}_x v\mid\gcase{\Put}{v}{\argument}{p}{x}{q}\\
                &\qquad\mid\pcase{v}{\brks{x,y} \mto p}\mid\oname{init} v\mid\case{v}{\inl x\mto p}{\inr y\mto q}\\
                &\qquad\mid  v w\mid\mbind{x\gets p}{q}\mid\handle{x}{p}{q}\mid \handleit{v}{y}{p}  \\
           t &\Coloneqq \ret v, \tau \mid \oname{raise}_x v, \tau \mid \pi \qquad (\tau\in\nat^\star, \pi\in\nat^\omega)
        \end{align*}

\smallskip
\textit{Rules:}\\[-3ex]
  \begin{align*}
    \anonrule{case-inl}{%
      q[\star/x] \Downarrow u, \tau
    }{%
      \gcase{\Put}{v}{\argument}{p}{x}{q}\Downarrow 
      u, \brks{v} \pp \tau
    }
    &&
    \anonrule{raise}{%
    }{%
      \oname{raise}_x v \Downarrow \oname{raise}_x v,\brks{}
    }    
  \end{align*}
  \\[-5ex]
  \begin{align*}
    \anonrule{case-inl}{%
      q[\star/x] \Downarrow \pi
    }{%
      \gcase{\Put}{v}{\argument}{p}{x}{q}\Downarrow 
      \brks{v} \pp \pi
    }
    &&
    \anonrule{lambda}{%
      p[v/x] \Downarrow t
    }{%
      (\lambda x.\, p)\, v \Downarrow t
    }
  \end{align*}
\\[-5.5ex] 
  \begin{align*}
    \anonrule{ret-infty}{%
    }{%
      \Pred(\zero) \Downarrow \ret\inl\star, \brks{}
    }
    &&
    \anonrule{ret-infty}{%
    }{%
      \Pred(\Succ(v)) \Downarrow \ret\inj_2 v, \brks{}
      }
  \end{align*}
\\[-5ex]
  \begin{align*}
    \anonrule{case-inl}{%
      p[v/x] \Downarrow t
    }{%
      \case{\inl v}{\inl x \mto p}{\inr y \mto q} \Downarrow t
    }
    &\quad&
    \anonrule{case-inr}{%
      q[w/y] \Downarrow t
    }{%
      \case{\inr w}{\inl x \mto p}{\inr y \mto q} \Downarrow t
    }
  \end{align*}
\\[-5ex]
  \begin{align*}
    \anonrule{ret-bind}{%
      p \Downarrow \ret v, \tau \qquad q[v/x] \Downarrow u, \tau'
    }{%
      \mbind{x \gets p}{q} \Downarrow u, \tau \pp \tau'
    }
    &~~&\anonrule{ret-bind}{%
      p \Downarrow \ret v, \tau \qquad q[v/x] \Downarrow \pi
    }{%
      \mbind{x \gets p}{q} \Downarrow \tau \pp \pi
    }
  \end{align*}
\\[-5ex]
  \begin{flalign*}
&&
    \anonrule{ret-infty}{%
      p \Downarrow \oname{raise}_{x} v, \tau
    }{%
      \mbind{x \gets p}{q} \Downarrow \oname{raise}_{x} v, \tau
    }
&&
    \anonrule{ret-infty}{%
      p \Downarrow \pi
    }{%
      \mbind{x \gets p}{q} \Downarrow \pi
    }
&&
    \anonrule{ret-handle}{%
      p \Downarrow \ret v, \tau
    }{%
      \handle{x}{p}{q} \Downarrow \ret v, \tau
    }
&&
  \end{flalign*}
\\[-5ex]
  \begin{flalign*}
   \anonrule{put-handle-other}{%
      p \Downarrow \oname{raise}_{y} v, \tau 
    }{%
      \handle{x}{p}{q} \Downarrow \oname{raise}_{y} v, \tau
    }\quad (x \neq y)
    &~&
    \anonrule{put-handle}{%
      p \Downarrow \oname{raise}_{x} v, \tau \qquad q[v/x] \Downarrow u, \tau'
    }{
      \handle{x}{p}{q} \Downarrow u, \tau \pp \tau'
    }  
  \end{flalign*}
\\[-5ex]
  \begin{flalign*}
    &&
    \anonrule{put-handle-inf}{%
      p \Downarrow \pi
    }{%
      \handle{x}{p}{q} \Downarrow \pi
    }\!\!
    &&
    \anonrule{put-handle-inf2}{%
      p \Downarrow \oname{raise}_x v, \tau\quad q[v/x] \Downarrow \pi
    }{%
      \handle{x}{p}{q} \Downarrow \tau \pp \pi
    }\!\!
    &&
  \end{flalign*}
\\[-5ex]
  \begin{align*}
   \anonrule{ret-triv}{%
    }{%
      \ret v \Downarrow \ret v, \brks{}
    }
    &\quad&
    \anonrule{put-handleit}{%
      v_0 = v\quad q[v_0/x] \Downarrow \oname{raise}_{x} v_1, \tau_1\qquad \ldots\qquad
      q[v_{n-1}/x] \Downarrow u, \tau_{n}
    }{%
      \handleit{v}{x}{q} \Downarrow u, \tau_1 \pp \cdots \pp \tau_{n}
    }
  \end{align*}
\\[-5ex]
  \begin{flalign*}
    \anonrule{case-prod}{%
      q[v/x,w/y] \Downarrow t
    }{%
      \pcase{\brks{v,w}}{\brks{x,y} \mto q} \Downarrow t
    }
    &&
    \anonrule{put-handleit-inf}{%
      v_0 = v\qquad q[v_0/x] \Downarrow \oname{raise}_{x} v_1, \tau_1\quad \ldots \quad
      q[v_{n-1}/x] \Downarrow \pi
    }{%
      \handleit{v}{x}{q} \Downarrow \tau_1 \pp \cdots
      \pp \tau_{n-1} \pp \pi
    }
  \end{flalign*}
\\[-5ex]
  \begin{align*}
    \anonrule{put-handleit}{%
           v_0 = v\qquad q[v_0/x] \Downarrow \oname{raise}_{x} v_1, \tau_1\qquad q[v_1/x] \Downarrow \oname{raise}_{x} v_2, \tau_2\quad \ldots \quad
    }{%
      \handleit{v}{x}{q} \Downarrow \tau_1 \pp \tau_2
      \pp \cdots
    }\quad (\forall i.\,\tau_i\neq\brks{})
  \end{align*}
          \caption{Operational semantics.}\label{fig:oper}
      }}
\end{figure}
We fix $TX=(X\times\nat^{\star}) \cup \nat^{\omega}$ for the rest of
the section.  Let us spell out the details of the structure of~$\BBT$,
which is in fact an instance of Example~\ref{expl:resump} under
$T=\Id$, $\Sigma=\nat\times(\argument)$. The unit
of $\BBT$ sends $x$ to $(x,\brks{})$. Given $f\c X\to TY$, we have
\begin{align*}
  f^{\klstar}(x, \tau) =
  \begin{cases}(y, \tau \pp \tau') &\text{if } f(x) = (y, \tau'),\\
    \tau \pp \pi &\text{if } f(x) = \pi,
  \end{cases}&&\qquad
  f^{\klstar}(\pi) = \pi.
\end{align*}
for $x\in X$, $\tau\in\nat^\star$, $\pi\in\nat^\omega$ with $\pp$ denoting
concatenation of a finite trace with a possibly infinite one.
Guardedness for $\BBT$ is defined as follows: $f\c X\to_2 (Y+Z)\times\nat^\star\cup\nat^\omega$
if for every $x\in X$, either $f(x)\in\nat^\omega$ or $f(x) = (\inj_1 y,\tau)$ for 
some $y\in Y$, $\tau\in\nat^\star$ or $f(x) = (\inj_2 z,\tau)$ for some $z\in Z$, 
$\tau\in\nat^\mplus$. Finally, given $f\c X\to_2 T(Y+X)$, 
\begin{align*}
  f^\istar(x) = 
\begin{cases}
  (y,\tau_1\pp\cdots\pp\tau_n) &\text{if } f(x) = (\inj_2 x_1, \tau_1),\ldots, f(x_n) = (\inj_1 y, \tau_n),\\
  \tau_1\pp\cdots\pp\tau_{n-1}\pp\pi &\text{if } f(x) = (\inj_2 x_1, \tau_1),\ldots, f(x_n) = \pi,\\
  \tau_1\pp\cdots &\text{if } f(x) = (\inj_2 x_1, \tau_1),\ldots
\end{cases}
\end{align*}
where the first clause addresses the situation when iteration finishes after 
finitely many steps, the second one addresses the situation when we hit divergence
witnessed by some $x_n\in X$ reachable after finitely many iterations,
and the third clause addresses the remaining situation of divergence
via unfolding the loop infinitely many times. In the latter case, the guardedness 
assumption for~$f$ is crucial, as it ensures that each $\tau_i$ is nonempty, and 
therefore the resulting trace $\tau_1\pp\tau_2\pp\cdots$ is indeed
infinite.

Operationally, guardedness in the above sense is modelled by cutting the control flow 
with the $\Put$ command, which is the only command contributing to the traces. 
Concretely, let $\oname{Base}=\{\nat\}$,
$\Sigma_v=\{\zero\c 1 \to \nat, \Succ\c \nat \to \nat\}$ and
$\Sigma_c=\{\Pred\c \nat \to (1 + \nat)[0],\Put\c\nat\to~0[1]\}$ (note
that while $\Pred$ does not cause any side effects, it does perform a
computation and therefore needs to be in~$\Sigma_c$ rather
than~$\Sigma_v$). The operational semantics over these data is given
in Fig.~\ref{fig:oper}, where $\Pred(v)$ is a shortcut for
$\gcase{\Pred}{v}{x}{\ret x}{y}{\oname{init} y}$, and similarly for
$\Put(v)$.  The judgement $p \Downarrow t$ relates programs $p$ with
\emph{terminals} $t$, which can consist of either a finite trace
$\tau$ together with a result value $\ret v$ or an exceptional value
$\oname{raise}_x v$, or an infinite trace $\pi$. The traces correspond
to the natural numbers written explicitly using the
operation~$\Put$. %

\begin{figure}
{\parbox{\textwidth}{\small
\begin{flalign*}
  \sem{x}_{\rho}        =\;& \rho(x)
  & \sem{zero}_{\rho}   =\;& 0
  & \sem{\inl v}_{\rho} =\;& \inj_1 \sem{v}_{\rho}\\
  \sem{\star}_{\rho}      =\;& \star
  & \sem{\Succ u}_{\rho}  =\;& \sem{u}_{\rho} + 1 
  & \sem{\inr v}_{\rho}   =\;& \inj_2 \sem{v}_{\rho}\\
  \sem{\brks{v,w}}_{\rho} =\,& \brks{\sem{v}_{\rho}, \sem{w}_{\rho}}
  & \sem{\lambda x.\,p}_{\rho} =\,& \xi^{\mone}(\lambda a.\,\sem{p}_{\rho[a/x]})
  & \sem{v w}_{\rho} =\,& \xi(\sem{v}_{\rho})(\sem{w}_{\rho})
\end{flalign*} 
\begin{flalign*}
  \sem{\Pred(v)}_{\rho} =&
  \begin{cases}(\inj_1 \inj_1 \star, \brks{})
    &\text{if } \sem{v}_{\rho} = 0\\
    (\inj_1 \inj_2 n, \brks{}) &\text{if } \sem{v}_{\rho} = n+1
  \end{cases}
\end{flalign*}
\begin{flalign*}
 \qquad \sem{\ret v}_{\rho} =\;& (\inj_1 \sem{v}_{\rho}, \brks{})
 & \sem{\oname{raise}_x v}_{\rho} =\;& (\inj_2 \inj_x \sem{v}_{\rho}, \brks{})
 \qquad
\end{flalign*}
\begin{flalign*}
  \sem{\pcase{v}{\brks{x,y} \mto p}}_{\rho} 
    =\;& \sem{p}_{[\rho, u/x, w/y]} \text{~~ where ~~} \sem{v}_{\rho} = \brks{u, w} \\[1ex]
  ~~\sem{\gcase{\Put}{v}{\argument}{p}{x}{q}}_{\rho} =\;&
  \begin{cases}
    (t, \brks{\sem{v}_{\rho}} \pp \tau) &\text{if } \sem{q}_{\rho[\star/x]} = (t, \tau)\\
    (\brks{\sem{v}_{\rho}} \pp \pi) &\text{if } \sem{q}_{\rho[\star/x]} =\pi 
  \end{cases}\\[1ex]
  \sem{\case{v}{\inl x \mto p}{\inr y \mto q}}_{\rho}  
    =\;& \begin{cases}
      \sem{p}_{\rho[u/x]} &\text{~ if ~~} \sem{v}_{\rho} = \inj_1 u \\
      \sem{q}_{\rho[w/y]} &\text{~ if ~~} \sem{v}_{\rho} = \inj_2 w 
  \end{cases} 
\end{flalign*} 
\begin{flalign*}
\sem{\mbind{x \gets p}{q}}_{\rho} =\;&
  \begin{cases}
    (t, \tau \pp \tau') & \text{if } \sem{p}_{\rho}
    = (\inj_1 a, \tau) \text{ and } \sem{q}_{\rho[a/x]} = (t, \tau') \\
    \tau \pp \pi & \text{if } \sem{p}_{\rho}
    = (\inj_1 a, \tau) \text{ and } \sem{q}_{\rho[a/x]} = \pi \\
    (\inj_2 b, \tau) & \text{if } \sem{p}_{\rho} = (\inj_2 b, \tau) \\
    \pi & \text{if } \sem{p}_{\rho} = \pi \\
  \end{cases} \\[1ex]
\sem{\handle{x}{p}{q}}_{\rho} =\;&
  \begin{cases}
    (\inj_1 t, \tau) & \text{if } \sem{p}_{\rho} = (\inj_1 t, \tau) \\
    (\inj_2 \inj_e t, \tau) & \text{if } \sem{p}_{\rho} = (\inj_2 \inj_e t, \tau) \text{ and } x\neq e\\
    (b, \tau \pp \tau') & \text{if } \sem{p}_{\rho}
    = (\inj_2 \inj_x a, \tau)\text{ and }\sem{q}_{\rho[a/x]} = (b, \tau') \\
    \tau \pp \pi & \text{if } \sem{p}_{\rho}
    = (\inj_2 \inj_x a, \tau)\text{ and }\sem{q}_{\rho[a/x]} = \pi \\
    \pi & \text{if } \sem{p}_{\rho} = \pi 
  \end{cases} \\[1ex]
\sem{\handleit{v}{x}{q}}_{\rho} =\;&
  \begin{cases}
    (\inj_1 w, \tau_0 \pp \cdots \pp \tau_k) &\text{if } 
      v_0 = \sem{v}_{\rho}, \\    
      &\sem{q}_{[\rho, v_i/x]} = (\inj_2 \inj_x v_{i+1}, \tau_{i}), \\
      &\text{and~~}\, \sem{q}_{[\rho,v_k/x]} = (\inj_1 w, \tau_{k}) \\
    (\inj_2 \inj_y w, \tau_0 \pp \cdots \pp \tau_k) &\text{if } v_0 = \sem{v}_{\rho}, \\ 
      &\sem{q}_{[\rho,v_i/x]} = (\inj_2 \inj_x v_{i+1}, \tau_{i}), \\
      &\text{and~~}\, \sem{q}_{[\rho,v_k/x]} = (\inj_2 \inj_y w, \tau_{k}) \\ 
    \tau_0 \pp \cdots \pp \tau_{k-1} \pp \pi &\text{if } v_0 = \sem{v}_{\rho}, \\ 
      &\sem{q}_{[\rho,v_i/x]} = (\inj_2 \inj_x v_{i+1}, \tau_{i}), \\
      &\text{and~~}\, \sem{q}_{[\rho,v_k/x]} = \pi \\
    \tau_0 \pp \tau_1 \pp\cdots & \text{if } v_0 = \sem{v}_{\rho}, \\ 
      &\sem{q}_{[\rho,v_i/x]} = (\inj_2 \inj_x v_{i+1}, \tau_{i}), \\
      &\text{for all~~} i\in\nat
  \end{cases}
\end{flalign*}
  \caption{Denotational semantics over $TX=(X\times\nat^{\star}) \cup \nat^{\omega}$.}\label{fig:sem-trace}
}}
\end{figure}
In Fig.~\ref{fig:sem-trace} we give a full account of the denotational
semantics in an appropriate set-based notation for the concrete choice
of the monad $\BBT$ as above. We omit contexts and types and moreover
we index the semantic brackets with a \emph{valuation}~$\rho$ sending
each variable $x\c A$ from the context $\Gamma$ to a corresponding
element of the set~$\ul A$\lsnote{Fix the figure to do just that. SG:
  It seems that they already do just that. LS: Right. Maybe someone
  fixed it?}. That is, we assume the following equations
\begin{align*}
\sem{\Gamma\vctx p\c A}_{\rho} = \sem{\Gamma\vctx p\c A} \comp\rho, &\quad&
\sem{\Delta\csep\Gamma\cctx p\c A}_{\rho} = \sem{\Delta\csep\Gamma\cctx p\c A} \comp\rho.
\end{align*}
(that is, composition with $\rho$ is written as indexing by $\rho$.)

As usual, we have a substitution lemma saying that substitution of
terms can be replaced by calculating values of terms and
correspondingly updating the valuation. We write substitution in
postfix notation, and assume the standard notion of capture-avoiding
substitution.

\begin{lemma}[Substitution Lemma]\label{lem:subst} 
  Let $\sigma$ be a substitution sending each variable $x_i\c A_i$ from
  the context $\Gamma$ to a term $\Gamma'\vctx v_i\c A_i$, and
  let~$\rho$ be a valuation for the variables in~$\Gamma'$. Then
\begin{align*}
  \sem{\Gamma'\vctx v\sigma\c A}_\rho = \sem{\Gamma\vctx v\c A}_{\ul\sigma},&\quad&
  \sem{\Delta\csep\Gamma'\cctx p\sigma\c A}_\rho = \sem{\Delta\csep\Gamma \cctx p\c A}_{\ul\sigma}.
\end{align*}
where the valuation~$\ul\sigma$ sends each $x_i$ to
$\sem{\Gamma'\vctx v_i\c A_i}_\rho$.
\end{lemma}
\begin{proof}
  Straightforward induction over the term structure.
\end{proof}

\noindent We now can state the main result of this section as follows.
\begin{theorem}[Soundness and Adequacy]\label{thm:adequacy}
  Let $\Delta\csep - \cctx p\c B$. Then
  \begin{enumerate}%
    \item $p \Downarrow \ret v, \tau$ iff\/ $\sem{\Delta\csep - \cctx p\c B} = (\inj_1\sem{v},\tau)\in (B+\Delta)\times\nat^\star$.
    \item $p \Downarrow \oname{raise}_x v, \tau$ and $x\c E^\gtag$ is in $\Delta$ iff\/ $\sem{\Delta\csep - \cctx p\c B} = (\inj_2 \inj_x v, \tau)\in (B+\Delta)\times\nat^\mplus$.
    \item $p \Downarrow \oname{raise}_x v, \tau$ and $x\c E^\utag$ is in $\Delta$ iff\/ $\sem{\Delta\csep - \cctx p\c B} = (\inj_2 \inj_x v, \tau)\in (B+\Delta)\times\nat^\star$.
    \item $p \Downarrow \pi$ iff\/ $\sem{\Delta\csep - \cctx p\c B} = \pi\in\nat^\omega$.
  \end{enumerate}
\end{theorem}
\noindent Each clause of Theorem~\ref{thm:adequacy} is an
iff-statement in which the left-to-right direction stands for
\emph{soundness} and the right-to-left direction stands for
\emph{adequacy}. This view of soundness and adequacy in not entirely
standard and we compare it to the more established one. Suppose
that we give a big-step semantics to a deterministic language in a
system where every computation $p$ either provably evaluates to some
value~$v$ via $p\Downarrow v$, indicated by writing $p\Downarrow$, or
$p\Downarrow v$ is not provable for any~$v$, indicated by writing
$p\Uparrow$.
Denotationally, the former situation corresponds to $\sem{p}=v$ and
the latter to $\sem{p}=\bot$ for a designated divergence
constant~$\bot$. Soundness then means that $p\Downarrow v$ implies
$\sem{p}=v$, while adequacy means that $p\Uparrow$ implies
$\sem{p}=\bot$. Equivalently, by contraposition, adequacy amounts to
the implication from $\sem{p}\neq\bot$ to~$p\Downarrow$. Now,
$\sem{p}\neq\bot$ is the same as $\sem{p}=v$ for some value $v$ and
$p\Downarrow$ means that $p\Downarrow w$ for a possibly different
value~$w$.  Using soundness of the denotational semantics, we obtain
$w=v$; thus, adequacy amounts to the implication from $\sem{p}=v$
to~$p\Downarrow v$, i.e.\ the perfect converse of soundness. We argue
that the obtained reformulation of adequacy is advantageous in two
respects: it does not hinge on contraposition, which is equivalent to
excluded middle, and it does not depend on the presence of only one
type of divergence~$\bot$ -- e.g.\ in our present semantics there are
as many types of divergence as infinite traces.

We prove Theorem~\ref{thm:adequacy} analogously to~\cite{GeronLevy16}
by showing a stronger type-indexed property used as an induction
invariant in the style of Tait~\cite{Tait67}. Specifically, let us
define a predicate $\mathfrak{P}$ over all terms that are typable with
empty variable context $(\argument)$ by induction over their return types as
follows:
\begin{itemize} 
  \item if $- \vctx v\c 1$ or $- \vctx v\c \nat$ then $\mathfrak{P}(v)$;
  \item if $- \vctx v\c A$ then $\mathfrak{P}(\inl v)$ if $\mathfrak{P}(v)$;
  \item if $- \vctx v\c A$ then $\mathfrak{P}(\inr v)$ if $\mathfrak{P}(v)$;
  \item if $- \vctx v\c A$ and $- \vctx w\c B$ then $\mathfrak{P}(\brks{v,w})$ if $\mathfrak{P}(v)$ and $\mathfrak{P}(w)$;
  \item if $- \vctx \lambda x.\,p\c A\to_{\Delta} B$ then $\mathfrak{P}(\lambda x.\,p)$ if $\mathfrak{P}(v)$ implies $\mathfrak{P}(p[v/x])$ for all $- \vctx v\c A$;
  \item if $\Delta\csep- \cctx p\c A$ then $\mathfrak{P}(p)$ if one of the following
        clauses applies
  \begin{enumerate}
  \item\label{P-cl1} $\sem{\Delta\csep - \cctx p\c A} = (\inj_1 \sem{- \vctx v\c A},\tau)$, 
    $\mathfrak{P}(v)$, and $p \Downarrow \ret v, \tau$  with
    $\tau\in\nat^\star$;
  \item\label{P-cl2} $\sem{\Delta\csep - \cctx p\c A} = (\inj_2\inj_x \sem{- \vctx v\c A},\tau)$, 
    $\mathfrak{P}(v)$ and $p \Downarrow \oname{raise}_x v, \tau$
    with $\tau\in\nat^\mplus$ and $x\c E^\gtag$ in $\Delta$;
  \item\label{P-cl3} $\sem{\Delta\csep - \cctx p\c A} = (\inj_2\inj_x \sem{- \vctx v\c A},\tau)$,
    $\mathfrak{P}(v)$ and $p \Downarrow \oname{raise}_x v, \tau$ with
    $\tau\in\nat^\star$ and $x\c E^\utag$ in $\Delta$;
  \item\label{P-cl4} $\sem{\Delta\csep - \cctx p\c A} = \pi$ and $p \Downarrow \pi$ with
    $\pi\in\nat^\omega$.
\end{enumerate}
\end{itemize}
Our main technical task is to prove the following lemma:
\begin{lemma}\label{lem:adeq}
\begin{enumerate}
\item Whenever $x_1\c B_1,\ldots, x_n\c B_n\vctx v\c A$ and $-\vctx w_i\c B_i$
  such that $\mathfrak{P}(w_i)$ for $i=1,\dots,n$, then
  $\mathfrak{P}(v[w_1/x_1,\ldots,w_n/x_n])$.
\item Whenever $\Delta\csep x_1\c B_1, \ldots, x_n\c B_n\cctx p\c A$ and
  $-\vctx w_i\c B_i$ such that $\mathfrak{P}(w_i)$ for $i=1,\dots,n$,
  then $\mathfrak{P}(p[w_1/x_1,\ldots,w_n/x_n])$.
\end{enumerate}
\end{lemma}
\noindent Using Lemma~\ref{lem:adeq}, Theorem~\ref{thm:adequacy} is obtained straightforwardly: 
\begin{proof}[Proof of Theorem~\ref{thm:adequacy}]
Lemma~\ref{lem:adeq} implies that $\mathfrak{P}$ is totally true on all closed value and 
computation terms, and thus we are done by definition of $\mathfrak{P}$.
\end{proof}

\begin{proof}[Proof of Lemma~\ref{lem:adeq}]
  \lsnote{Check that this is not impacted by the new handleit rule} We
  proceed by induction over the structure of values and
  computations. We write $\sigma = [w_1/x_1,\ldots,w_n/x_n]$. During
  the proof, we make extensive use of the substitution lemma
  (Lemma~\ref{lem:subst}) without notice.  Consider the value terms.
  \begin{citemize}
  \item $v = x_i$: since $x_i\sigma = w_i$, $\mathfrak{P}(v\sigma)$ holds by assumption;
  \item for $v$ of type $1$ or $\nat$, i.e.\ $v = \star$, $v = zero$, $v = \Succ u$, 
    $\mathfrak{P}(v\sigma)$ holds by definition;
  \item for $v = \inl u$ or $v = \inr u$, $\mathfrak{P}(v\sigma)$ reduces to $\mathfrak{P}(u\sigma)$
    by induction;
  \item for $v = \brks{u,w}$, $\mathfrak{P}(v\sigma)$ reduces to $\mathfrak{P}(u\sigma)$ and 
    $\mathfrak{P}(w\sigma)$ by induction;
  \item if $v = \lambda x.\,p$ then we need to show that for every $-\vctx u\c A$
    satisfying $\mathfrak{P}$, $\mathfrak{P}(p\sigma[u/x])$ is true, and the latter follows by induction.
  \end{citemize}
Next, we analyse computation terms. 
  \begin{citemize}
  \item If $p = \ret v$ then we are done straightforwardly by induction.
  \item If $p = \Pred(v)$ then $v\sigma$ can either be $zero$ or $\Succ
    v'$. In both cases, $p \Downarrow \ret u, \brks{}$ for some $u$, and
    $\sem{\Delta\csep -\cctx \Pred(v\sigma)\c\nat} = (\inj_1 u, \brks{})$, so the first clause from the
    definition of $\mathfrak{P}$ applies.
  \item With $p = \oname{raise}_x v$ we are done immediately by induction.
  \item If $p = \gcase{\Put}{v}{\argument}{q}{x}{r}$ then by induction
    $\mathfrak{P}(r[\star/x]\sigma)$. The latter must follow from one of the four
    clauses in the definition of $\mathfrak{P}$. E.g.\ if it follows from the first
    clause then $\sem{\Delta\csep\Gamma\cctx r[\star/x]\sigma\c A}=(\inj_1 w,\tau)$
    and $r\sigma[\star/x]\Downarrow\ret w,\tau$, hence $p\sigma\Downarrow \ret
    w,\brks{v}\pp\tau$, $\sem{\Delta\csep\Gamma\cctx p[\star/x]\sigma\c A}=(\inj_1
    w,\brks{v}\pp\tau)$, and therefore $\mathfrak{P}(p\sigma)$, again by the Clause~\ref{P-cl1}
    in the definition of~$\mathfrak{P}$. The remaining three alternatives are
    checked analogously.
  \item Let $p = \pcase{v}{\brks{x,y} \mto q}$ and let $v\sigma = \brks{u,w}$.
    By induction, $\mathfrak{P}(q\sigma[u/x,w/y])$ and further analysis runs analogously
    to the previous case.
  \item Let $p = \case{v}{\inl x \mto q}{\inr y \mto r}$. Since $v\sigma$ is either 
    of the form $\inl w$ or of the form $\inr u$, by induction, in the corresponding 
    cases either $\mathfrak{P}(q[w/x])$ or $\mathfrak{P}(r[u/y])$. Each of these cases
    is analyzed analogously to the previous two clauses.
  \item If $p = \oname{init} v$ then $v\sigma$ must have $0$ as the return type,
    but there are no values of this type. Therefore, $\mathfrak{P}(p\sigma)$ is vacuously true.
  \item Let $p = (\lambda x.\,q)\, w$. Assuming that $x\sigma=x$, note that 
    $p\sigma = (\lambda x.\,q\sigma)\,  w\sigma$. By induction, $\mathfrak{P}(w\sigma)$,
    and thus, in turn, also by induction, $\mathfrak{P}(q\sigma [w\sigma/x])$. Now, on the one 
    hand
    \begin{align*}
      \sem{\Delta\csep -& \cctx p\sigma \c A} \\*
       =&\; \sem{\Delta\csep - \cctx (\lambda x.\,q\sigma) w\sigma \c A} \\
       =&\; \xi\bigl(\xi^{\mone}(\lambda a.\,\sem{\Delta\csep x\c B \cctx q\sigma \c A}_{[a/x]})\bigr)\sem{\Delta\csep - \vctx w\sigma \c B}\\
       =&\; \sem{\Delta\csep x\c B \cctx q\sigma \c A}_{[\sem{\Delta\csep - \vctx w\sigma \c B}/x]}\\
       =&\;\sem{\Delta\csep - \cctx q\sigma[w\sigma/x]\c A},
    \end{align*} 
    and on the other hand $p\sigma$ and $q\sigma [w\sigma/x]$ reduce to the same 
    terminal. Therefore $\mathfrak{P}(p\sigma)$ is equivalent to $\mathfrak{P}(q\sigma [w\sigma/x])$,
    i.e.\ true.
  \item $p = \mbind{x \gets q}{r}$. By induction hypothesis, $\mathfrak{P}(q\sigma)$.
    Depending on how $q$ reduces, we have the following cases to cover.
    \begin{itemize}[label=\textscale{.6}{$\blacklozenge$}, leftmargin=.9cm,itemindent=.6cm] 
    \item $q\sigma \Downarrow \ret v, \tau$, $\sem{\Delta\csep-\cctx q\sigma\c B} = (\inj_1\sem{v},
        \tau)$ and $\mathfrak{P}(v)$. By induction, $\mathfrak{P}(r\sigma[v/x])$. 
        Observe that either $r\sigma[v/x]\Downarrow t,\tau'$ and $p\sigma\Downarrow t,\tau\pp\tau'$
        or $r\sigma[v/x]\Downarrow\pi$ and $p\sigma\Downarrow \tau\pp\pi$ for suitable $t$, $\pi$, $\tau'$
        and analogously, either $\sem{\Delta\csep -\cctx r\sigma[v/x]\c A}=(t,\tau')$ and 
        $\sem{\Delta\csep -\cctx p\sigma\c A} = (t,\tau\pp\tau')$ or $\sem{\Delta\csep -\cctx r\sigma[v/x]\c A}=\pi$ and 
        $\sem{\Delta\csep -\cctx p\sigma\c A} = \tau\pp\pi$.
        By further case distinction over the Clauses~\ref{P-cl1}-\ref{P-cl4} in the definition
        of $\mathfrak{P}$, we conclude that $\mathfrak{P}(p\sigma)$ is equivalent 
        to $\mathfrak{P}(r\sigma[v/x])$ and therefore true.
    \item $q\sigma \Downarrow \oname{raise}_e v, \tau$, $\sem{\Delta\csep-\cctx q\sigma\c A} =
        (\inj_2 \inj_e v, \tau)$ and $\mathfrak{P}(v)$. This case is analysed completely 
        analogously to the previous one. 
    \item $q\sigma \Downarrow \pi$. By the respective operational semantic rule, 
        also $p\sigma\Downarrow \pi\c A$. Also, by definition, $\sem{\Delta\csep-\cctx p\sigma} =
        \pi$, hence $\mathfrak{P}(v)$ follows from Clause~\ref{P-cl4} in the definition of $\mathfrak{P}(v)$.
    \end{itemize}
  \item $p = \handle{x}{q}{r}$. Again, we have multiple subcases, which are analogous to the 
        cases for $p = \mbind{x \gets q}{r}$, as considered previously, with an important distinction 
        that we now have to process the exception context $\Delta$.
    \begin{itemize}[label=\textscale{.6}{$\blacklozenge$}, leftmargin=.9cm,itemindent=.6cm] 
    \item $q\sigma \Downarrow \ret v, \tau$, $\sem{\Delta,x\c E^\utag\csep-\cctx q\sigma\c A} = (\inj_1\sem{v},
        \tau)$ and~$\mathfrak{P}(v)$. By the derivation rule, we obtain $p\sigma \Downarrow
        \ret v, \tau$ and by definition, $\sem{{\Delta\csep-\cctx p\sigma\c A}} = (\inj_1\sem{v}, \tau)$,
        whence $\mathfrak{P}(p\sigma)$ holds by Clause~\ref{P-cl1} in the definition of $\mathfrak{P}$.
    \item $q\sigma \Downarrow \oname{raise}_e v, \tau$, $\sem{\Delta,x\c E^\utag\csep-\cctx q\sigma\c A} = (\inj_2
        \inj_e \sem{v}, \tau)$ and $\mathfrak{P}(v)$. If $e \neq x$, then from the
        respective operational semantic rule we know that $p\sigma \Downarrow \oname{raise}_e v, \tau$.
        Moreover, $\sem{\Delta\csep-\cctx p\sigma\c A} = (\inj_2\inj_e \sem{v}, \tau)$, and hence
        Clause~\ref{P-cl3} of the definition~$\mathfrak{P}$ can be applied to obtain $\mathfrak{P}(p\sigma)$. 
        Let us proceed under the assumption that $e = x$. Then either $r\sigma[v/x] \Downarrow \ret v', \tau'$,
        or $r\sigma[v/x] \Downarrow \oname{raise}_{e'} v', \tau'$ or $r\sigma[v/x] \Downarrow \pi$, and therefore,
        respectively, either $p\sigma\Downarrow \ret v',\tau\pp\tau'$, or $p\sigma\Downarrow \oname{raise}_{e'} v',\tau\pp\tau'$,
        or $p\sigma\Downarrow \tau\pp\pi$. Since by induction $\mathfrak{P}(r\sigma[v/x])$, in the respective
        cases we obtain that $\sem{\Delta\csep-\cctx p\sigma\c A}$ is either 
        $(\inj_1\sem{v'}, \tau\pp\tau')$ or $(\inj_2\inj_{e'}\sem{v'}, \tau\pp\tau')$ or
        $\tau\pp\pi$. Now, $\mathfrak{P}(p\sigma)$ follows by further analysis
        into the Clauses~\ref{P-cl1}-\ref{P-cl4} in the definition of $\mathfrak{P}$.
    \item $q\sigma \Downarrow \pi$. Analogous to the case for $\oname{do}$.
    \end{itemize}
  \item $p = (\handleit{v}{x}{q})$. Let $v_0=\sem{v}_\sigma$ and consider the sequence $v_0,\ldots$
      formed as follows: $\sem{\Delta, x\c E^\gtag \csep - \cctx q\sigma [v_i/x]\c A} = (\inj_2 \inj_x \sem{v_{i+1}}, \tau_{i})$. This sequence
      can either be infinite or terminate according to three different scenarios. Depending on this we proceed by case distinction. 
    \begin{itemize}[label=\textscale{.6}{$\blacklozenge$}, leftmargin=.9cm,itemindent=.6cm] 
      \item Suppose that the sequence $v_0,\ldots$ is infinite. Then, by induction $\mathfrak{P}(q\sigma [v_i/x])$ 
      for every $i$ and therefore also $q\sigma [v_i/x]\Downarrow\oname{raise}_x v_{i+1},\tau_{i+1}$
      by Clause~\ref{P-cl2} in the definition of $\mathfrak{P}$ where each $\tau_i$ is from $\nat^\mplus$.
      Now $p\sigma\Downarrow\tau_1\pp\tau_2\pp\ldots $ and we are done by Clause~\ref{P-cl4} in the definition 
      of $\mathfrak{P}$.
   \item Suppose that the sequence $v_0,\ldots$ ends with $v_k$ such that
      $\sem{\Delta, x\c E^\gtag \csep - \cctx q\sigma [v_k/x]\c A} = (\inj_2\inj_z \sem{w}, \tau_k)$ or 
      $\sem{\Delta, x\c E^\gtag \csep - \cctx q\sigma [v_k/x]\c A} = (\inj_1 \sem{w}, \tau_k)$.
      By induction, $\mathfrak{P}(q\sigma [v_i/x])$ for every $i\leq k$ and therefore 
      $q\sigma[v_k/x] \Downarrow \oname{raise}_z w, \tau_k$ or $z \neq x$ or 
      $q\sigma[v_k/x] \Downarrow \ret w, \tau_k$. By the same considerations as in the 
      previous clause, we obtain $\mathfrak{P}(p\sigma)$.
    \item The case of the sequence $v_0,\ldots$ ending with $v_k$ such that
      $\sem{\Delta, x\c E^\gtag \csep - \cctx q\sigma [v_k/x]\c A} = \pi$ is handled
      analogously to the above.
\qed
    \end{itemize}
  \end{citemize}
\noqed
\end{proof}

\section{Conclusions and Further Work}\label{sec:concl}
\noindent We have instantiated the notion of abstract
guardedness~\cite{GoncharovSchroderEtAl17,GoncharovSchroder18} to a
multivariable setting in the form of a metalanguage for guarded
iteration, which incorporates both monad-based encapsulation of
side-effects~\cite{Moggi91} and the fine-grain call-by-value
paradigm~\cite{LevyPowerEtAl02}. As a side product, this has
additionally resulted in a semantically justified unification of
(guarded) iteration and exception handling, extending previous work by
Geron and Levy~\cite{GeronLevy16}. Our denotational semantics is
\emph{generic}, and is parametrized by two orthogonal features: a
notion of computation, given as a strong monad, and a notion
of axiomatic guardedness, which serves to support guarded
iteration. The notion of guardedness can range from \emph{vacuous
  guardedness} (inducing trivial iteration, which unfolds at most
once) to \emph{total guardedness} (supported by monads equipped with a
total iteration operator, specifically \emph{Elgot monads}); the
latter case covers classical denotational semantics, since any monad
in a category of domains is Elgot~\cite{GoncharovSchroderEtAl18}. 

In contrast, our (big-step) operational semantics is specific and
addresses a concrete guarded iterative monad on $\Set$, for which we
have proved a soundness and adequacy result. This discrepancy in the
status of operational and denotational semantics is related to the
phenomenon that operational semantics generally appears to be harder
to generalize than denotational semantics. For one example, we note
that operational semantics needs to be completely reframed in a
constructive setting, where it must arguably be understood
coinductively rather than inductively~\cite{NakataUustalu15}.

In future research, we thus aim to use our present work for developing
operational accounts of computational phenomena from their
denotational models.
One prospective example is suggested by the mentioned work of Nakata
and Uustalu~\cite{NakataUustalu15}, who give a coinductive big-step
trace semantics for a while-language. We conjecture that this work has
an implicit guarded iterative monad~$\BBT$\!\textscale{.6}{$\BBR$}
under the hood, for which guardedness cannot be defined using the
standard argument based on a final coalgebra structure of the monad
because the objects $T$\!\textscale{.6}{$R$}\,$X$ are not final
coalgebras. The relevant notion of guardedness is thus to be
identified. More generally, we regard the generic denotational
semantics for our metalanguage as a guiding principle for identifying
semantic structures underlying computational phenomena found in the
wild, most importantly those that resist standard treatment, e.g.\ via
domain theory. A recent example of such identification is \emph{hybrid
  computation}, where the iterative behaviour can be organized in the
form of Elgot iteration on a suitable \emph{hybrid monad}, and the
notion of guardedness naturally corresponds to \emph{progressiveness}
of computations in time~\cite{GoncharovJakobEtAl18}.

Another direction for further research on generic soundness and
adequacy theorems is motivated by previous work on operational
semantics for languages parametrized by \emph{algebraic
  effects}~\cite{PlotkinPower01,JohannSimpsonEtAl10}, which provide
syntactic access to generic notions of side effect. We will pay
particular attention to the tension between iteration and general
recursion, of which iteration is conventionally viewed as a
light-weight counterpart. As the case of hybrid computation indicates,
in some models it is not quite clear what general recursion can mean,
and even formalizing total (unguarded) iteration presents considerable
difficulties. Nevertheless, we will explore connections between
guarded iteration and guarded recursion (in the sense of previous
work~\cite{GoncharovSchroder18}) whenever the latter can be
identified. Standardly, iteration is expressible as a combination of
recursion and second order types. We plan to explore conditions under
which this connection generalizes to the guarded setting. As a basis
for the prospective ``metalanguage for guarded recursion'' we plan to
use Levy's call-by-push-value as the most natural
candidate~\cite{Levy99}, into which fine-grain call-by-value embeds. In
view of this fact, our task can be seen as the task of enriching this
embedding with respective guarded fixpoints on both sides.

\section*{Acknowledgements}
We would like to thank anonymous referees for careful and thorough reading of
the initial submission.

\bibliographystyle{abbrvurl} %
\bibliography{monads}

\end{document}